\documentclass{article}
\usepackage{epsf}
\usepackage{natbib}
\usepackage{epubtk}
\usepackage{apjfonts}

\showlistoftablesfalse

\begin{document}

\title{Coronal Holes}

\author{\epubtkAuthorData{Steven R.\ Cranmer}{%
Harvard-Smithsonian Center for Astrophysics \\
60 Garden Street, Mail Stop 50 \\
Cambridge, MA 02138, U.S.A.}
{scranmer@cfa.harvard.edu}
{http://www.cfa.harvard.edu/~scranmer/}
}

\date{}
\maketitle

\begin{abstract}
Coronal holes are the darkest and least active regions of the
Sun, as observed both on the solar disk and above the solar limb.
Coronal holes are associated with rapidly expanding open magnetic
fields and the acceleration of the high-speed solar wind. 
This paper reviews measurements of the plasma properties in coronal
holes and how these measurements are used to reveal details about
the physical processes that heat the solar corona and accelerate
the solar wind.
It is still unknown to what extent the solar wind is fed by flux
tubes that remain open (and are energized by footpoint-driven
wave-like fluctuations), and to what extent much of the mass and
energy is input intermittently from closed loops into the
open-field regions.
Evidence for both paradigms is summarized in this paper.
Special emphasis is also given to spectroscopic and coronagraphic
measurements that allow the highly dynamic non-equilibrium
evolution of the plasma to be followed as the asymptotic conditions
in interplanetary space are established in the extended corona.
For example, the importance of kinetic plasma physics and turbulence
in coronal holes has been affirmed by surprising measurements from
the UVCS instrument on SOHO that heavy ions are heated to hundreds
of times the temperatures of protons and electrons.
These observations point to specific kinds of collisionless
Alfv\'{e}n wave damping (i.e., ion cyclotron resonance), but
complete theoretical models do not yet exist.
Despite our incomplete knowledge of the complex multi-scale plasma
physics, however, much progress has been made toward the goal
of understanding the mechanisms ultimately responsible for
producing the observed properties of coronal holes.
\end{abstract}

\epubtkKeywords{%
Alfv\'{e}n waves,
Corona,
Coronal heating,
Coronal holes,
Cyclotron resonance,
Kinetic and MHD theory,
Magnetohydrodynamic waves,
Magnetohydrodynamics,
Plasma heating,
Solar cycle,
Solar wind,
Solar wind plasma,
Solar wind turbulence,
Space plasmas,
Stellar winds,
Sun,
Turbulence,
UV radiation,
Wave-particle interactions,
Waves and instabilities}

%\newpage
%============================================================================
\section{Introduction}
\label{section:introduction}

Coronal holes are regions of low density plasma on the Sun
that have magnetic fields opening freely into the heliosphere.
Because of their low density, coronal holes tend to be the
regions of the outer solar atmosphere that are most prone to
behaving as a collisionless plasma.
Ionized atoms and electrons flow along the open magnetic fields
to form the highest-speed components of the solar wind.

The term ``coronal hole'' has come to denote several phenomena
that may not always refer to the same regions.
First, the darkest patches on the solar surface, as measured in
ultraviolet (UV) and X-ray radiation, are called coronal holes.
Second, the term also applies to the lowest-intensity regions
measured above the solar limb, seen either during a total solar
eclipse or with an occulting coronagraph.
Third, there is a more theoretical usage that equates
coronal holes with all open-field footpoints of time-steady
solar wind flows.
There are good reasons why these three ideas should be related to
one another, but the overlap between them is not complete.
To avoid possible confusion, this paper will mainly use the first
two \emph{observational} definitions, with the third one being only
partly applicable.

During times of low solar activity, when the Sun's magnetic field
is dominated by a rotationally-aligned dipole component, there
are large coronal holes that cover the north and south polar
caps of the Sun.
In more active periods of the solar cycle, coronal holes can exist
at all solar latitudes, but they may only persist for several solar
rotations before evolving into different magnetic configurations.

Despite not being as visually spectacular as active regions,
solar flares, or coronal mass ejections (CMEs), coronal holes are
of abiding interest for (at least) three main reasons.

\begin{enumerate}
\item
The extended corona and solar wind connected with coronal holes
tends to exist in an \emph{ambient time-steady} state, at least in
comparison with other regions.
This makes coronal holes a natural starting point for theoretical
modeling, since it often makes sense to begin with the simplest
regions before attempting to understand more complex and
variable structures.
\item
Coronal hole plasma has the lowest density, which makes it an
optimal testbed for studies of \emph{collisionless kinetic processes}
that are the ultimate dissipation mechanisms in many theories
of coronal heating.
Other regions tend to have higher densities and more rapid
Coulomb collisions, and thus the unique signatures of the kinetic
processes (in their particle velocity distributions) are not as
straightforward to measure as in coronal holes.
\item
Coronal holes and their associated high-speed wind streams are
also responsible for a fraction of major geomagnetic storms at 1~AU.
Corotating interaction regions (CIRs) form when fast and slow
wind streams collide with one another, and the subsequent
interaction between these structures and the Earth's magnetosphere
can give rise to long-lasting fluxes of energetic electrons.
\end{enumerate}

This paper reviews measurements of the plasma properties of
coronal holes and how these measurements have been used to put
constraints on theoretical models of coronal heating and solar
wind acceleration. 
There have been several earlier reviews that have focused mainly
on the topic of coronal holes, including
\citet{Zirker1977}, \citet{Suess1979}, \citet{HarveySheeley1979},
\citet{Parker1991}, \citet{KohlCranmer1999}, \citet{Hudson2002},
\citet{Cranmer2002a}, \citet{Ofman2005}, \citet{deTomaArge2005},
\citet{Jones2005}, and \citet{Wang2009}.
Interested readers are urged to survey these other reviews in
order to fill in any gaps in topical coverage in the present paper.

The remainder of this paper is organized as follows.
Section~\ref{section:history} gives a brief history of the
discovery and early years of research on coronal holes.
Section~\ref{section:ondisk} summarizes the observations and
derived plasma properties of ``on-disk'' coronal holes (i.e.,
primarily using the definition of holes as dark patches on
the solar surface at UV and X-ray wavelengths).
Section~\ref{section:offlimb} reviews the measurements of
``off-limb'' coronal holes and describes our current knowledge
of how these structures are linked to various kinds of
solar wind streams measured \emph{in~situ.}
Section~\ref{section:heataccel} discusses a broad range of
possible theoretical explanations for how the plasma in coronal
holes is heated and how the solar wind in these regions is
accelerated.
Section~\ref{section:summconc} concludes this paper with a
few words about how the study of coronal holes helps to improve
our wider understanding of heliophysics, astrophysics, and
plasma physics.

%\newpage
%============================================================================
\section{Historical Overview}
\label{section:history}

Although the term ``coronal hole'' was not first used until the
middle of the 20th century, people have reported the existence of
visible features associated with the Sun's corona -- seen during
total eclipses -- for centuries
\citep[see, e.g.,][]{WangSiscoe1980,Vaquero2003}.
A popular astronomy book from the first decade of the 20th century
\citep{Serviss1909} contained clear descriptions of coronal
streamers, eruptive prominences, and polar plumes in coronal holes.
The following description of the latter, from an eclipse in 1900,
conveys that early speculation may sometimes be prescient:

\emph{``The sheaves of light emanating from the poles look precisely
like the `lines of force' surrounding the poles of a magnet.
It will be noticed in this photograph that the corona appears to
consist of two portions: one comprising the polar rays just spoken
of, and the other consisting of the broader, longer, and less-defined
masses of light extending out from the equatorial and middle-latitude
zones. Yet even in this more diffuse part of the phenomenon one can
detect the presence of submerged curves bearing more or less
resemblance to those about the poles. Just what part electricity
or electro-magnetism plays in the mechanism of solar radiation it
is impossible to say, but on the assumption that it is a very
important part is based the hypothesis that there exists a direct
solar influence not only upon the magnetism, but upon the weather
of the earth''} \citep{Serviss1909}.

The first quantitative observations of coronal holes were
made by \citet{Waldmeier1956,Waldmeier1957} at the Swiss Federal
Observatory in Z\"{u}rich.
These features were identified as long-lived regions of
negligible intensity in coronagraphic (off-limb) images of the
5303 {\AA} green emission line
\citep[see also][]{Waldmeier1975,Waldmeier1981}.
Waldmeier called the features that appeared more-or-less
circular when projected onto the solar disk
\emph{L\"{o}cher} (holes), and the more elongated features
were called \emph{Kanal} (channels) or \emph{Rinne} (grooves).

In off-limb eclipse and coronagraph images, the darkest coronal hole
regions are surrounded by brighter and more complex \emph{streamers.}
These wispy structures appear to be connected to closed magnetic
field lines at the solar surface, but they are often stretched
upwards to an elongated cusp-like point, with thin ``stalks'' of
radial rays at the top.
For this reason their appearance was likened to a pointed German
helmet (or a brush-topped Greek or Roman helmet), and the common
phrase \emph{helmet streamers} is often seen.
The earliest studies of coronal morphology tended to
concentrate more on streamers than coronal holes because the
former are significantly easier to see than the latter
\citep[see, e.g.,][]{Miller1908,Mitchell1932,Newkirk1967,Pneuman1968}.
\citet{Piddington1972} outlined some early ideas about the global
structure of the ``quiet'' (i.e., solar minimum) corona.
Figure~\ref{figure:pidd_compare} compares an adaptation of
J.\  H.\  Piddington's sketch of the quiet corona to a more recent
photograph from another eclipse around solar minimum.

\epubtkImage{}{%
\begin{figure}[t]
  \def\epsfsize#1#2{0.80#1}
  \centerline{\epsfbox{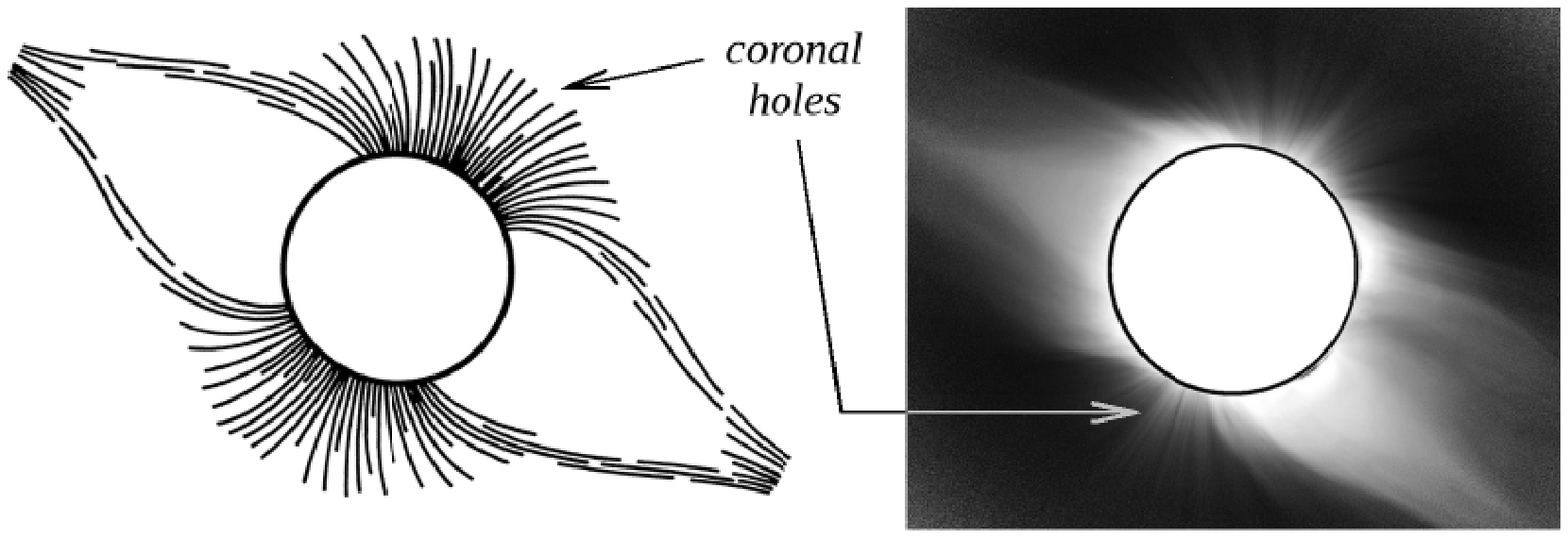}}
  \caption{Left: Adaptation of a sketch of the quiet solar
corona made by \citet{Piddington1972}, based on prior drawings
\citep{Waldmeier1955} and photographs \citep{Gold1955} of the
30 June 1954 eclipse.  Right: Contrast-adjusted eclipse image
taken with the POISE instrument on 26 February 1998, in Westpunt,
Cura\c{c}ao.  The original image was made available courtesy of
the High Altitude Observatory (HAO), University Corporation for
Atmospheric Research (UCAR), Boulder, Colorado.  UCAR is
sponsored by the National Science Foundation.}
  \label{figure:pidd_compare}
\end{figure}}

As the quality of the observations improved, coronal holes became
objects of study in their own right.
The largest coronal holes were observed to contain fine thread-like
\emph{polar plumes} that appear to follow the superradially expanding
open magnetic field lines above the solar limb
\citep{Saito1958,StoddardEtal1966,NewkirkHarvey1968}.
These elongated structures were found to correlate with bright
chromospheric faculae on the surface \citep[e.g.,][]{Harvey1965}
and with longer extensions for the small jet-like spicules
that continually rise and fall above the limb
\citep{Lippincott1957,Beckers1968}.

Coronal holes were essentially re-discovered in the late
1960s and early 1970s as discrete dark patches on the X-ray
and ultraviolet solar disk.
\citet{Newkirk1967} reviewed some of the earliest rocket-based
measurements in the extreme UV, and \citet{TouseyEtal1968}
discussed how the UV emission was ``usually weaker over the poles''
in images from a series of rocket flights between 1963 and 1967
(around solar minimum).
These regions on the solar disk came to be called coronal holes
in parallel with the earlier off-limb usage.
\citet{MunroWithbroe1972} analyzed {\em{OSO--4}} observations to
conclude that both the density and electron temperature were
lower in these dark regions.
In 1973 and 1974, solar instruments on the Apollo telescope mount
(ATM) on \emph{Skylab} confirmed many earlier ideas about coronal
holes with data significantly better in quantity and quality
\citep{HuberEtal1974,Kahler2000}.

In addition to the large north and south polar holes, there were
also found to be smaller coronal holes that exist at lower
latitudes (often at times other than solar minimum).
Sometimes the largest coronal holes can exhibit thin ``peninsulas''
that jut out from the main regions.
\citet{HarveyRecely2002} called these regions ``polar lobes.''
Notable examples have been the so-called ``Boot of Italy'' seen
by \emph{Skylab} in 1974 \citep[e.g.,][]{Zirker1977} and the
``Elephant's Trunk'' seen by \emph{SOHO} in 1996
\citep{DelZannaBromage1999}.
A third example, from December 2000, is shown in
Figure~\ref{figure:yohkoh_gorg}.

\epubtkImage{}{%
\begin{figure}[t]
  \def\epsfsize#1#2{0.37#1}
  \centerline{\epsfbox{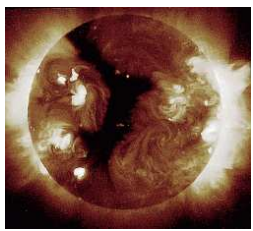}}
  \caption{X-ray corona (0.25--4.0 keV) observed by the
Soft X-ray Telescope (SXT) on Yohkoh, on 6 December 2000.
Yohkoh is a mission of the Institute of Space and Astronautical
Sciences in Japan, with participation from the U.S.\  and U.K.}
  \label{figure:yohkoh_gorg}
\end{figure}}

Additional insights came from the fusion of spectroscopy and
coronagraphic occultation.
Inspired by rocket-borne UV observations of the extended corona
during a solar eclipse in March 1970, \citet{KohlEtal1978}
developed a UV coronagraph spectrometer to measure the profile
shape of the bright H~I Lyman $\alpha$ emission line at 1216 {\AA}.
This line is sensitive to several key properties of the velocity
distribution of coronal protons, and thus these observations
could be used to begin distinguishing proton temperatures from
electron temperatures in the collisionless outer regions of
coronal holes (see Section~\ref{section:offlimb_uv}).
The rocket-borne UV coronagraph spectrometer was launched three
times (in 1979, 1980, and 1982), and the results included the
first direct evidence for proton heating and supersonic outflow
in coronal holes \citep{KohlEtal1980,StrachanEtal1993,KohlEtal2006}.

The fact that coronal holes coincide with regions of open
magnetic field that expands out into interplanetary space was
realized during the first decade of \emph{in~situ} solar wind
observations
\citep[e.g.,][]{Wilcox1968,AltschulerEtal1972,Hundhausen1972}.
\citet{Noci1973} made a theoretical argument,
on the basis of measured wave fluxes and heat conduction,
that coronal holes should have the largest solar wind kinetic
energy fluxes (i.e., the highest speeds).
\citet{Pneuman1973} argued that coronal holes need not have
lower energy deposition than closed-field regions (as is
suggested by the lower intensities of coronal holes) if the solar
wind carries away much of that energy.
\citet{KriegerEtal1973} utilized X-ray sounding rocket images
to identify a large coronal hole as the solar source of a
strong high-speed stream as measured by the \emph{Pioneer 6}
and \emph{Vela} spacecraft.
Around this time it was also realized that coronal holes and
high-speed wind streams are also responsible for a fraction of
the geomagnetic storms seen at 1~AU
\citep{BellNoci1973,NeupertPizzo1974,BellNoci1976}.
(See also \citet{TanskanenEtal2005} and \citet{ZhangEtal2007}.)
Although there is still no complete understanding of which
types of solar wind flow are connected with which types of
coronal structures, the causal link between the largest 
coronal holes and high-speed streams remains firm
(see also Section~\ref{section:offlimb_connect}).

%\newpage
%============================================================================
\section{Properties of On-Disk Coronal Holes}
\label{section:ondisk}

The traditional observational distinction between coronal holes
and their surroundings (i.e., active regions and ``quiet Sun'')
is that coronal holes have the lowest emission in the UV and X-ray.
This definition must be amended, however, to exclude \emph{filaments}
(which are often dark when projected against the solar disk)
that are cool magnetic structures and not part of the corona
\citep[see, e.g.,][]{SchollHabbal2008,KristaGallagher2009}.
Coronal hole magnetic fields are known to be more uniform and
unipolar than in other regions (see below).
The boundaries between coronal holes and surrounding regions
are sometimes sharp, sometimes diffuse, and sometimes filled
with many small loops \citep{Hudson2002}.

Coronal holes are more or less indistinguishable from their
surroundings in the photosphere and low chromosphere, and usually
one cannot see any significant intensity contrast between hole
and non-hole regions until the temperature exceeds about $10^5$ K.
Spectra, however, can help make the distinction clearer.
\citet{TeplitskayaEtal2007} found that central self-reversals in
the chromospheric Ca~II H and K lines are noticeably different in
coronal holes as opposed to surrounding quiet-Sun regions.
A frequently used observational diagnostic of coronal hole boundaries
is the He~I 10830 {\AA} near-infrared absorption line triplet
\citep{HarveyEtal1975,HarveyRecely2002}.
At these wavelengths, the absorption is weakest in coronal holes
(i.e., the intensity is highest) and spectroheliogram images show
the coronal holes quite clearly.
It is somewhat counterintuitive that a spectral line of a neutral
species (He$^{0}$) should be sensitive to the properties of the
hot corona.
However, the atomic level populations that determine the
He~I 10830 {\AA} source function are unusually responsive to
photoionization from UV wavelengths shortward of 504 {\AA}.
The overlying solar corona emits these wavelengths in abundance.
In coronal hole regions, though, the corona generally has a lower
density and temperature, and thus there is less intense UV and
X-ray emission to populate the upper levels of the
He$^{0}$ triplet state \citep[see, e.g.,][]{Zirin1975,
AndrettaJones1997,CentenoEtal2008}.
This gives rise to reduced absorption.
The He~I 10830 {\AA} lines are also good probes of supersonic
outflow velocities in distant \emph{stellar winds}
\citep[see, e.g.,][]{DupreeEtal2005,KwanEtal2007,SanzForcadaDupree2008}.

Another observational diagnostic of coronal holes is their elemental
abundance signature \citep[e.g.,][]{Feldman1998,FeldmanWiding2003}.
In the upper chromosphere, transition region, and low corona, holes
exhibit abundances very close to those measured in the photosphere.
This stands in contrast to other regions (quiet Sun and active
regions), which show significant enhancements in the abundances of
elements with low first ionization potential (FIP).
This selective fractionation is believed to occur in the upper
chromosphere, where low-FIP elements become ionized and high-FIP
elements remain more neutral.
These patterns extend into the heliosphere, where high-speed
flows associated with coronal holes are often identifiable from
their near-photospheric FIP abundances
\citep{vonSteigerEtal1995,ZurbuchenEtal2002}.
However, there is still no widespread agreement about the exact
physical processes that give rise to this preferential ionization.

The number, sizes, and heliographic locations of coronal holes vary
as a function of the solar activity cycle.
Large polar holes exist for about 7 years around solar minimum, and
are not present for about 3 or 4 years around solar maximum.
However, in the declining phase of activity soon after the maximum,
it is possible to see the gradual growth of the new-polarity polar
coronal holes.
This occurs as a number of smaller high-latitude holes eventually
collect together at the poles \citep{TimothyEtal1975,HarveyRecely2002}.
Taking this growth phase into account, there are only about 1 or 2
years at solar maximum without any distinct high-latitude coronal
hole presence.
Figure~\ref{figure:miralles_holes} illustrates the growth phase
around the peak of solar cycle 23 in 2001.
The growth and full ``reappearance'' of polar coronal holes at this
time was described by \citet{MirallesEtal2001b} and
\citet{McComasEtal2002}.
The post-maximum growth of new polar coronal holes lasts about twice
as long as their disappearance in the rising phase of the next
maximum \citep{Waldmeier1981,FisherSime1984}.

\epubtkImage{}{%
\begin{figure}[t]
  \def\epsfsize#1#2{0.77#1}
  \centerline{\epsfbox{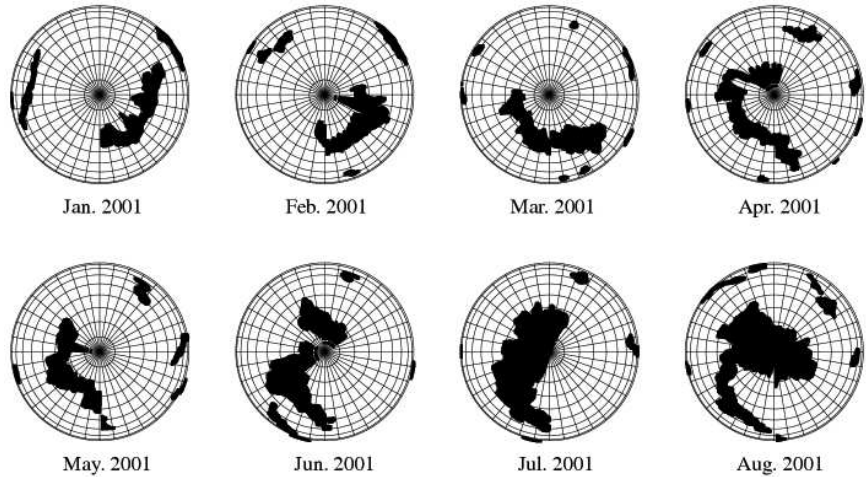}}
  \caption{Polar view of the development of the north polar
coronal hole from January to August 2001 (e.g., Carrington rotations
1972 to 1979), using reconstructed coronal hole boundaries from
Kitt Peak He~I 10830 {\AA} maps.
The maximum of solar activity occurred between late 2000 and
early 2001.
Data from the National Solar Observatory/Kitt Peak were produced
cooperatively by NSF/NOAO, NASA/GSFC, and NOAA/SEL.}
  \label{figure:miralles_holes}
\end{figure}}

Many low-latitude coronal holes tend to be situated near the
edges of magnetically complex active regions.
Sometimes active regions emerge \emph{within} the coronal holes
themselves; these have been called ``sea anemone'' type
regions from their spiky, flower-like appearance
\citep{ShibataEtal1994,AsaiEtal2008}.
Evolving magnetic interactions between active regions and
coronal holes have been studied both as a means of enhancing the
mass flux of the associated solar wind on nearby open flux tubes
\citep[e.g.,][]{HabbalEtal2008,WangEtal2009}
and as a possible explanation for the nearly rigid rotation
of coronal holes \citep{WangEtal1996,AntiochosEtal2007}.
A burst of emerging magnetic flux in one of these active regions
may give rise to new systems of loop connections in the area
bordering the coronal hole, and thus cause the coronal hole to
decrease in size.
This kind of rapid topological evolution of the magnetic field
may be relevant in explaining extreme space weather events such
as ``the day the solar wind disappeared''
\citep[i.e., the dramatic drop in the \emph{in situ} density seen
on 11 May 1999; see][]{JanardhanEtal2008}.

Photospheric magnetograms show that coronal holes are more
\emph{unipolar} than other regions; i.e., they have a larger
degree of magnetic flux imbalance between the two polarities
\citep{Levine1982,Wang2009}.
For the large polar coronal holes, this appears to be the
long-term outcome of the decay of active-region magnetic fields
and their eventual diffusion up to the poles.
The unipolar nature of coronal holes is likely to be related
to their connection with open-field solar wind streams.
As described in Section~\ref{section:history}, one reason
why coronal holes are dark is that the solar wind carries away
both mass and energy, leaving a lower density and pressure.
In addition, \citet{AbramenkoEtal2006a} and \citet{HagenaarEtal2008}
found that the local rate of emergence of small-scale magnetic flux
(mostly in the form of ``ephemeral'' small-scale bipoles) is
substantially lower in unipolar regions than in more mixed or
balanced regions of positive and negative magnetic polarity.
In most theories of coronal heating of closed loops, the total
flux and the overall complexity of the field both drive the
total heating rate.
Thus, the lower emergence rate of new flux elements in coronal
holes may be another factor determining why they have lower
densities and pressures \citep[i.e., less coronal heating;
see, e.g.,][]{RosnerEtal1978} and thus why they are dark.

Over the past few decades, the magnetic connection between
coronal holes and the fast solar wind has been traced down to 
ever smaller spatial scales.
We now can say with some certainty that much of the plasma
that eventually becomes the time-steady solar wind seems to
originate in thin magnetic flux tubes (with observed sizes
of order 50--200 km) observed mainly in the dark lanes between
the $\sim$1000 km size photospheric granulation cells.
These strong-field (1--2 kG) flux tubes have been called
G-band bright points and network bright points, and groups
of them have been sometimes termed ``solar filigree''
\citep[e.g.,][]{DunnZirker1973,Spruit1984,BergerTitle2001,
TsunetaEtal2008}.
These flux tubes are concentrated most densely in the
supergranular network (i.e., the bright lanes between
the larger $\sim$30,000 km size supergranulation cells).
Somewhere in the low chromosphere, the thin flux tubes
expand laterally to the point where they merge with one
another into a more-or-less homogeneous network field
distribution of order $\sim$100 G.
This merged (mainly vertically oriented) field is associated
mainly with the lanes and vertices between supergranular cells.
Because this field does not fill the entire coronal volume,
it is still susceptible to an additional stage of lateral
expansion and weakening.
Thus, at a larger height in the chromosphere, these network
flux bundles are believed to undergo further broadening
into so-called ``funnels'' \citep{Gabriel1976,DowdyEtal1986}.
However, it is still unclear to what extent the closed fields
in the supergranular cell centers affect the canopy-like
regions between funnels \citep[e.g.,][]{SchrijverTitle2003}.
Figure~\ref{figure:cvb_cartoon} illustrates the
successive merging of unipolar field in coronal holes.

\epubtkImage{}{%
\begin{figure}[t]
  \def\epsfsize#1#2{0.89#1}
  \centerline{\epsfbox{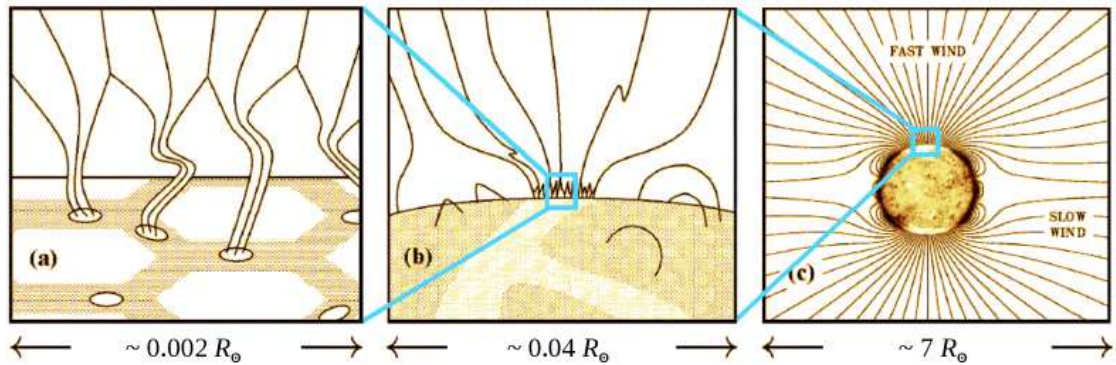}}
  \caption{Summary of the largely unipolar magnetic field
structure of polar coronal holes, with the fields of view
successively widening from flux tubes in intergranular lanes (a),
to a ``funnel'' rooted in a supergranular network lane (b),
and finally to the extended corona (c).  Adapted from Figure 1
of \citet{CranmerBallegooijen2005}.}
  \label{figure:cvb_cartoon}
\end{figure}}

Observations of blueward Doppler shifts in supergranular network
lanes and vertices, especially in large coronal holes,
may be evidence for either the ``launching'' of the solar wind
itself or for upward-going waves that are linked
to wind acceleration processes \citep[e.g.,][]{HasslerEtal1999,
PeterJudge1999,AiouazEtal2005,TuEtal2005}.
These measurements are consistent with several models of the
dynamic, multi-species solar wind in superradially expanding
funnels \citep{ByhringEtal2008,MarschEtal2008}.
However, this interpretation of the data is still not definitive,
because there are other observational diagnostics that have shown
more of a blueshift in the supergranular cell-centers \emph{between}
funnels \citep[e.g., He I 10830 {\AA};][]{DupreeEtal1996,
MalanushenkoJones2004}.
There may be subtle radiative transfer effects that preferentially
brighten regions of upflow or downflow
\citep[see, e.g.,][]{ChaeEtal1997,Avrett1999}, and these may need
to be taken into account in order to understand the meaning of
the measured Doppler shifts in these regions.

Lastly, it is important to mention the phenomenon of \emph{transient
coronal holes} (sometimes known as ``coronal dimmings'').
These are rapid reductions in the UV and X-ray intensity in
discrete patches that appear to be spatially and temporally
correlated with the liftoff of CME plasma
\citep[e.g.,][]{Rust1983,ThompsonEtal2000,YangEtal2009}.
Spectroscopically, these regions are seen to exhibit
similar characteristics as normal coronal holes, including
Doppler blueshifts \citep{HarraEtal2007} and large amplitudes
of nonthermal wave broadening \citep{McIntosh2009}.
UV coronal dimmings are beginning to be used as diagnostics
for the amount of plasma released -- i.e., the total mass -- in the
CME event \citep[e.g.,][]{AschwandenEtal2009}.
It should be noted that transient coronal holes represent just one
kind of observed dimming that is associated with time-dependent
flare/CME activity; there are other types of dimmings that do
not resemble coronal holes \citep{Hudson2002}.

%\newpage
%============================================================================
\section{Properties of Off-Limb Coronal Holes}
\label{section:offlimb}

Coronal holes observed above the solar limb usually trace out the
same regions that are identified as dark coronal-hole ``patches''
directly on the solar disk.
Thus, the lower plasma density that more or less defines the
off-limb coronal hole is directly related to the lower density
measured on the disk.
Section~\ref{section:offlimb_connect} briefly discusses how these
regions are believed to be connected magnetically with the broader
heliosphere.
The observations of off-limb coronal holes made with visible-light
imaging and polarimetry (Section~\ref{section:offlimb_white}) and
ultraviolet spectroscopy (Section~\ref{section:offlimb_uv})
are also summarized below.

\subsection{Magnetic connectivity with the solar wind}
\label{section:offlimb_connect}

Although the magnetic field in the solar corona is generally too
weak to be measured directly, the overall morphology of the field
lines can be extrapolated from magnetograms taken at the level
of the photosphere.
One popular technique is the ``potential field source surface'' (PFSS)
method, which assumes the corona is current-free out to a spherical
surface (set typically at a radius between 2.5 and 3.5 solar radii,
or $R_{\odot}$), above which the field is radial
\citep[e.g.,][]{SchattenEtal1969,AltschulerNewkirk1969,
HoeksemaScherrer1986}.
The PFSS method has been shown to create a relatively good mapping
between the Sun and the heliosphere \citep{ArgePizzo2000,
LuhmannEtal2002,WangSheeley2006,Wang2009},
although the results can be problematic for regions dominated
by stream-stream interactions \citep{PoduvalZhao2004}.

By far, the strongest causal link between a specific type of
coronal structure (measured via remote sensing) and a particular
type of quasi-steady solar wind flow (measured \emph{in situ}) is
the connection between large coronal holes and high-speed streams
\citep{Wilcox1968,KriegerEtal1973}.
The general interpretation of this correlation -- together with the
results of magnetic extrapolation models such as PFSS -- is that
coronal holes represent a bundle of open flux tubes that flare out
horizontally as distance increases.
In other words, the coronal hole flux tubes expand \emph{superradially.}
Although there are some observations that appear to support other
interpretations \citep{WooEtal1999,HabbalEtal2001,Woo2005,
WooDruckmullerova2008}, the preponderance of evidence seems
clearly to support the idea that fast solar wind streams emerge
mainly from superradially expanding coronal holes
\citep[e.g.,][]{GuhathakurtaEtal1999a,CranmerEtal1999b, Jones2005,
WangSheeley2006,WangEtal2007}.

In contrast to the rather definitive correlation between large
coronal holes and the fast solar wind, the coronal sources of the
more chaotic \emph{slow-speed} solar wind are not as well understood
\citep[see][]{Schwenn2006}.
Two regions that are frequently cited as sources of slow wind are:
(1) boundaries between coronal holes and streamers, and
(2) narrow plasma sheets that extend out from the tops of streamer
cusps \citep{WangEtal2000,StrachanEtal2002,SusinoEtal2008}.
These regions tend to dominate around solar minimum.
Note that the former type of boundary region tends to contain flux
tubes that may be classified as coronal holes when using the
theoretical definition (i.e., footpoints of field lines that are
open; see Section~\ref{section:introduction}) but would not be
defined as such when using the observational definitions (i.e.,
low emission or low density).

During more active phases of the solar cycle, there is
evidence that slow solar wind streams also emanate from small
coronal holes \citep[e.g.,][]{NolteEtal1976,NeugebauerEtal1998,
ZhangEtal2003} and active regions
\citep{HickEtal1995,LiewerEtal2004,SakaoEtal2007}.
During the rising phase of solar activity, there seems to be a
relatively abrupt ($< 6$ month) change in the locations of slow wind
footpoints: from the high-latitude hole/streamer boundaries to
the low-latitude active region and small coronal hole regions
\citep{LuhmannEtal2002}.
The ability of many of these kinds of regions to produce slow wind
was modeled by \citet{CranmerEtal2007} and \citet{WangEtal2009};
see also Section~\ref{section:heataccel}.

\subsection{Visible light observations}
\label{section:offlimb_white}

Measurements of the plasma properties in the \emph{extended corona}
(i.e., $r \approx 1.5$ to 10 $R_{\odot}$, where the main solar
wind acceleration occurs) require the bright solar disk
to be occulted.
The coronal emission is many orders of magnitude less bright
than the emission from the solar disk, so even the vast majority
of ``stray light'' that diffracts around the occulting edge
must be eliminated.
Visible-light coronagraphs that combine stray-light rejection
with linear polarimetry have the ability to measure the
Thomson-scattered polarization brightness ($pB$) in the corona.
The use of $pB$ rather than the total coronal brightness
eliminates the contribution from the dust-scattered F-corona,
which is believed to be unpolarized up to distances of
about 5 $R_{\odot}$.
Because the coronal plasma is optically thin to the
Thomson-scattered photons, $pB$ is proportional to the
line-of-sight integral of the electron density $n_e$,
multiplied by a known scattering function.
Methods for inverting this integral to derive $n_e$ as a
function of position in various coronal structures have been
developed and improved over the years \citep[e.g.,][]{vandeHulst1950,
AltschulerPerry1972,MunroJackson1977,GuhathakurtaHolzer1994,
FrazinEtal2007}.
For coronal holes, the LASCO (Large Angle and Spectrometric
Coronagraph) instrument on \emph{SOHO} has also been used to
probe the superradial expansion of open magnetic flux tubes
\citep{DeForestEtal1997,DeForestEtal2001} and the evolution
of transient polar jets \citep{WangEtal1998,WoodEtal1999}.
The White Light Coronagraphs on \emph{Spartan 201}
\citep{FisherGuhathakurta1995} and on the UVCS (Ultraviolet
Coronagraph Spectrometer) instrument aboard \emph{SOHO}
\citep[e.g.,][]{KohlEtal1995,RomoliEtal2002} have provided
electron densities between 1.5 and 5 $R_{\odot}$ in coronal holes.

Figure~\ref{figure:xne} illustrates a selection of visible-light
measurements of the electron density in coronal holes and
compares them to similar measurements of streamers and to a
semi-empirical model of the chromosphere, transition region,
and low corona \citep{AvrettLoeser2008}.
The blue coronal hole curves were adapted from the results of
\citet{FisherGuhathakurta1995} (dotted),
\citet{CranmerEtal1999b} (solid), \citet{DoyleEtal1999} (dashed),
and \citet{GuhathakurtaEtal1999b} (dot-dashed).
The red curves for equatorial helmet streamers were adapted from
the results of \citet{SittlerGuhathakurta1999} (solid) and
\citet{GibsonEtal1999} (dashed).

\epubtkImage{}{%
\begin{figure}[t]
  \def\epsfsize#1#2{0.70#1}
  \centerline{\epsfbox{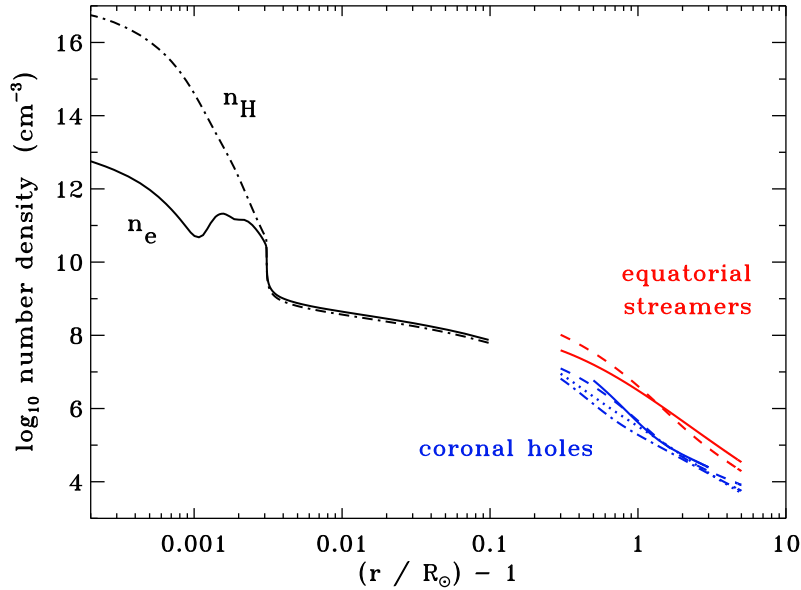}}
  \caption{Comparison of empirically determined densities in
the upper solar atmosphere.  \citet{AvrettLoeser2008} values of
electron number density (solid black curve) and total hydrogen
number density (dot-dashed black curve) are compared with various
visible-light $pB$ electron number densities for coronal holes
(blue curves) and streamers (red curves); see text for details.}
  \label{figure:xne}
\end{figure}}

Note that streamers are denser than coronal holes by about a factor
of 10, but the hole measurements themselves can often exhibit
variations in the electron density by factors of the order of 2--3.
Much of this spread is likely to be the result of different
lines of sight passing through regions that contain varying
numbers of polar plumes
\citep[see, e.g.,][]{CranmerEtal1999b,CranmerEtal2008}.
Some of the cited $pB$ observations were optimized to avoid bright
concentrations of plumes, and others have been purposefully
averaged over the full range of coronal hole substructure.
It is also possible that absolute calibration uncertainties
may still exist between the different instruments used to
determine $pB$ and $n_e$, and this could compound the reported
range of variation in coronal hole electron densities.

For a steady-state solar wind outflow, the conservation of mass
demands that the product of density, flow speed, and cross-sectional
area of the flux tube remain constant.
Thus, if the magnetic geometry and the electron density are known,
mass conservation allows the solar wind \emph{outflow speed} to
be computed.
\citet{KohlEtal2006} used the representative values of $n_e$
shown in Figure~\ref{figure:xne} together with
a range of estimates for the superradial flux-tube expansion of
coronal holes to determine outflow speeds in coronal holes.
Figure 41a of \citet{KohlEtal2006} illustrates the result of
this process, which shows a large range of values reflecting the
uncertainties in both $n_e$ and the flux-tube area factor.
Despite these uncertainties, though, the electron densities that
became available in the 1990s demonstrated that the fast solar
wind accelerates \emph{rapidly} in coronal holes -- probably
reaching half of its asymptotic terminal speed
($u_{\infty} \approx 700$--800 km s$^{-1}$) by heights no larger
than 2--4 $R_{\odot}$.

The increased sensitivity of the LASCO instrument over earlier
coronagraphs revealed a nearly continual release of low-contrast
density inhomogeneities, or ``blobs,'' from the cusps of helmet
streamers \citep{SheeleyEtal1997,TappinEtal1999,WangEtal2000,
ChenEtal2009}.
These features were seen to accelerate up to speeds of order
300--400 km s$^{-1}$ by the time they reached the outer edge of the
LASCO field of view ($r \approx 30 \, R_{\odot}$); see also
Figure~\ref{figure:Uall} below.
The blobs are typically only about 10\% to 15\% brighter or
dimmer than the surrounding streamer material.
Because of this low contrast, these features do not appear to
comprise a large fraction of the mass flux of the slow solar wind.
However, it is still unclear whether blobs are passive ``tracers''
that flow with the solar wind speed, or whether they are wavelike
fluctuations that propagate relative to the background solar wind
reference frame.
This diagnostic tool has been much more difficult to apply in coronal
holes than it has in the bright streamers, so no firm measurements
of the fast wind acceleration yet exist from this technique.

Visible light measurements have also revealed evidence for
compressive magnetohydrodynamic (MHD) waves that propagate along
open field lines in coronal holes.
Intensity oscillations measured by the UVCS and EIT instruments on
\emph{SOHO} were found to have periodicities between about 10 and
30 minutes and are consistent with being upwardly propagating
slow-mode magnetosonic waves \citep{DeForestGurman1998,
OfmanEtal1999,OfmanEtal2000}.
The relative amplitude of the density fluctuations
($\delta n / n_{0}$) for these waves was found to range
between about 0.03 and 0.15 \citep[see][]{Cranmer2004a}.
This is consistent with measurements of the density fluctuation
amplitudes made at larger distances via radio scintillations
\citep{Spangler2002} and \emph{in situ} instruments
\citep{TuMarsch1994}.
There have also been claims that low-frequency oscillations
have been measured in H~{\small I} Ly$\alpha$ emission
\citep{MorganEtal2004,BemporadEtal2008,TelloniEtal2009}.
In these cases, however, it is extremely important to take
into account all of the relevant instrumental effects.
These measurements still appear to be provisional.

As seen in Section~\ref{section:history} above, coronal holes
have long been observed as the sites of thin, ray-like
polar plumes.
The earliest measurements of polar plume properties were made
in broad-band visible light, and these dense strands are often
seen to stand out distinctly from the ambient interplume corona.
It is not clear, though, to what extent off-limb observations
(which integrate over long optically thin lines of sight)
ever capture only the ``pure'' plume or interplume plasmas.
Space-based observations from, e.g., \emph{Spartan 201} and
\emph{SOHO} improved our ability to measure the physical
properties in and between plumes
\citep[e.g.,][]{FisherGuhathakurta1995,DeForestEtal1997,
CranmerEtal1999b,DeForestEtal2001}.
Although the brightest plumes are still discernible at the
uppermost heights observed by LASCO (i.e., 30--40 $R_{\odot}$),
the plume/interplume density contrast becomes too low to
measure clearly in interplanetary space ($r > 60 \, R_{\odot}$).
However, indirect and possibly plume-related signatures in
the \emph{in situ} data have been reported by 
\citet{ThiemeEtal1990}, \citet{ReisenfeldEtal1999}, and
\citet{YamauchiEtal2002}.
The disappearance of plumes probably is the result of some
combination of transverse pressure balance
\citep[i.e., dense plumes expanding to fill more of the
available volume; see][]{DelZannaEtal1998}
and MHD instabilities that can mix the two components
\citep{ParhiEtal1999,AndriesGoossens2001}.

\subsection{Ultraviolet spectroscopy}
\label{section:offlimb_uv}

Ultraviolet spectroscopy of the corona is a powerful tool for
obtaining detailed empirical descriptions of solar plasma
conditions \citep[see, e.g.,][]{KohlWithbroe1982,WithbroeEtal1982}.
Coronal holes, being the lowest density regions of the outer
solar atmosphere, exhibit a complex array of plasma parameters
due to their nearly collisionless nature.
As a result, every ion species tends to have its own unique
temperature, its own type of departure from a Maxwellian velocity
distribution, and its own outflow speed.
After a brief summary of near-limb measurements made with the SUMER
instrument on \emph{SOHO}, this section mainly describes results
at larger heights (in the more clearly collisionless extended
corona) from UVCS.

The un-occulted SUMER (Solar Ultraviolet Measurements of Emitted
Radiation) spectrometer has observed off-limb regions up to heights
of approximately 1.3 $R_{\odot}$ in coronal holes
\citep{WilhelmEtal1995,WilhelmEtal2000,WilhelmEtal2004,WilhelmEtal2007}.
In polar regions at solar minimum, ion temperatures exceed
electron temperatures even at $r \sim 1.05 \, R_{\odot}$, where
densities were presumed to be so high as to ensure rapid
collisional coupling and thus equal temperatures for all species
\citep{SeelyEtal1997,TuEtal1998,LandiCranmer2009}.
Spectroscopic evidence is also mounting for the presence
of transverse Alfv\'{e}n waves propagating into the corona
\citep{BanerjeeEtal1998,BanerjeeEtal2009,
DollaSolomon2008,LandiCranmer2009}.

Electron temperatures derived from line ratios
\citep{HabbalEtal1993,DavidEtal1998,DoschekEtal2001,Wilhelm2006,
Landi2008} exhibit relatively low values in the off-limb coronal
holes ($T_{e} \sim 800,000$ K) that are not in agreement with
higher temperatures derived from ``frozen-in'' \emph{in situ}
charge states \citep{KoEtal1997}.
It is difficult to reconcile these observations with one another
in the absence of either non-Maxwellian electron velocity
distributions or strong differential flows between different ion
species near the Sun \citep{EsserEdgar2000}.
However, some models consistent with both the SUMER temperatures
and the \emph{in situ} charge states are being produced
\citep[e.g.,][]{LamingLepri2007}.

Figure~\ref{figure:Tall} displays a range of temperatures measured
in coronal holes and high-speed wind streams, and it shows how
the SUMER electron temperatures ($T_e$) are noticeably lower than
the heavy ion temperatures (e.g., shown for the O$^{+5}$ ions that
correspond to the bright O~VI 1032, 1037 {\AA} spectral lines)
even in the low corona.
The degree of agreement between the spectroscopic measurements
and one-fluid models of the low corona (i.e., the semi-empirical
model of \citet{AvrettLoeser2008} and the theoretical model of
\citet{CranmerEtal2007}) depends on the height of the
sharp transition region between chromospheric (10$^4$ K) and
coronal (10$^6$ K) temperatures.

\epubtkImage{}{%
\begin{figure}[t]
  \def\epsfsize#1#2{0.70#1}
  \centerline{\epsfbox{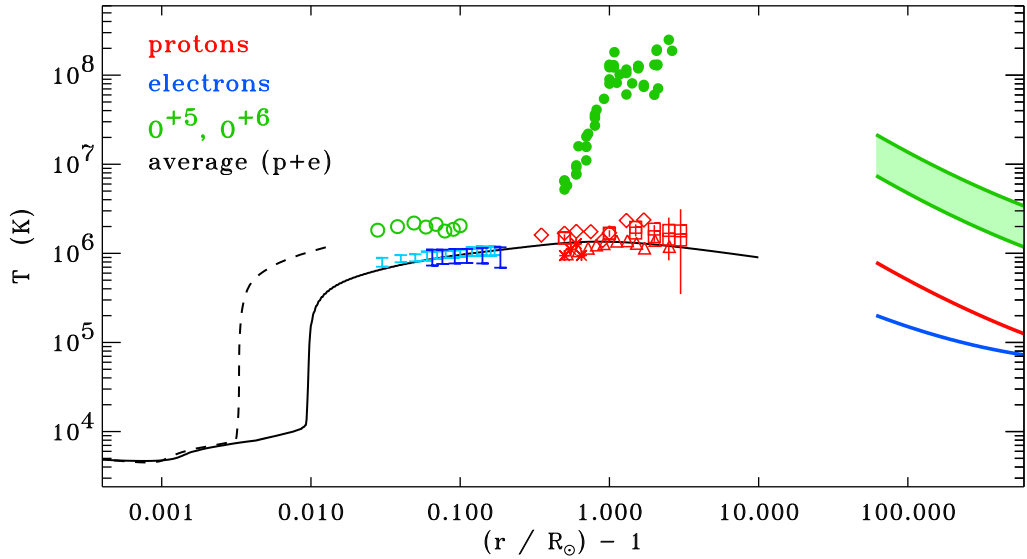}}
  \caption{Radial dependence of empirical and model temperatures
in polar coronal holes and fast wind streams.
Mean plasma temperatures from a semi-empirical model
\citep[dashed black curve;][]{AvrettLoeser2008} and from a
turbulence-driven coronal heating model
\citep[solid black curve;][]{CranmerEtal2007}.
$T_e$ from off-limb SUMER measurements made by \citet{Wilhelm2006}
(dark blue bars) and \citet{Landi2008} (light blue bars),
$T_p$ from UVCS measurements assembled by \citet{Cranmer2004b}
(see text), and perpendicular O$^{+5}$ ion temperatures from
\citet{LandiCranmer2009} (open green circles) and
\citet{CranmerEtal2008} (filled green circles).
In~situ proton and electron temperatures in the fast wind
($r > 60 \, R_{\odot}$) are from \citet{CranmerEtal2009}.}
  \label{figure:Tall}
\end{figure}}

The UVCS instrument on \emph{SOHO} is a combination of an
ultraviolet spectrometer and a linearly occulted coronagraph
that observes a 2.5 $R_{\odot}$ long swath of the extended corona,
oriented tangentially to the radial direction at heliocentric
distances ranging between about 1.5 and 10 $R_{\odot}$
\citep{KohlEtal1995,KohlEtal1997,KohlEtal1998,KohlEtal2006}.
In coronal holes, UVCS measurements have allowed many key details
about the velocity distributions of H$^0$, O$^{+5}$, and Mg$^{+9}$
to be derived.
For the resonantly scattered emission lines seen at large heights
with UVCS, the most straightforward plasma diagnostic is to use
the Doppler-broadened \emph{line width} as a sensitive probe of the
overall variance of random particle motions along the line of sight.
In other words, measuring the line width provides a constraint
on the so-called ``kinetic temperature'' (i.e., a combination of
microscopic stochastic motions and macroscopic [but unresolved]
motions due to waves or turbulence) along the direction
perpendicular to the (nearly radial) magnetic field lines.

In the ionized solar corona, a given hydrogen nucleus spends most
of its time as a free proton, and only a small fraction of time as
a bound H$^0$ atom.
Thus, the measured plasma properties of neutral hydrogen are
considered to be valid proxies of the proton properties below
about 3 $R_{\odot}$ \citep{AllenEtal2000}.
\emph{Spartan 201} and UVCS observations of the
H~{\small I} Ly$\alpha$ emission line in coronal holes indicated
rather large proton kinetic temperatures in the direction
perpendicular to the magnetic field ($T_{p \perp} \sim 3$ MK)
and also the possibility of a mild temperature anisotropy (with
$T_{p \perp} > T_{p \parallel}$) above heights of 2--3 $R_{\odot}$
\citep{KohlEtal1997,CranmerEtal1999b,AntonucciEtal2004,KohlEtal2006}.

UVCS observations indicated that the O$^{+5}$ ions are much more
strongly heated than protons in coronal holes, with perpendicular
temperatures in excess of 200 MK
(see Figure~\ref{figure:uvcs_classic}).
This exceeds the temperature at the \emph{central core} of the
Sun by an order of magnitude!
The UVCS measurements also provided signatures of temperature
anisotropies possibly greater than
$T_{\perp i} / T_{\parallel i} \approx 10$
\citep[e.g.,][]{CranmerEtal1999b,CranmerEtal2008}.
The measured kinetic temperatures of O$^{+5}$ and Mg$^{+9}$
are significantly greater than mass-proportional when
compared with protons, with $T_{i}/T_{p} > m_{i}/m_{p}$
\citep[see also][]{KohlEtal1999,KohlEtal2006}.
The surprisingly ``extreme'' properties of heavy ions in
coronal holes have led theorists to develop a number of new
ideas regarding the heating and acceleration of the solar wind;
these are discussed further in Section
\ref{section:heataccel_kinetic}.

\epubtkImage{}{%
\begin{figure}[t]
  \def\epsfsize#1#2{0.84#1}
  \centerline{\epsfbox{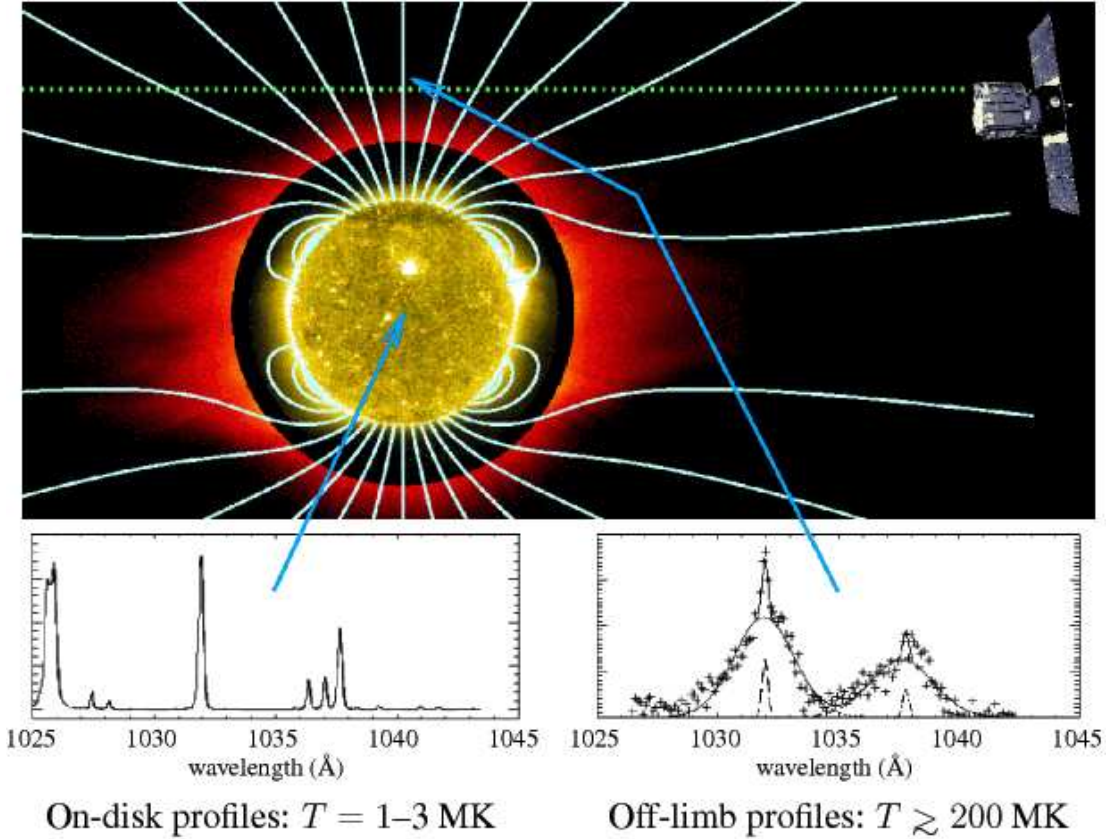}}
  \caption{Combined image of the solar corona from
17 August 1996, showing the solar disk in Fe~XII 195 {\AA}
intensity from EIT (yellow inner image) and the extended corona
in O~VI 1032 {\AA} intensity from UVCS (red outer image).
Axisymmetric field lines are from the solar-minimum model of
\citet{BanaszkiewiczEtal1998}, and O~VI emission line profiles
(bottom) are from SUMER \citep[left]{WarrenEtal1997} and UVCS
\citep[right]{KohlEtal1997}.}
  \label{figure:uvcs_classic}
\end{figure}}

Figure~\ref{figure:Tall} shows UVCS perpendicular temperatures
for protons and O$^{+5}$ ions in coronal holes.
The O$^{+5}$ data points were taken from the recent re-analysis
of solar minimum data from 1996--1997 by \citet{CranmerEtal2008}.
The proton temperature data were assembled by \citet{Cranmer2004b}
from a number of individual measurements of the H~I Ly$\alpha$
profile at solar minimum.
The sources of these measurements are:
\citet{CranmerEtal1999b} (squares),
\citet{EsserEtal1999} (diamonds),
\citet{ZangrilliEtal1999} (asterisks),
and \citet{AntonucciEtal2000} (triangles).
The \emph{kinetic} proton temperatures are of order 2--3.5 MK,
but in Figure~\ref{figure:Tall} we attempted to remove the
contribution of nonthermal wave broadening.
The semi-empirical model of \citet{CranmerBallegooijen2005}
was used to specify the amplitude of transverse Alfv\'{e}n waves
as a function of height, and their contribution to the line widths
was subtracted.
The remaining ``microscopic'' $T_{\perp p}$ does not show as
clear a signal of ``preferential'' proton heating as would be
apparent from the larger kinetic temperature.
Although one can still marginally see that $T_{p} > T_{e}$,
the existing measurements of $T_p$ and $T_e$ do not fully overlap
with one another in radius.
Improved measurements are needed in order to better constrain
the proton and electron heating rates in the corona.

The UVCS emission line data contain information about the Doppler
motions of atoms and ions \emph{along} the magnetic field (i.e.,
transverse to the line of sight).
The so-called ``Doppler dimming'' diagnostic technique provides
constraints on both the bulk outflow speed along the field
and the parallel kinetic temperature
\citep[for more details, see][]{KohlWithbroe1982,NociEtal1987,
KohlEtal2006}.
In coronal holes, Doppler-dimmed line intensities from UVCS are
consistent with the outflow velocity for O$^{+5}$ being larger than
the outflow velocity for protons by as much as a factor of two
at large heights \citep{KohlEtal1998,LiEtal1998,CranmerEtal1999b}.
Figure~\ref{figure:Uall} illustrates the outflow speeds measured
by UVCS in coronal holes, and compares with the theoretical model
of the fast solar wind presented by \citet{CranmerEtal2007}.
Also shown for comparison are observational and theoretical
data for the slow solar wind associated with equatorial helmet
streamers at solar minimum.

\epubtkImage{}{%
\begin{figure}[t]
  \def\epsfsize#1#2{0.70#1}
  \centerline{\epsfbox{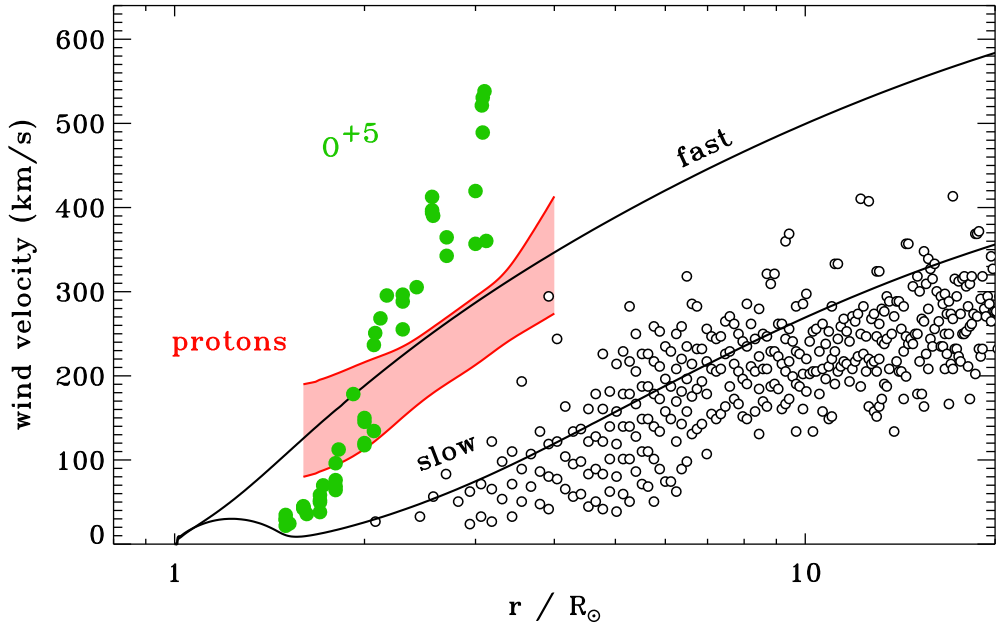}}
  \caption{Radial dependence of solar wind outflow speeds.
UVCS Doppler dimming determinations for protons
\citep[red;][]{KohlEtal2006} and O$^{+5}$ ions
\citep[green;][]{CranmerEtal2008} are shown for polar coronal
holes, and are compared with theoretical models of the polar
and equatorial solar wind at solar minimum
\citep[black curves;][]{CranmerEtal2007} and the speeds of
``blobs'' measured by LASCO above equatorial streamers
\citep[open circles;][]{SheeleyEtal1997}.}
  \label{figure:Uall}
\end{figure}}

In contrast to many prior analyses of UVCS data, which concluded
that there must be both intense preferential heating of the
O$^{+5}$ ions and a strong field-aligned anisotropy,
\citet{RaouafiSolanki2004}, \citet{RaouafiSolanki2006}, and
\citet{RaouafiEtal2007} reported that there may not be a
compelling need for O$^{+5}$ anisotropy depending on the
assumptions made about the other plasma properties of the
coronal hole (e.g., electron density).
However, \citet{CranmerEtal2008} performed a detailed re-analysis
of these observations and concluded that there remains strong
evidence in favor of both preferential O$^{+5}$ heating and
acceleration and significant O$^{+5}$ ion anisotropy (in the
sense $T_{\perp i} > T_{\parallel i}$) above
$r \approx 2.1 \, R_{\odot}$ in coronal holes.
In determining these properties, it was found to be important
to search the full range of possible ion temperatures and flow
speeds, and not to make arbitrary assumptions about any given
subset of the parameters.

The UVCS results discussed above are similar in character to
\emph{in situ} measurements made in the fast solar wind, but they
imply more extreme departures from thermodynamic equilibrium in
the extended corona.
For example, proton velocity distributions measured in the fast
solar wind between 0.3 and 1~AU have anisotropic cores with
$T_{p \perp} > T_{p \parallel}$, and their magnetic moments
increase with increasing distance; this implies net input of
perpendicular energy on kinetic scales
\citep[e.g.,][]{MarschEtal1982,Marsch2006}.
Many heavy ions appear to flow faster than the protons by about
the local Alfv\'{e}n speed \citep{HeftiEtal1998,ReisenfeldEtal2001},
and in the fastest solar wind streams they are also heated
preferentially in the same sense as in the corona
\citep{CollierEtal1996}.

In the years since the solar minimum of 1996--1997, UVCS observed
a large number of other coronal holes that appeared throughout
the maximum of solar cycle 23 and the new-millennium solar minimum
of 2007--2009.
UVCS tends to observe only the largest coronal holes, since when
the smallest holes rotate to the solar limb their UV line profiles
tend to be contaminated by emission from streamers in the foreground
and background.
This selection effect naturally screens out small coronal holes
that have been correlated with slow solar wind streams at 1~AU
\citep{NolteEtal1976,NeugebauerEtal1998}.
In the cases where UVCS and \emph{in situ} measurements were
made for the same regions associated with large coronal holes,
high speeds in excess of 600 km s$^{-1}$ were inevitably seen
in interplanetary space \citep{MirallesEtal2004,MirallesEtal2006}.
However, the O$^{+5}$ outflow speeds measured in the extended
corona via Doppler dimming have showed a substantial range of
variation.
For example, \citet{MirallesEtal2001a} found that the outflow
speeds at 2--3 $R_{\odot}$ in an equatorial coronal hole were
approximately three times lower than those measured in the
polar coronal holes from 1996--1997 at similar heights
\citep[see also][]{PolettoEtal2002}.
This implies that the range of coronal heights over which the
acceleration of the solar wind occurs can vary greatly, even when
the wind at 1~AU ends up similarly fast.

UVCS also has measured the plasma properties in bright polar plumes.
The densest concentrations of polar plumes along the line of sight
are seen to exhibit narrower line widths -- i.e., lower kinetic
temperatures -- than the lower-density interplume regions
\citep{KohlEtal1997,NociEtal1997,KohlEtal2006}.
Similarly, plumes are seen to have lower outflow speeds than
the interplume regions \citep{GiordanoEtal2000,TeriacaEtal2003,
RaouafiEtal2007}, although at low heights the data are not as
clear-cut \citep[e.g.,][]{GabrielEtal2003}.
UVCS also has put constraints on the plume/interplume density
contrast and the filling factor of polar plumes in coronal holes.
\citet{CranmerEtal1999b} used a large number of UVCS synoptic
measurements to determine statistically that, at heights around
$r \approx 2 \, R_{\odot}$, the plume/interplume density ratio
is approximately 2, and polar coronal holes are comprised of
about 25\% plume and 75\% interplume plasma (corresponding to
$\sim$40 individual plumes distributed throughout the coronal hole).
Earlier measurements made closer to the limb showed a higher
density contrast and a smaller filling factor, so the UVCS data
are generally consistent with \emph{lateral expansion} of polar
plumes with increasing distance.

It is still relatively unknown how much of the mass, momentum,
and energy flux of the fast solar wind comes from polar plumes.
Despite that uncertainty, though, there have been several
reasonably successful models of polar plume formation.
\citet{Wang1994,Wang1998} presented models of polar plumes as the
extensions of flux tubes with concentrated bursts of added coronal
heating at the base -- presumably via nanoflare-like reconnection
events in X-ray bright points \citep[see also][]{DeForestEtal2001}.
In these models, the extra basal heat input is balanced
by conductive losses to produce the larger plume density.
The heating rate in the extended corona is not affected
by the basal burst, but the larger density in the flux tube
gives rise to less heating \emph{per particle} at all heights,
which leads to lower temperatures in the extended corona and a
smaller gas pressure force for solar wind acceleration.
This model is consistent with the smaller temperatures and
outflow speeds measured in plumes with UV spectroscopy.

UVCS made the first spectroscopic measurements of \emph{polar jets}
in coronal holes \citep{DobrzyckaEtal2000,DobrzyckaEtal2006}.
The events observed by EIT, LASCO, and UVCS during the 1996--1997
solar minimum tended to be ``cool jets'' with higher densities,
lower temperatures, and faster outflows than the surrounding
coronal holes \citep[see also][]{WangEtal1998}.
More recently, \emph{Hinode} has observed a new population of
``hot'' X-ray jets in coronal holes
\citep{CulhaneEtal2007,CirtainEtal2007,ShimojoEtal2007,
FilippovEtal2009}.
UVCS found that some of these events persist up to heights of
at least 1.7 $R_{\odot}$ and that the jet protons remain hotter
than the surrounding coronal hole \citep{MirallesEtal2007}.
Thus, there appear to exist two distinct kinds of polar jets
(cool and hot), with differences possibly related to the relative
degrees of heating and adiabatic expansion of the jet parcels.
Jets and plumes have roughly similar angular sizes and
intensity contrasts near the solar limb, and there is growing
evidence that they share a common origin
\citep[e.g.,][]{RaouafiEtal2008}.
It is possible that the only substantial difference between the
two phenomena is the \emph{duration} of the bursts of basal heating;
i.e., jets seem to be the result of short-lived bursts of heating,
whereas plumes may be the product of base-heating events that
last longer than several hours.

%\newpage
%============================================================================
\section{Coronal Heating and Solar Wind Acceleration}
\label{section:heataccel}

Despite more than a half-century of study, the basic physical
processes that are responsible for heating the million-degree
corona and accelerating the supersonic solar wind are not known.
This section broadens the topic of this paper a bit beyond just
coronal holes, since an understanding of solar wind acceleration
naturally encompasses not only the question of why fast solar
wind streams are fast, but also why (various kinds of) slow
solar wind streams are slow.
Section~\ref{section:heataccel_debate} summarizes some of the
major issues regarding coronal energy deposition.
The next two subsections describe two alternate views of solar
wind acceleration via waves and turbulence in open flux tubes
(Section~\ref{section:heataccel_wtd}) and reconnection between
open and closed flux tubes (Section~\ref{section:heataccel_rlo}).
Lastly, Section~\ref{section:heataccel_kinetic} reviews how
collisionless kinetic effects in coronal holes (i.e., preferential
ion heating and temperature anisotropies) can be used to more
conclusively identify the detailed physical processes that produce
the solar wind.

\subsection{Sources of energy}
\label{section:heataccel_debate}

Different physical mechanisms for heating the corona probably
govern active regions, closed loops in the quiet corona, and
the open field lines that give rise to the solar wind
\citep[see other reviews by][]{Marsch1999,HollwegIsenberg2002,
Longcope2004,Gudiksen2005,Aschwanden2006,Klimchuk2006}.
The ultimate source of the energy is the solar convection zone
\citep[e.g.,][]{AbramenkoEtal2006b,McIntoshEtal2007}.
A key aspect of solving the ``coronal heating problem'' is thus
to determine how a small fraction of that mechanical energy is
transformed into magnetic free energy and thermal energy above
the photosphere.
It seems increasingly clear that loops in the low corona are
heated by small-scale, intermittent magnetic reconnection that
is driven by the continual stressing of their magnetic footpoints.
However, the extent to which this kind of impulsive energy
addition influences the acceleration of the solar wind is not
yet known.

Intertwined with the coronal heating problem is the heliophysical
goal of being able to make accurate predictions of
how both fast and slow solar wind streams are accelerated.
Empirical correlation techniques have become more sophisticated
and predictively powerful \citep[e.g.,][]{WangSheeley1990,
WangSheeley2006,ArgePizzo2000,LeamonMcIntosh2007,CohenEtal2007,
VrsnakEtal2007} but they are limited because they do
not identify or utilize the physical processes actually responsible
for solar wind acceleration.
There seem to be two broad classes of physics-based models that
attempt to self-consistently answer the question:
\emph{``How are fast and slow wind streams heated and accelerated?''}

\begin{enumerate}
\item
In {\bf wave/turbulence-driven (WTD) models,}
it is generally assumed that the convection-driven jostling of magnetic
flux tubes in the photosphere drives wave-like fluctuations that
propagate up into the extended corona.
These waves (usually Alfv\'{e}n waves) are often proposed to
partially reflect back down toward the Sun, develop into strong
MHD turbulence, and dissipate over a range of heights.
These models also tend to explain the differences between fast and
slow solar wind \emph{not} by any major differences in the lower
boundary conditions, but instead as an outcome of different rates of
lateral flux-tube expansion over several solar radii as the wind
accelerates
\citep[see, e.g.,][]{Hollweg1986,WangSheeley1991,MatthaeusEtal1999,
Cranmer2005,Suzuki2006,SuzukiInutsuka2006,CranmerEtal2007,
VerdiniVelli2007,VerdiniEtal2009}.
\item
In {\bf reconnection/loop-opening (RLO) models,}
the flux tubes feeding the solar wind are assumed to be influenced
by impulsive bursts of mass, momentum, and energy addition in the
lower atmosphere.
This energy is usually assumed to come from magnetic reconnection
between closed, loop-like magnetic flux systems (that are in the
process of emerging, fragmenting, and being otherwise jostled by
convection) and the open flux tubes that connect to the solar wind.
These models tend to explain the differences between fast and slow
solar wind as a result of qualitatively different rates of flux
emergence, reconnection, and coronal heating at the basal footpoints
of different regions on the Sun
\citep[see, e.g.,][]{AxfordMcKenzie1992,AxfordMcKenzie1997,
FiskEtal1999,RyutovaEtal2001,MarkovskiiHollweg2002,
MarkovskiiHollweg2004,Fisk2003,SchwadronMcComas2003,WooEtal2004,
FiskZurbuchen2006}.
\end{enumerate}

It is notable that both the WTD and RLO models have recently passed
some basic ``tests'' of comparison with observations.
Both kinds of model have been shown to be able to produce
fast ($v > 600$ km/s), low-density wind from coronal holes and
slow ($v < 400$ km/s), high-density wind from streamers rooted
in quiet regions.
Both kinds of model also seem able to reproduce the observed
\emph{in~situ} trends of how frozen-in charge states and the
FIP effect vary between fast and slow wind streams.

The fact that both sets of ideas described above seem to mutually
succeed at explaining the fast/slow solar wind \emph{could}
imply that a combination of both ideas would work best.
However, it may also imply that the existing models do not yet
contain the full range of physical processes -- and that once these
are included, one or the other may perform noticeably better than
the other.
It also may imply that the comparisons with observations have
not yet been comprehensive enough to allow the true differences
between the WTD and RLO ideas to be revealed.

Several recent observations have pointed to the importance of
understanding the relationships and distinctions between the
WTD and RLO models.
The impulsive polar jets discussed in Section~\ref{section:offlimb}
may be evidence that that magnetic reconnection drives some
fraction of the fast solar wind \citep[see also][]
{Fisk2005,MorenoInsertisEtal2008,PariatEtal2009}.
Also, direct observations of Alfv\'{e}n waves above the solar limb
indicate the highly intermittent nature of how kinetic energy is
distributed in spicules, loops, and the open-field corona
\citep{DePontieuEtal2007,TomczykEtal2007,TomczykMcIntosh2009}.
Spectroscopic observations of blueshifts in the chromospheric
network have long been interpreted as the launching points of
solar wind streams, but it remains unclear how nanoflare-like
events or loop-openings contribute to the interpretation of these
diagnostics \citep{HeEtal2007,AschwandenEtal2007,McIntoshEtal2007}.
Even out in the \emph{in~situ} solar wind -- far above the roiling
``furnace'' of flux emergence at the Sun -- there remains evidence
for ongoing reconnection \citep{GoslingEtal2005,GoslingSzabo2008}.
There is also evidence that the dominant range of turbulence
timescales measured in interplanetary space (i.e., tens of minutes
to hours) is related to the timescale of flux cancellation in
the low corona \citep{Hollweg1990,Hollweg2006}.

Determining whether the WTD or RLO paradigm -- or some combination
of the two -- is the dominant cause of global solar wind variability
is a key prerequisite to building physically realistic predictive
models of the heliosphere.
Many of the widely-applied global modeling codes
\citep[e.g.,][]{RileyEtal2001,RoussevEtal2003,TothEtal2005,
UsmanovGoldstein2006,FengEtal2007}
continue to utilize relatively simple empirical prescriptions
for coronal heating in the energy conservation equation.
Improving the identification and characterization of the
key physical processes will provide a clear pathway for
inserting more physically realistic coronal heating ``modules''
into three-dimensional MHD codes.

\subsection{The Wave/Turbulence-Driven (WTD) solar wind idea}
\label{section:heataccel_wtd}

There has been substantial work over the past few decades devoted
to exploring the idea that the plasma heating and wind acceleration
along open flux tubes may be explained as a result of wave damping
and turbulent cascade.
No matter the relative importance of reconnections and
loop-openings in the low corona, we do know that waves and turbulent
motions are present everywhere from the photosphere to the heliosphere,
and it is important to determine how they affect the mean state
of the plasma.
A review of the observational evidence for waves and turbulence in
the solar wind is beyond the scope of this paper, but several
recent reviews of the remote-sensing and \emph{in~situ} data
include \citet{TuMarsch1995}, \citet{MullanYakovlev1995},
\citet{GoldsteinEtal1997}, \citet{Roberts2000}, \citet{Bastian2001},
and \citet{Cranmer2002a,Cranmer2004a,Cranmer2007}.
Although this subsection mainly describes recent work by the author,
these results would not have been possible without earlier work
on wave/turbulent heating by, e.g.,
\citet{Coleman1968}, \citet{Hollweg1986}, \citet{HollwegJohnson1988},
\citet{Isenberg1990}, \citet{LiEtal1999}, \citet{MatthaeusEtal1999},
\citet{DmitrukEtal2001,DmitrukEtal2002}, and many others.

\citet{CranmerEtal2007} described a set of models in which the
time-steady plasma properties along a one-dimensional magnetic
flux tube are determined.
These model flux tubes are rooted in the solar photosphere and
are extended into interplanetary space.
The numerical code developed in that work, called ZEPHYR,
solves the one-fluid equations of mass, momentum,
and energy conservation simultaneously with transport
equations for Alfv\'{e}nic and acoustic wave energy.
ZEPHYR is the first code capable of producing self-consistent
solutions for the photosphere, chromosphere, corona, and solar wind
that combine:
(1) shock heating driven by an empirically
guided acoustic wave spectrum, (2) extended heating from
Alfv\'{e}n waves that have been partially reflected, then damped
by anisotropic turbulent cascade, and (3) wind acceleration
from gradients of gas pressure, acoustic wave pressure, and
Alfv\'{e}n wave pressure.

The only input ``free parameters'' to ZEPHYR are the photospheric
lower boundary conditions for the waves and the radial dependence
of the background magnetic field along the flux tube.
The majority of heating in these models comes from the
turbulent dissipation of partially reflected Alfv\'{e}n waves
\citep[see also][]{MatthaeusEtal1999,DmitrukEtal2002,
VerdiniVelli2007,ChandranEtal2009a}.
Photospheric measurements of the horizontal motions of
strong-field intergranular flux concentrations (i.e.,
G-band bright points) were used to constrain the Alfv\'{e}n
wave power spectrum at the lower boundary.
This empirically determined power spectrum is dominated by wave
periods of order 5--10 minutes.
It is important to note, however, that radio and \emph{in~situ}
measurements find that most of the fluctuation power in the solar
wind is at lower frequencies (i.e., periods of hours).
We still do not yet know (1) if the shape of the power spectrum
evolves significantly between the lower solar atmosphere and
interplanetary space, or (2) if some low-frequency power is
missed by the existing measurements of G-band bright point motions.
In any case, as seen below, the resulting wave reflection and
turbulent dissipation that comes from just the 5--10 minute periods
appear to be sufficient to explain the observed levels of coronal
heating and solar wind acceleration.

Non-WKB wave transport equations were solved to determine
the degree of linear reflection at heights above the
photospheric base \citep[see][]{CranmerBallegooijen2005}.
The resulting values of the Elsasser amplitudes $Z_{\pm}$, which
denote the energy contained in upward ($Z_{-}$) and downward
($Z_{+}$) propagating waves, were then used to constrain the
energy flux in the cascade.
\citet{CranmerEtal2007} used a phenomenological form for the damping
rate that has evolved from studies of Reduced MHD and
comparisons with numerical simulations.
The resulting heating rate (in units of erg s$^{-1}$ cm$^{-3}$)
is given by
\begin{equation}
  Q \, = \, \rho \, \left(
  \frac{1}{1 + [ t_{\rm eddy} / t_{\rm ref} ]^{n}} \right) \,
  \frac{Z_{-}^{2} Z_{+} + Z_{+}^{2} Z_{-}}{4 L_{\perp}}
  \label{eq:Qturb}
\end{equation}
where $\rho$ is the mass density and $L_{\perp}$ is an
effective perpendicular correlation length of the turbulence
\citep[see, e.g.,][]{HossainEtal1995,ZhouMatthaeus1990,
BreechEtal2008,PodestaBhattacharjee2009,BeresnyakLazarian2009}.
\citet{CranmerEtal2007} used a standard assumption that
$L_{\perp}$ scales with the cross-sectional width of the
flux tube \citep{Hollweg1986}.
The term in parentheses above is an efficiency factor that
accounts for situations in which the cascade does not have time to
develop before the waves or the wind carry away the energy
\citep{DmitrukMatthaeus2003}.
In open field regions, the cascade is ``quenched''
when the nonlinear eddy time scale $t_{\rm eddy}$ becomes much
longer than the macroscopic wave reflection time scale $t_{\rm ref}$.
In closed field regions, the correction factor may behave in an
opposite sense as it does for open field regions
\citep[see, e.g.,][]{GomezEtal2000,RappazzoEtal2008}.
In most of the solar wind models, though, \citet{CranmerEtal2007} 
used $n=1$ in equation (\ref{eq:Qturb}) based on
analytic and numerical results
\citep{DobrowolnyEtal1980,OughtonEtal2006}, but they also
tried $n=2$ to explore a stronger form of the quenching effect.

\epubtkImage{}{%
\begin{figure}[t]
  \def\epsfsize#1#2{0.84#1}
  \centerline{\epsfbox{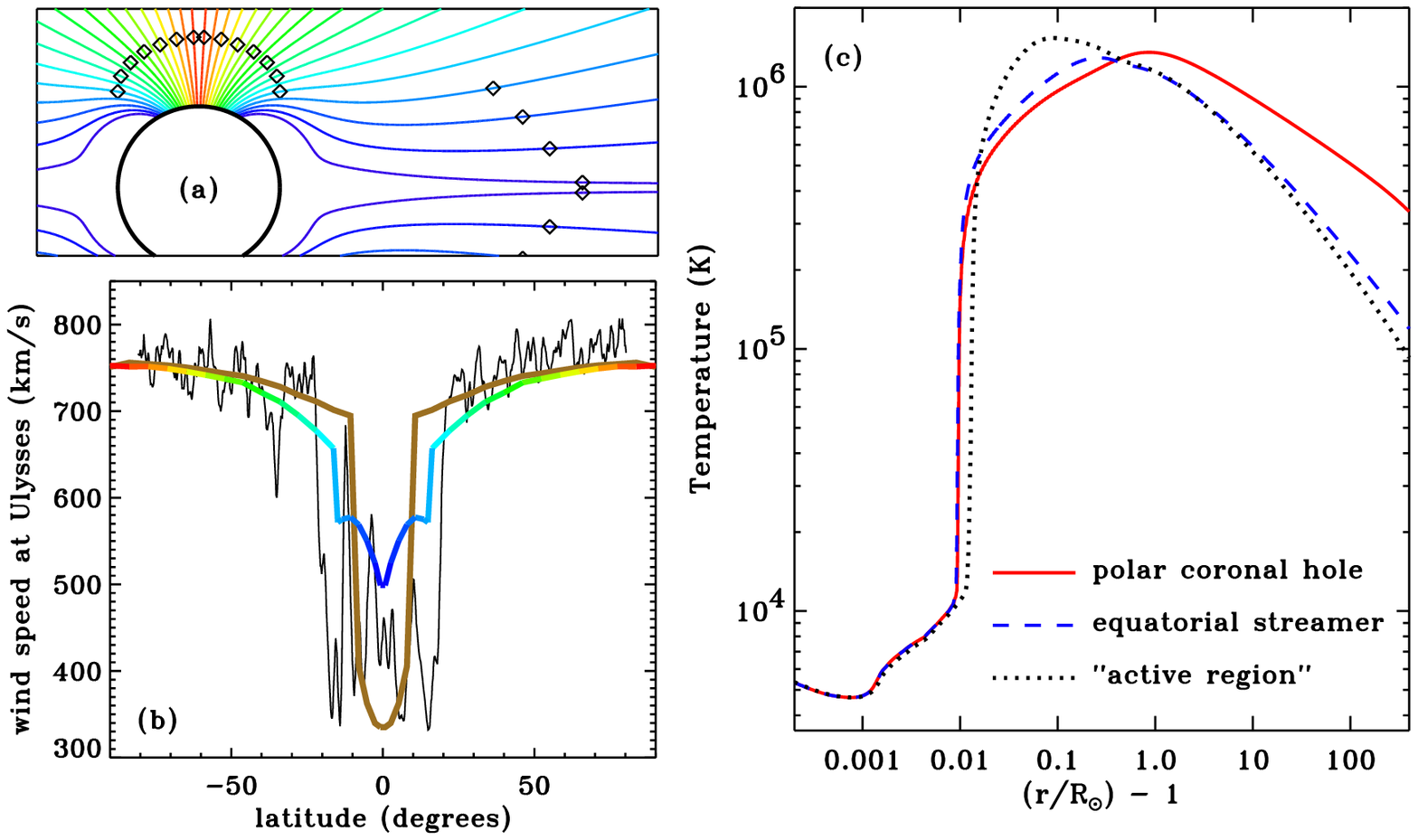}}
  \caption{Summary of \citet{CranmerEtal2007} models:
{\bf (a)} The adopted solar-minimum field geometry of
\citet{BanaszkiewiczEtal1998},
with radii of wave-modified critical points marked by symbols.
{\bf (b)} Latitudinal dependence of wind speed at $\sim$2~AU
for models with $n=1$ (multi-color curve) and $n=2$ (brown curve),
compared with data from the first \emph{Ulysses} polar pass in
1994--1995 \citep[black curve;][]{GoldsteinEtal1996}.
{\bf (c)} $T(r)$ for polar coronal hole
(red solid curve), streamer edge (blue dashed curve), and
strong-field active region (black dotted curve) models.}
  \label{figure:wtd2007_srt}
\end{figure}}

Figure~\ref{figure:wtd2007_srt}
summarizes the results of varying the magnetic field
properties while keeping the lower boundary conditions fixed.
For a single choice for the photospheric wave properties, the
models produced a realistic range of slow and fast solar wind
conditions.
A two-dimensional model of coronal holes and streamers at solar
minimum reproduces the latitudinal bifurcation of slow and fast
streams seen by \emph{Ulysses.}
An active-region-like enhancement of the magnetic field strength
in the low corona generates a high mass flux and a slow wind speed,
in agreement with observations of open field lines connected with
active regions \citep[see also][]{WangEtal2009}.
As predicted by earlier studies, a larger coronal ``expansion
factor'' naturally gives rise to a slower and denser wind, higher
temperature at the coronal base, and lower-amplitude Alfv\'{e}n
waves at 1~AU.

In these models, the radial gradient of the Alfv\'{e}n speed affects
where the waves are reflected and damped, and thus whether energy
is deposited below or above the \citet{Parker1958a} critical point.
Early studies of solar wind energetics
\citep[e.g.,][]{LeerHolzer1980,Pneuman1980,LeerEtal1982}
showed that if there is substantial heating below the critical point,
its primary impact is to ``puff up'' the hydrostatic scale height,
drawing more particles into the accelerating wind and thus
producing a slower and more massive wind.
If most of the heating occurs at or above the critical point,
the subsonic atmosphere is relatively unaffected, and the local
increase in energy flux has nowhere else to go but into the
kinetic energy of the wind (leading to a faster and less dense
outflow).
The ZEPHYR results shown in Figure~\ref{figure:wtd2007_srt}
display this kind of dichotomy because the superradial expansion
creates a much higher critical point over the equatorial regions
than over the poles.
Additional studies of how and where the mass flux and wind speed are
determined include \citet{Withbroe1988}, \citet{HansteenLeer1995},
\citet{HansteenEtal1997}, \citet{JanseEtal2007}, and
\citet{WangEtal2009}.

Perhaps more surprisingly, varying the coronal expansion factor
in the models shown in Figure~\ref{figure:wtd2007_srt} also
produces correlative trends that are in good agreement with
\emph{in~situ} measurements
of commonly measured ion charge state ratios (e.g.,
O$^{7+}$/O$^{6+}$) and FIP-sensitive abundance ratios (e.g., Fe/O).
\citet{CranmerEtal2007} showed that the slowest solar wind
streams -- associated with active-region fields at the base -- can
produce a factor of $\sim$30 larger frozen-in ionization-state ratio of
O$^{7+}$/O$^{6+}$ than high-speed streams from polar coronal holes,
despite the fact that the temperature at 1~AU is lower in
slow streams than in fast streams.
Furthermore, when elemental fractionation is modeled using a theory
based on preferential wave-pressure acceleration \citep{Laming2004,
Laming2009,BryansEtal2009},
the slow wind streams exhibit a substantial relative buildup of
elements with low FIP with respect to the high-speed streams.
Although the WTD models utilize identical photospheric lower boundary
conditions for all of the flux tubes, the self-consistent solutions
for the upper chromosphere, transition region, and low corona are
qualitatively different.
Feedback from larger heights (i.e., from variations in the
flux tube expansion rate and the resulting heating rate) extends
downward to create these differences.

Another empirical ``marker'' of heliospheric stream structure is
the proton specific entropy, or entropy per proton, which is
often approximated as being proportional to
$\ln (T_{p} / n_{p}^{\gamma - 1})$, where $\gamma \approx 1.5$
is an empirical adiabatic index for solar wind protons
\citep[e.g.,][]{BurlagaEtal1990,PagelEtal2004}.
When measured in regions of the (non-CME) heliosphere where
corotating interaction regions have not yet formed shocks, this
quantity is seen to clearly distinguish slow wind streams from
fast wind streams.
Figure~\ref{figure:zentropy} shows how the specific entropy is
positively correlated with wind speed, both in measurements made
by the Solar Wind Electron Proton Alpha Monitor (SWEPAM)
instrument on \emph{ACE} \citep{McComasEtal1998} and in the
\citet{CranmerEtal2007} ZEPHYR models discussed above.
Each model data point was computed independently of the others.
The models had identical lower boundary conditions at the
photosphere, and they differed from one another only by having a
different radial dependence of the magnetic field.
Because entropy should be conserved in the absence of significant
small-scale dissipation, the quantity that is measured at 1~AU
may be a long-distance proxy for the near-Sun locations of
strong coronal heating.
In other words, the comparison of measured and modeled solar
wind entropy variations may be a key way to discriminate between
competing explanations of solar wind acceleration.

\epubtkImage{}{%
\begin{figure}[t]
  \def\epsfsize#1#2{0.84#1}
  \centerline{\epsfbox{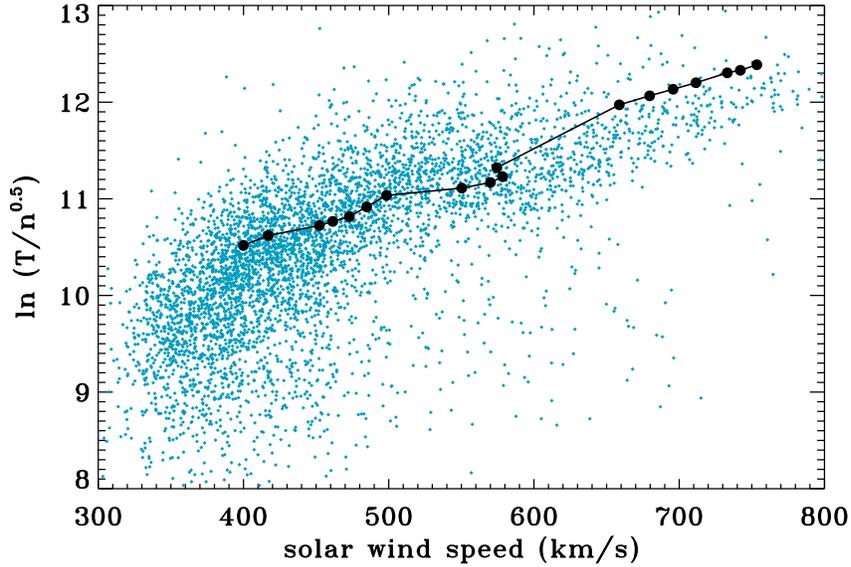}}
  \caption{Solar wind specific entropy plotted as a function
of solar wind speed, computed for both the ZEPHYR models at 1~AU
(black symbols, curve) and from ACE/SWEPAM data
(blue points).}
  \label{figure:zentropy}
\end{figure}}

Although equation (\ref{eq:Qturb}) describes the plasma heating
rate in terms of the local properties of MHD turbulence, it is also
possible to see that this expression gives a heating rate proportional
to the mean \emph{magnetic flux density} at the coronal base.
As illustrated above in Figure~\ref{figure:cvb_cartoon}, the
mean field strength in the low corona is determined by both the
photospheric field strength in the intergranular bright points
and the total number of bright points that eventually merge their
fields together in the low corona.
The field strength at this merging height can thus be estimated as
$B \approx f_{\ast} B_{\ast}$,
where $B_{\ast} \approx 1500$~G is the (nearly universal)
photospheric bright-point field strength and $f_{\ast}$ is the
area filling factor of bright points in the photosphere.
The latter quantity appears to vary by more than an order of
magnitude in different regions on the Sun, from about 0.002 (at
low latitudes at solar minimum) to $\sim$0.1 (in active regions).
If the regions below the merging height can be treated using
approximations from ``thin flux tube theory''
\citep[e.g.,][]{Spruit1981,CranmerBallegooijen2005}, then it
is possible to express each term in equation (\ref{eq:Qturb})
as a function of $f_{\ast}$ and the photospheric properties.
For example, $B \propto \rho^{1/2}$ applies to thin flux tubes
in pressure equilibrium, and thus $\rho$ at the merging height
can be estimated as $f_{\ast}^{2} \rho_{\ast}$ (where
$\rho_{\ast}$ is the photospheric density).
For Alfv\'{e}n waves at low heights, $Z_{\pm} \propto \rho^{-1/4}$,
and so $Z_{\pm}$ at the merging height scales like
$Z_{\pm \ast} / f_{\ast}^{1/2}$.
Also, we assumed that $L_{\perp} \propto B^{-1/2}$.
If the quenching factor in parentheses in equation (\ref{eq:Qturb})
is neglected, then
\begin{equation}
  Q \, \approx \, \frac{\rho Z^3}{L_{\perp}} \approx
  \left( \frac{\rho_{\ast} Z_{\ast}^3}{L_{\perp \ast}} \right)
  \frac{f_{\ast}^{2} f_{\ast}^{-3/2}}{f_{\ast}^{-1/2}}
  \, \approx \, Q_{\ast} f_{\ast}  \,\, .
  \label{eq:Qflux}
\end{equation}
Equivalently, equation (\ref{eq:Qflux}) implies that
$Q / Q_{\ast} \approx B / B_{\ast}$, and thus that the heating
in the low corona scales directly with the mean magnetic field
strength there.
In a more highly structured field, the latter is equivalent to
the magnetic ``flux density'' averaged over a given region.
Observational evidence for such a linear scaling has been found
for both a variety of solar regions and other stars as well
\citep[see, e.g.,][]{PevtsovEtal2003,
SchwadronEtal2006,Suzuki2006,KojimaEtal2007,PintoEtal2009}.

\subsection{The Reconnection/Loop-Opening (RLO) solar wind idea}
\label{section:heataccel_rlo}

It is clear from observations of the Sun's highly dynamical
``magnetic carpet'' \citep{SchrijverEtal1997,TitleSchrijver1998,
HagenaarEtal1999} that much of coronal heating is driven by the
continuous interplay between the emergence, separation, merging,
and cancellation of small-scale magnetic elements.
Reconnection seems to be the most likely channel for the injected
magnetic energy to be converted to heat \citep[e.g.,][]{PriestForbes2000}.
Only a small fraction of the photospheric magnetic flux is in
the form of \emph{open} flux tubes connected to the heliosphere
\citep{CloseEtal2003}.
Thus, the idea has arisen that the dominant source of energy for open
flux tubes is a series of stochastic reconnection events between
the open and closed fields \citep[e.g.,][]{FiskEtal1999,
RyutovaEtal2001,Fisk2003,SchwadronMcComas2003,FeldmanEtal2005,
SchwadronEtal2006,SchwadronMcComas2008,FiskZhao2009}.

The natural appeal of the RLO idea is evident from the fact that
open flux tubes are always rooted in the vicinity of closed loops
\citep[e.g.,][]{DowdyEtal1986}
and that all layers of the solar atmosphere
seem to be in continual motion with a wide range of timescales.
In fact, observed correlations between the lengths of closed loops
in various regions, the electron temperature in the low corona, and
the wind speed at 1~AU  \citep{FeldmanEtal1999,GloecklerEtal2003}
are highly suggestive of a net transfer
of Poynting flux from the loops to the open-field regions that may
be key to understanding the macroscopic structure of the solar wind.
The proposed RLO reconnection events may also be useful in
generating energetic particles and cross-field diffusive transport
throughout the heliosphere \citep[e.g.,][]{FiskSchwadron2001}.

Testing the RLO idea using theoretical models seems to be more difficult
than testing the WTD idea because of the complex multi-scale nature
of magnetic reconnection.
It can be argued that one needs to create fully three-dimensional
models of the coronal magnetic field (arising from multiple
magnetic elements on the surface) to truly assess the full range of
closed/open flux interactions.
The idea of modeling the coronal field via a collection of discrete
magnetic sources (referred to in various contexts as
``magneto-chemistry,'' ``tectonics,'' or ``magnetic charge topology'')
has been used extensively to study the evolution of the closed-field
corona \citep[e.g.,][]{Longcope1996,SchrijverEtal1997,
LongcopeKankelborg1999,SturrockEtal1999,PriestEtal2002,
BeveridgeEtal2003,BarnesEtal2005,Parnell2007,NgBhattacharjee2008}
but applications to open fields and the solar wind have been rarer
\citep[see, however,][]{Fisk2005,TuEtal2005}.

In order to develop the RLO paradigm to the point where it can be
tested more quantitatively, several key questions remain to be
answered.
For example, how much magnetic flux actually \emph{opens up} in
the magnetic carpet?
Also, what is the time and space distribution of reconnection-driven
energy addition into the (transiently) open flux tubes?
Lastly, how is the reconnection energy distributed into various
forms (e.g., bulk kinetic energy in ``jets,'' thermal energy,
waves, turbulence, and energetic particles) that each affect the
accelerating solar wind in different ways?
Combinations of simulations, analytic scaling relations, and
observations are needed to make further progress.

\subsection{Kinetic microphysics}
\label{section:heataccel_kinetic}

The theoretical models discussed in the previous two subsections
mainly involved a ``one-fluid'' or MHD approach to the coronal
heating and solar wind acceleration.
However, at large heights in coronal holes, the collisionless
divergence of plasma parameters for protons, electrons, and heavy
ions allows the multi-fluid kinetic processes to be distinguished
in a more definitive way.
The UVCS measurements of strong O$^{+5}$ preferential heating,
preferential acceleration, and temperature anisotropy have
spurred a great deal of theoretical work in this direction
\citep[see reviews by][]{HollwegIsenberg2002,Cranmer2002a,
Marsch2005,Marsch2006,KohlEtal2006}.
Specifically, the observed ordering of $T_{i} \gg T_{p} > T_{e}$
and the existence of anisotropies of the form
$T_{\perp} > T_{\parallel}$ in coronal holes led to a resurgence
of interest in models of \emph{ion cyclotron resonance.}

The ion cyclotron heating mechanism is a classical resonance
between left-hand polarized Alfv\'{e}n waves and the Larmor
gyrations of positive ions around the background magnetic field.
If the wave frequency and the natural ion gyrofrequency are
equal, then in the rest frame of the ion the oscillating
electric and magnetic fields of the wave are no longer felt
by the ion to be oscillating.
The ion in such a frame senses a constant DC electric field, and
it can secularly gain or lose energy depending on the relative
phase between the ion's velocity vector and the electric field
direction.
In a wave field with random phases, an ion will undergo a random
walk in energy.
Thus, on average the ions can be considered to ``diffuse'' into
faster (i.e., wider) Larmor orbits with larger perpendicular
energy \citep[see][]{RowlandsEtal1966,GalinskyShevchenko2000,
Isenberg2001,Cranmer2001,IsenbergVasquez2009}.

In the actual solar corona, however, it is not likely that the
situation is as straightforward as summarized above.
Instead of a population of pre-existing, linear cyclotron waves
that are dissipated, there may be a rich variety of nonlinear
plasma mechanisms at play.
The observed ion heating is likely to be just the final stage
of a multi-step process of energy conversion between waves,
turbulent motions, reconnection structures, and various kinds
of distortions in the particle velocity distributions.
Figure~\ref{figure:ion_jungle} surveys the field of suggested
possibilities, and the remainder of this section discusses these
ideas in more detail.

\epubtkImage{}{%
\begin{figure}[t]
  \def\epsfsize#1#2{0.95#1}
  \centerline{\epsfbox{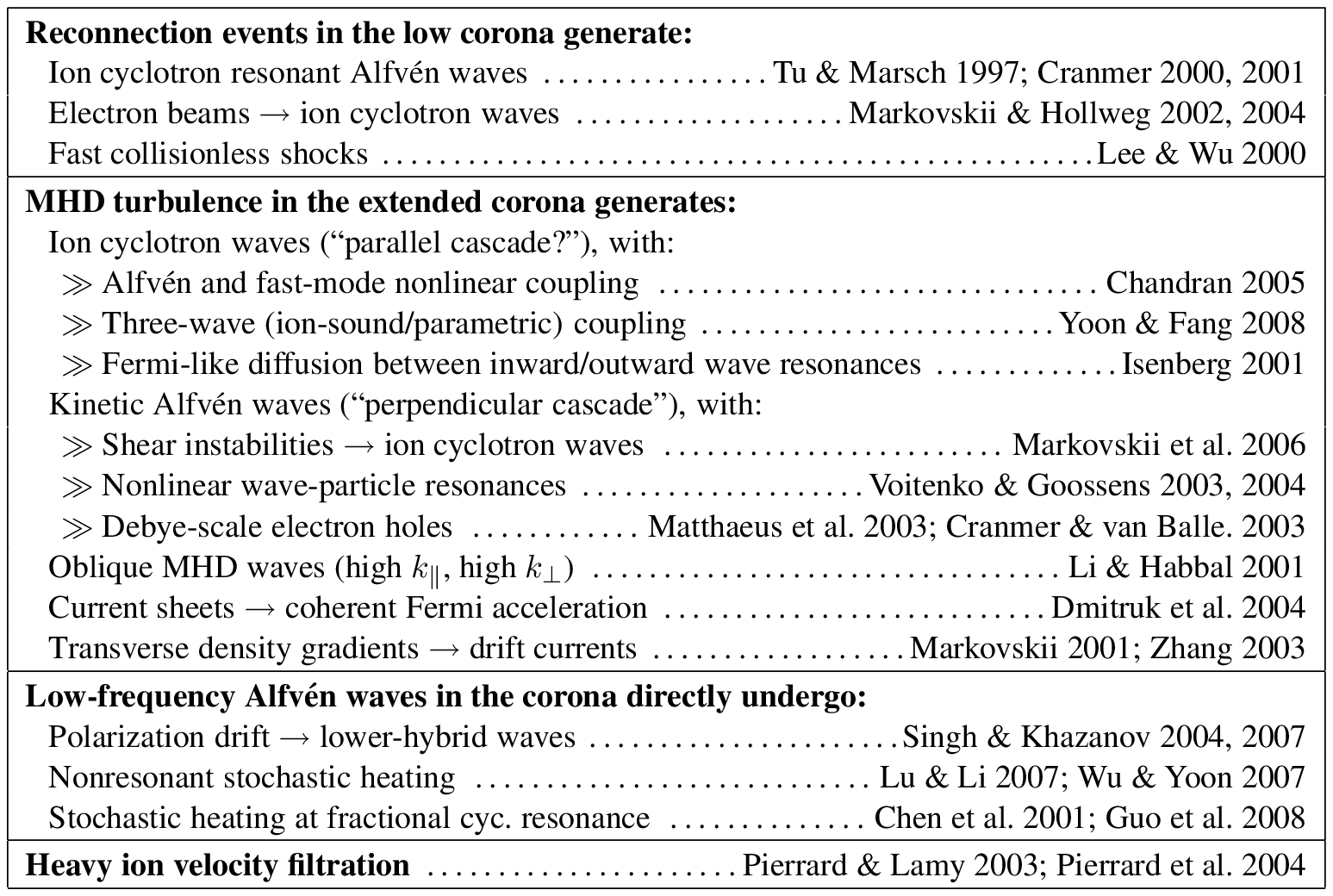}}
  \caption{Tabular outline of suggested physical processes for
preferentially heating and accelerating minor ions in coronal holes.}
  \label{figure:ion_jungle}
\end{figure}}
%FS source?
% convert to table with linked refs!

One potential obstacle to the idea of ion cyclotron heating
is that the required gyroresonant wave frequencies in the
corona are of order $10^2$ to $10^4$ Hz, whereas the dominant
frequencies of Alfv\'{e}n waves believed to be emitted by the
Sun are thought to be much lower (i.e., less than 0.01 Hz,
corresponding to periods of minutes to hours).
\citet{AxfordMcKenzie1992} suggested that the right kinds of
high-frequency waves may be generated in small-scale reconnection
events in the chaotic ``furnace'' of the supergranular network.
These waves could propagate upwards in height -- and downwards in
magnetic field strength -- until they reached a location where
they became cyclotron resonant with the local ions, and thus
would damp rapidly to provide the ion heating
\citep[see also][]{SchwartzEtal1981,TuMarsch1997}.

The above scenario of ``basal generation'' of ion cyclotron
waves has been called into question for several reasons.
\citet{Cranmer2000,Cranmer2001} argued that the passive
sweeping of a pre-existing fluctuation spectrum would involve
ions with low gyrofrequencies (i.e., small ratios of charge
to mass; $q_{i} / m_{i} \approx 0.1$ to 0.2 in proton units)
encountering waves of a given frequency at lower heights than the
ions that have been observed to exhibit preferential heating,
like O$^{+5}$ ($q_{i} / m_{i} = 0.31$) and Mg$^{+9}$
($q_{i} / m_{i} = 0.37$).
Thus, the resonances of many minor ion species may be strong
enough to damp out a base-generated spectrum of waves before
they can become resonant with the observed species.
Furthermore, \citet{Hollweg2000} found that a base-generated
spectrum of ion cyclotron waves would exhibit a very different
appearance in interplanetary radio scintillations than the observed
radio data.
There remains some uncertainty about these criticisms of a
basal spectrum of ion cyclotron waves, and definitive
conclusions cannot yet be made \citep[see discussions in][]
{TuMarsch2001,HollwegIsenberg2002}.

There are several other interesting consequences of the
\citet{AxfordMcKenzie1992} idea of rapid reconnection events
at the coronal base.
It is possible, for example, that such microflaring activity
would give rise to intermittent bursts of parallel electron
beams that propagate up into the extended corona.
Sufficiently strong beams may be unstable to the growth
of wave power at the ion cyclotron frequencies
\citep{MarkovskiiHollweg2002,MarkovskiiHollweg2004,
VoitenkoGoossens2002}, and these waves would then go on to
heat the ions.
Also, \citet{LeeWu2000} suggested that small-scale reconnection
events could produce fast collisionless shocks in the extended
corona.
For shocks sufficiently thin and strong (i.e., with a bulk velocity
jump of at least $\sim$0.3 times the Alfv\'{e}n speed), ions that
cross from one side of the shock to the other remain
``nondeflected'' by the rapid change in direction of the
magnetic field.
Thus, they can convert some of their parallel motion into
perpendicular gyration.
\citet{MancusoEtal2002} suggested this mechanism may be applied
to understanding UVCS measurements of ion heating in shocks
associated with CMEs.
However, it is unclear to what extent the open magnetic regions
in coronal holes are filled with sufficiently strong shocks
to enable this process to occur
\citep[see also][]{HollwegIsenberg2002}.

In contrast to the ideas of base-generation of ion cyclotron waves,
there have been several proposed mechanisms for ``gradual
generation'' of these waves over a range of distances in the
corona and solar wind.
A natural way to produce such an extended source of fluctuations
is MHD \emph{turbulent cascade,} which continually transports
power at large scales to small scales via the stochastic shredding
of transient eddies.
A strong turbulent cascade is certainly present in interplanetary
space \citep[see reviews by][]{TuMarsch1995,GoldsteinEtal1997}.
It is well known, though, that in both numerical simulations and
analytic descriptions of Alfv\'{e}n-wave turbulence (with a
strong background ``guide field'' like in the corona) the cascade
from large to small length scales (i.e., from small to large
wavenumbers) occurs most efficiently for modes that do not
increase in frequency.
In other words, the cascade acts most rapidly to increase the
perpendicular wavenumber $k_{\perp}$ while leaving the
parallel wavenumber $k_{\parallel}$ largely unchanged
\citep[e.g.,][]{Strauss1976,ShebalinEtal1983,
GoldreichSridhar1995,ChoEtal2002,OughtonEtal2004}.
This type of cascade is expected to generate so-called kinetic
Alfv\'{e}n waves (KAWs) with $k_{\perp} \gg k_{\parallel}$,
but not ion cyclotron waves.

Under typical ``low plasma beta'' conditions in the corona and
fast solar wind, the linear dissipation of KAWs would lead to
the preferential parallel heating of electrons
\citep{LeamonEtal1999,CranmerBallegooijen2003,GaryBorovsky2008}.
This is essentially the \emph{opposite} of what has been
observed with UVCS.
However, there have been several suggestions for more complex
(nonlinear or multi-step) processes that may be responsible for
ions to receive perpendicular heating from KAW-type fluctuations.

\begin{enumerate}
\item
\citet{MarkovskiiEtal2006} discussed how a perpendicular turbulent
cascade produces increasingly strong shear motions transverse
to the magnetic field, and that this shear may eventually be unstable
to the generation of cyclotron resonant waves that can in turn
heat protons and ions \citep[see also][]{MikhailenkoEtal2008}.
This effect may also produce a steepening in the power spectrum of
the magnetic field fluctuations that agrees with the observed
``dissipation range'' \citep{SmithEtal2006}.
\item
\citet{VoitenkoGoossens2003,VoitenkoGoossens2004}
suggested that high-$k_{\perp}$ KAWs with sufficiently large amplitudes
could begin to exhibit nonlinear resonance effects (``demagnetized
wave phases'') leading to rapid ion perpendicular heating.
There are definite thresholds in the KAW amplitude that must be exceeded
for these mechanisms to be initiated, and it is unclear whether the
actual coronal turbulent spectrum has enough power in the relevant
regions of wavenumber space \citep[see also][]{WuYang2006,WuYang2007}.
\item
The damping of low-frequency KAWs may give rise to substantial
parallel electron acceleration.
If the resulting electron velocity distributions were sufficiently
\emph{beamed,} they could become unstable to the generation of
parallel Langmuir waves.
In turn, the evolved Langmuir wave trains may exhibit a periodic
electric potential-well structure in which some of the beam
electrons can become trapped.
Adjacent potential wells can then merge with one another to form
isolated ``electron phase space holes'' of saturated potential.
\citet{ErgunEtal1999}, \citet{MatthaeusEtal2003}, and
\citet{CranmerBallegooijen2003} described how these tiny
(Debye-scale) electrostatic structures can heat ions
perpendicularly via Coulomb-like quasi-collisions.
\item
Obliquely propagating MHD waves with large perpendicular \emph{and}
parallel wavenumbers -- including KAWs and fast-mode waves -- can
interact resonantly with positive ions via channels that are not
available when either $k_{\parallel}$ or $k_{\perp}$ are small.
\citet{LiHabbal2001} found that oblique fast-mode waves with
large wavenumbers may be even more efficient than Alfv\'{e}n
waves at heating ions under coronal conditions.
\citet{HollwegMarkovskii2002} discussed how the higher-order
cyclotron resonances become available to obliquely propagating
waves with large wavenumbers, and how these can lead to stochastic
velocity-space diffusion for ions.
\item
On the smallest spatial scales, the plasma in numerical simulations
of MHD turbulence is seen to develop into a collection of narrow
current sheets undergoing oblique magnetic reconnection (i.e.,
with the strong ``guide field'' remaining relatively unchanged).
\citet{DmitrukEtal2004} performed test-particle simulations in
a turbulent plasma and found that protons can become
perpendicularly accelerated around the guide field because of
coherent forcing from the perturbed fields associated with the
current sheets \citep[see also][]{ParasharEtal2009}.
It remains to be seen whether this process could lead to
more than mass-proportional energization for minor ions.
\item
If the plasma contains sufficiently small-scale density
gradients transverse to the magnetic field ($\nabla_{\perp}\rho$),
then drift currents can be excited
that are unstable to the generation of high-frequency waves
\citep{Markovskii2001,Zhang2003,VranjesPoedts2008,MecheriMarsch2008}.
These instabilities depend on both the amplitudes and scale
lengths of $\nabla_{\perp}\rho$.
To measure the latter, it is important to take into account both
remote-sensing measurements of coronal density inhomogeneities
\citep[e.g.,][]{Woo2006,PasachoffEtal2007} and constraints from
radio scintillation power spectra at larger distances
\citep{Bastian2001,Spangler2002,HarmonColes2005}.
\end{enumerate}

Despite the fact that theory predicts a predominantly perpendicular
cascade, there is some evidence that the turbulent fluctuations
in the solar wind have some energy that extends up to large
$k_{\parallel}$ values in a power-law tail
\citep[see, e.g.,][]{BieberEtal1996,DassoEtal2005,MacBrideEtal2008}.
Whether or not this means that true ``parallel cascade'' occurs
in the corona and solar wind is still not known.
However, some progress has been made using a phenomenological
approach to modeling the cascade as a combination of advection
and diffusion in wavenumber space.
In the model of \citet{CranmerBallegooijen2003}, the relative
strengths of perpendicular advection and diffusion determine the
slope of the power-law spectrum in $k_{\parallel}$, and thus they
specify the amount of wave energy that is available at the ion
cyclotron frequencies \citep[see also][]{CranmerEtal1999a,
LandiCranmer2009,JiangEtal2009}.

There have also been proposals for additional mechanisms that
could allow a parallel cascade to occur in the corona and solar wind.
Nonlinear couplings between the dominant Alfv\'{e}n waves and
other modes such as fast magnetosonic waves
\citep{Chandran2005,LuoMelrose2006} and ion-acoustic waves
\citep{YoonFang2008} have the potential to enhance the
wave power at high frequencies.
It has also been known for some time that nonlinear coupling
between Alfv\'{e}n and fast-mode waves may help explain why the
measured \emph{in~situ} magnetic field magnitude $|{\bf B}|$
remains roughly constant while its direction varies strongly
\citep{BarnesHollweg1974,VasquezHollweg1996,VasquezHollweg1998}.

If MHD waves have sufficiently large amplitudes, they may
undergo nonlinear wave steepening, which leads to density
variations as well as oscillations in the parallel components
of the velocity and magnetic field.
These may generate progressively smaller scales along the
magnetic field \citep[e.g.,][]{Medvedev2000,SuzukiEtal2007}.
There is also a ``bootstrap'' kind of effect for ion cyclotron
wave generation that was discussed by \citet{IsenbergEtal2001}.
If some outward-propagating cyclotron waves exist, the resonant
diffusion may act to produce proton velocity distributions that
are unstable to the generation of inward-propagating cyclotron waves.
In response, the proton distributions would become further deformed
and thus could become unstable to the growth of both inward and
outward waves.
It is not yet known if this process could reach the point of
becoming self-sustaining, but if so, it may also serve as an
extended generation mechanism for high-$k_{\parallel}$ waves.

In addition to the above ideas that involve large wavenumbers and
kinetic effects, there have been other suggested physical processes
that do not require high-$k$ resonances to be \emph{initially}
present in order to heat the ions.

\begin{enumerate}
\item
Particles in large-amplitude Alfv\'{e}n waves
exhibit both ${\bf E} \times {\bf B}$ drift motions (i.e., their
standard velocity amplitude) and a polarization drift
velocity $V_{\rm pol}$ that is smaller than the former by the ratio
$\omega / \Omega_{i}$, where $\omega$ is the wave frequency and
$\Omega_i$ is the ion cyclotron frequency.
A sufficiently large $V_{\rm pol}$ can lead to
cross-field currents unstable to the generation of high-frequency waves,
and to eventual equipartition between $V_{\rm pol}$ and the ion
thermal speed \citep{SinghKhazanov2004,SinghEtal2007,
KhazanovSingh2007}.
It is not yet known whether the effective $V_{\rm pol}$ for the
coronal fluctuation spectrum is large enough to provide a significant
fraction of the ion thermal speeds.
\item
Recently there have been suggested some completely
\emph{nonresonant} mechanisms that depend on the stochasticity
of MHD turbulence to produce an effective increase in random
ion motions \citep{LuLi2007,WuYoon2007,BourouaineEtal2008}.
Questions still remain, though, concerning the spatial scales over
which one should refer to particle motions as ``heating'' versus
``wave sloshing.''
This energization mechanism may be just a more chaotic form of the
standard velocity amplitude that an ion feels when in the presence
of a spectrum of Alfv\'{e}n waves \citep[see, e.g.,][]{WangWu2009}.
In this case, the maximum amount of heating from this process would
provide mass-proportional heating for minor ions and protons (i.e.,
$T_{i}/T_{p} = m_{i}/m_{p}$), and it is clear that the UVCS
measurements for O$^{+5}$ show heating in excess of this amount
(see Section~\ref{section:offlimb_uv}).
\item
Both numerical and analytic studies of Alfv\'{e}n waves show that,
at sufficiently large amplitudes, there can be gyroresonance-like
ion energization for sets of frequencies at specific fractions of
the local ion cyclotron resonance frequency
\citep[e.g.,][]{ChenEtal2001,GuoEtal2008}.
Like several other processes listed above, this effect becomes active
only above certain thresholds of wave amplitude.
Also, \citet{MarkovskiiEtal2009} showed that mildly nonlinear
Alfv\'{e}n waves -- with frequencies slightly below the local proton
gyrofrequency and power in both the upward and downward directions
along the field -- can also undergo additional modes of dissipation
and proton heating that are not anticipated in linear ion
cyclotron resonance theories.
\end{enumerate}

Finally, there has been some development of the so-called
\emph{velocity filtration} theory, which requires neither direct
heating nor wave damping in order to energize coronal ions.
Spacecraft measurements of plasma velocity distributions, both in the
solar wind and in planetary magnetospheres and magnetosheaths, have
revealed that ``suprathermal'' power-law tails are quite common.
These observations led to the suggestion by
\citet{Scudder1992a,Scudder1994} of an an alternative to
theories that demand explicit energy deposition
in the low corona \citep[see also][]{Parker1958b,Levine1974}.
A velocity distribution having a suprathermal tail will
become increasingly dominated by its high-energy particles
at larger distances from the solar gravity well.
Thus an effective ``heating'' occurs as a result of
particle-by-particle conservation of energy.
The major unresolved issue is whether suprathermal tails
of the required strength can be produced and maintained
in the upper chromosphere and transition region -- where
Coulomb collisions are traditionally believed to be
strong enough to rapidly drive velocity distributions
toward Maxwellians.
Whether the solar atmosphere actually plays host to strong
nonthermal tails is still under debate, with some evidence
existing in favor of their presence
\citep[e.g.,][]{EsserEdgar2000,RalchenkoEtal2007}
and other evidence against them
\citep{AndersonEtal1996,KoEtal1996,FeldmanEtal2007}.

The original \citep{Scudder1992a,Scudder1992b}
ideas about suprathermal velocity filtration were applied only
to the primary (proton and electron) coronal plasma.
More recently, \citet{PierrardLamy2003} and \citet{PierrardEtal2004}
have shown that this mechanism can produce extremely high temperatures
for heavy ions in the corona -- providing they had suprathermal tails
in the chromosphere.
The primary quantities presented in these papers,
however, were integrated isotropic temperatures $T$.
No information was given about the predicted sense of the
temperature anisotropy for the minor ions.
For a collisionless exospheric model, there is a suspicion that
a combination of several effects (e.g., the initial velocity
filtration and subsequent magnetic moment conservation) would
result in velocity distributions with $T_{\parallel} \gg T_{\perp}$,
which is not what is observed.

Since it is obvious that not all of the proposed mechanisms described
above (and shown in Figure~\ref{figure:ion_jungle}) can be the
dominant cause of the collisionless ion energization in coronal
holes, there is a great need to ``cut through the jungle'' and
assess the validity of each of these processes.
For many of these suggested ideas, further theoretical development
is required so that specific observational predictions can be made.
However, there are also several types of measurement that have not
been widely recognized or utilized as constraints on theoretical models.
A prime example is the use of radio sounding (i.e., interplanetary
scintillations and Faraday rotation) to measure the fine structure
of the corona and solar wind in density, velocity, and magnetic field
strength \citep[see, however,][]{HollwegIsenberg2002,Spangler2002,
HarmonColes2005,ChandranEtal2009b}.
Another example is the use of high-resolution UV spectral line
profiles to probe departures from Maxwellian or bi-Maxwellian
ion velocity distributions \citep[e.g.,][]{Cranmer2001}.

%\newpage
%============================================================================
\section{Summary and Conclusions}
\label{section:summconc}

The last decade has seen significant progress toward identifying
and characterizing the processes that produce coronal holes.
As remote-sensing plasma measurements have become possible in the
extended solar corona (i.e., the region of primary acceleration of
the solar wind), the traditional gap between solar physics and
\emph{in situ} space physics has become narrower.
However, there are still many unanswered questions:
How and where in the solar atmosphere are the relevant waves and
turbulent motions generated?
Which kinds of fluctuation modes (i.e., linear or nonlinear;
Alfv\'{e}n, fast, or slow; high $k_{\parallel}$ or high
$k_{\perp}$) are most important?
What frequencies dominate the radially evolving power spectrum?
What fraction of the interplanetary solar wind comes from
filamentary structures such as polar plumes and polar jets?
Are there relatively simple ``scaling laws'' that will allow us to
use only the measured properties at the solar surface to predict the
resulting amount of coronal heating and solar wind acceleration?

Answering the above questions involves moving forward in both the
theoretical and observational directions.
Section~\ref{section:heataccel} described the large number of
suggested physical processes for energizing the plasma in coronal
holes.
The validity of many of these processes still needs to be assessed,
and their relative contributions to the heating and acceleration of
the actual solar corona need to be quantified.
If, at the end of this process, there are still a number of mutually
inconsistent theories that are still viable, the only way forward
is to determine what \emph{future} measurements would best put the
remaining models to the test.
These activities are ongoing with the planning of inner heliospheric
missions such as \emph{Solar Probe} \citep{McComasEtal2007}
and \emph{Solar Orbiter} \citep{MarsdenFleck2007},
as well as next-generation ultraviolet coronagraph spectroscopy
missions that would follow up on the successes of UVCS/{\em{SOHO}}
\citep[see, e.g.,][]{Cranmer2002b,GardnerEtal2003,KohlEtal2006}.

The development of more physically sound models of the solar wind
feeds back in many ways to a wider understanding of stellar
outflows and star/planet evolution.
Figure~\ref{figure:atp_evolution} shows some of the the early
stages of evolution for a representative solar-type star.
At all ages, cool stars are inferred to exhibit some kind of wind
or jet-like outflow \citep{LamersCassinelli1999,Wood2004,
Gudel2007,Cranmer2008a}.
Young stars first become visible as dust-obscured cloud cores and
protostars \citep[e.g.,][]{Lada1985,Hartmann2000}, and these objects
are often associated with bipolar, collimated jets.
These outflows indicate some kind of transfer of energy from the
accretion disk's orbital motion to torqued magnetic fields (rooted
on the stellar surface) that relieve the buildup of angular
momentum and eject plasma out the poles
\citep[e.g.,][]{BlandfordPayne1982,vanBallegooijen1994}.
As the accretion rates decrease over time, protostars become visible
as \emph{classical T Tauri stars} (CTTS), and there remains ample
evidence for polar outflows in the form of both ``disk winds'' and
true stellar winds \citep{HartiganEtal1995,FerreiraEtal2006,
Cranmer2008b}.
The primordial accretion disk is dissipated gradually as the star
enters the \emph{weak-lined T Tauri star} (WTTS) phase, with a likely
transition to a protoplanetary dust/debris disk.
Strong stellar magnetic activity remains evident during these stages
from, e.g., X-rays \citep{FeigelsonMontmerle1999}.
Many ``post~T~Tauri'' stars, once they reach the zero-age main
sequence (ZAMS), remain rapidly rotating, and for young ZAMS stars
such as AB Dor there is evidence for a range of X-ray emitting
plasma from dark polar spots (which probably do \emph{not}
correspond to open magnetic field regions like coronal holes)
to huge ``slingshot prominences'' extending over several
stellar radii \citep[e.g.,][]{GudelEtal2001,
GudelEtal2003,JardineBallegooijen2005}.

\epubtkImage{}{%
\begin{figure}[t]
  \def\epsfsize#1#2{0.85#1}
  \centerline{\epsfbox{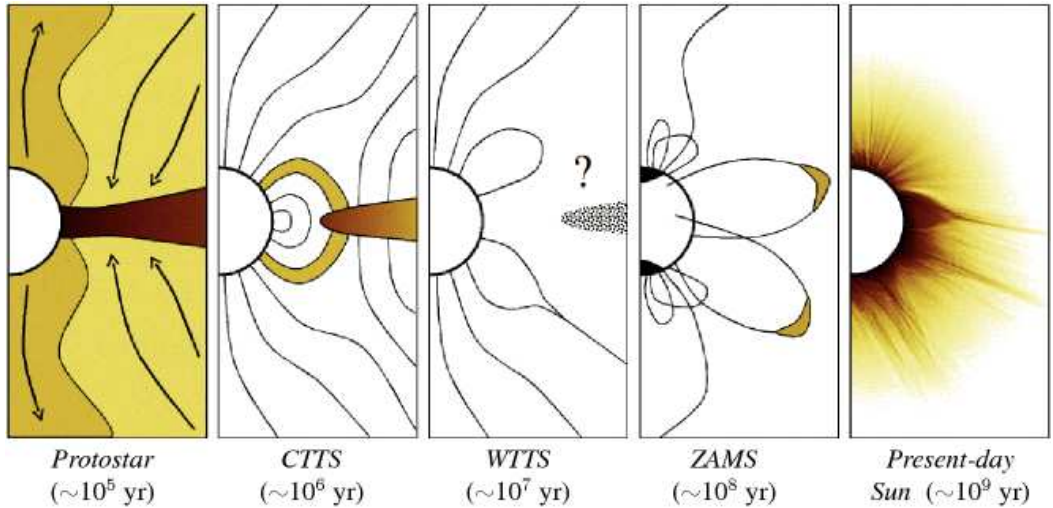}}
  \caption{Illustration of the evolving circumstellar environment
of a solar-mass star (see text), showing various kinds of open-field
structures that may be analogous to present-day coronal holes.}
  \label{figure:atp_evolution}
\end{figure}}

Learning about the fundamental physics responsible for solar
coronal holes has relevance that reaches into other areas of study
besides astrophysics, including plasma physics, space physics,
and astronautical engineering.
The practical benefits of improving long-term predictions for the
conditions of the Earth's local space environment are manifold
\citep[see, e.g.,][]{FeynmanGabriel2000,Eastwood2008}.
In addition, parallel research into the expansion of the
\emph{polar wind} from the Earth's ionosphere has led to an
improved understanding of kinetic processes in plasmas on the
boundary between collisional and collisionless conditions
\citep{LemairePierrard2001,BarakatSchunk2006}.
Finally, a growing realization that strict topical
compartmentalization is often a hindrance to making progress
has given rise to greater interest in interdisciplinary studies
of ``universal processes in heliophysics''
\citep{Crooker2004,DavilaEtal2009}.

%\newpage
%============================================================================
\section{Acknowledgements}
\label{section:acknowledgements}

I would like to thank the editors of \emph{Living Reviews in Solar
Physics} for the invitation to prepare this review.
I gratefully acknowledge John Kohl, Adriaan van Ballegooijen,
Leonard Strachan, John Raymond,
William Matthaeus, Philip Isenberg, Joseph Hollweg,
S.\  Peter Gary, Eckart Marsch, Marco Velli, Mari Paz Miralles,
Larry Gardner, and Enrico Landi
for many valuable discussions and collaborations.
I would especially like to thank the colleagues with whom I have
engaged in more adversarial debates over the years, such as
Ian Axford, Chuanyi Tu, Nathan Schwadron, and Nour-Eddine Raouafi.
\emph{``As brothers fight ye!''}

This work was supported by the National Aeronautics and Space
Administration (NASA) under grants NNG04GE77G,
NNX06AG95G, NNX07AL72G, and NNX09AB27G
to the Smithsonian Astrophysical Observatory.
\emph{SOHO} is a project of international cooperation between
ESA and NASA.  This research made extensive use of NASA's
Astrophysics Data System (ADS) and the John G.\  Wolbach
Library at the Harvard-Smithsonian Center for Astrophysics.

%\newpage
%============================================================================

% To use the bibtex bibliography in 'LivRevSolar.bib' do:
% 'latex LivRevSolar'
% 'bibtex LivRevSolar'
% 'latex LivRevSolar'
% 'latex LivRevSolar'

\bibliography{cranmer_hole2009}

\begin{thebibliography}{438}
\expandafter\ifx\csname natexlab\endcsname\relax\def\natexlab#1{#1}\fi
\expandafter\ifx\csname url\endcsname\relax
  \def\url#1{{\tt #1}}\fi
\expandafter\ifx\csname urlprefix\endcsname\relax\def\urlprefix{URL }\fi
\providecommand{\eprint}[2][]{\url{#2}}
\catcode`\% 12

\bibitem[Abramenko {\it et~al.\/}(2006a)]{AbramenkoEtal2006a}
Abramenko, V.I., Fisk, L.A., Yurchyshyn, V.B., 2006a, ``The Rate of Emergence
  of Magnetic Dipoles in Coronal Holes and Adjacent Quiet-Sun Regions'', {\it
  Astrophys. J. Lett.\/}, {\bf 641}, L65--L68. \newline ADS:
  \url{http://adsabs.harvard.edu/abs/2006ApJ...641L..65A}

\bibitem[Abramenko {\it et~al.\/}(2006b)]{AbramenkoEtal2006b}
Abramenko, V.I., Pevtsov, A.A., Romano, P., 2006b, ``Coronal Heating and
  Photospheric Turbulence Parameters: Observational Aspects'', {\it Astrophys.
  J. Lett.\/}, {\bf 646}, L81--L84. \newline ADS:
  \url{http://adsabs.harvard.edu/abs/2006ApJ...646L..81A}

\bibitem[Aiouaz {\it et~al.\/}(2005)]{AiouazEtal2005}
Aiouaz, T., Peter, H., Lemaire, P., 2005, ``The correlation between coronal
  Doppler shifts and the supergranular network'', {\it Astron. Astrophys.\/},
  {\bf 435}, 713--721. \newline ADS:
  \url{http://adsabs.harvard.edu/abs/2005A&A...435..713A}

\bibitem[Allen {\it et~al.\/}(2000)]{AllenEtal2000}
Allen, L.A., Habbal, S.R., Li, X., 2000, ``Thermal coupling of protons and
  neutral hydrogen with anisotropic temperatures in the fast solar wind'', {\it
  J. Geophys. Res.\/}, {\bf 105}, 23\,123--23\,134. \newline ADS:
  \url{http://adsabs.harvard.edu/abs/2000JGR...10523123A}

\bibitem[Altschuler and Newkirk(1969)]{AltschulerNewkirk1969}
Altschuler, M.D., Newkirk, G., 1969, ``Magnetic fields and the structure of the
  solar corona, I: Methods of calculating coronal fields'', {\it Solar
  Phys.\/}, {\bf 9}, 131--149. \newline ADS:
  \url{http://adsabs.harvard.edu/abs/1969SoPh....9..131A}

\bibitem[Altschuler and Perry(1972)]{AltschulerPerry1972}
Altschuler, M.D., Perry, R.M., 1972, ``On Determining the Electron Density
  Distribution of the Solar Corona from K-Coronameter Data'', {\it Solar
  Phys.\/}, {\bf 23}, 410--428. \newline ADS:
  \url{http://adsabs.harvard.edu/abs/1972SoPh...23..410A}

\bibitem[Altschuler {\it et~al.\/}(1972)]{AltschulerEtal1972}
Altschuler, M.D., Trotter, D.E., Orrall, F.Q., 1972, ``Coronal Holes'', {\it
  Solar Phys.\/}, {\bf 26}, 354--365. \newline ADS:
  \url{http://adsabs.harvard.edu/abs/1972SoPh...26..354A}

\bibitem[Anderson {\it et~al.\/}(1996)]{AndersonEtal1996}
Anderson, S.W., Raymond, J.C., van Ballegooijen, A., 1996, ``Ultraviolet
  Emission-Line Intensities and Coronal Heating by Velocity Filtration:
  Collisionless Results'', {\it Astrophys. J.\/}, {\bf 457}, 939--948. \newline
  ADS: \url{http://adsabs.harvard.edu/abs/1996ApJ...457..939A}

\bibitem[Andretta and Jones(1997)]{AndrettaJones1997}
Andretta, V., Jones, H.P., 1997, ``On the Role of the Solar Corona and
  Transition Region in the Excitation of the Spectrum of Neutral Helium'', {\it
  Astrophys. J.\/}, {\bf 489}, 375--394. \newline ADS:
  \url{http://adsabs.harvard.edu/abs/1997ApJ...489..375A}

\bibitem[Andries and Goossens(2001)]{AndriesGoossens2001}
Andries, J., Goossens, M., 2001, ``Kelvin-Helmholtz instabilities and resonant
  flow instabilities for a coronal plume model with plasma pressure'', {\it
  Astron. Astrophys.\/}, {\bf 368}, 1083--1094. \newline ADS:
  \url{http://adsabs.harvard.edu/abs/2001A&A...368.1083A}

\bibitem[Antiochos {\it et~al.\/}(2007)]{AntiochosEtal2007}
Antiochos, S.K., DeVore, C.R., Karpen, J.T., Miki{\'{c}}, Z., 2007, ``Structure
  and dynamics of the Sun's open magnetic field'', {\it Astrophys. J.\/}, {\bf
  671}, 936--946. \newline ADS:
  \url{http://adsabs.harvard.edu/abs/2007ApJ...671..936A}

\bibitem[Antonucci {\it et~al.\/}(2000)]{AntonucciEtal2000}
Antonucci, E., Dodero, M., Giordano, S., 2000, ``Fast Solar Wind Velocity in a
  Polar Coronal Hole during Solar Minimum'', {\it Solar Phys.\/}, {\bf 197},
  115--134. \newline ADS:
  \url{http://adsabs.harvard.edu/abs/2000SoPh..197..115A}

\bibitem[Antonucci {\it et~al.\/}(2004)]{AntonucciEtal2004}
Antonucci, E., Dodero, M.A., Giordano, S., Krishnakumar, V., Noci, G., 2004,
  ``Spectroscopic measurement of the plasma electron density and outflow
  velocity in a polar coronal hole'', {\it Astron. Astrophys.\/}, {\bf 416},
  749--758. \newline ADS:
  \url{http://adsabs.harvard.edu/abs/2004A&A...416..749A}

\bibitem[Arge and Pizzo(2000)]{ArgePizzo2000}
Arge, C.N., Pizzo, V.J., 2000, ``Improvement in the prediction of solar wind
  conditions using near-real time solar magnetic field updates'', {\it J.
  Geophys. Res.\/}, {\bf 105}, 10\,465--10\,480. \newline ADS:
  \url{http://adsabs.harvard.edu/abs/2000JGR...10510465A}

\bibitem[Asai {\it et~al.\/}(2008)]{AsaiEtal2008}
Asai, A., Shibata, K., Hara, H., Nitta, N.V., 2008, ``Characteristics of
  Anemone Active Regions Appearing in Coronal Holes Observed with the Yohkoh
  Soft X-Ray Telescope'', {\it Astrophys. J.\/}, {\bf 673}, 1188--1193.
  \newline ADS: \url{http://adsabs.harvard.edu/abs/2008ApJ...673.1188A}

\bibitem[Aschwanden(2006)]{Aschwanden2006}
Aschwanden, M.J., 2006, {\it Physics of the Solar Corona: An Introduction with
  Problems and Solutions, 2nd ed.\/}, Springer-Praxis Books in Geophysical
  Sciences, Springer; Praxis, Berlin; New York; Chichester

\bibitem[Aschwanden {\it et~al.\/}(2007)]{AschwandenEtal2007}
Aschwanden, M.J., Winebarger, A., Tsiklauri, D., Peter, H., 2007, ``The Coronal
  Heating Paradox'', {\it Astrophys. J.\/}, {\bf 659}, 1673--1681. \newline
  ADS: \url{http://adsabs.harvard.edu/abs/2007ApJ...659.1673A}

\bibitem[Aschwanden {\it et~al.\/}(2009)]{AschwandenEtal2009}
Aschwanden, M.J., Nitta, N.V., Wuelser, J.P., Lemen, J.R., Sandman, A.,
  Vourlidas, A., Colaninno, R.C., 2009, ``First Measurements of the Mass of
  Coronal Mass Ejections from the EUV Dimming Observed with STEREO EUVI A$+$B
  Spacecraft'', {\it Astrophys. J.\/}, submitted

\bibitem[Avrett(1999)]{Avrett1999}
Avrett, E.H., 1999, ``Combined Effects of Mass Flow and Particle Diffusion on
  the Ionization Structure of the Solar Transition Region'', in {\it Plasma
  Dynamics and Diagnostics in the Solar Transition Region and Corona\/}, (Eds.)
  Vial, J.-C., Kaldeich-Sch\"{u}mann, B., Proceedings of the 8th SOHO Workshop,
  22--25 June 1999, Paris, France, vol. SP-446 of ESA Conference Proceedings,
  pp. 141--147, ESA Publications Division, Noordwijk. \newline ADS:
  \url{http://adsabs.harvard.edu/abs/1999ESASP.446..141A}

\bibitem[Avrett and Loeser(2008)]{AvrettLoeser2008}
Avrett, E.H., Loeser, R., 2008, ``Models of the Solar Chromosphere and
  Transition Region from SUMER and HRTS Observations: Formation of the
  Extreme-Ultraviolet Spectrum of Hydrogen, Carbon, and Oxygen'', {\it
  Astrophys. J. Suppl. Ser.\/}, {\bf 175}, 229--276. \newline ADS:
  \url{http://adsabs.harvard.edu/abs/2008ApJS..175..229A}

\bibitem[Axford and McKenzie(1992)]{AxfordMcKenzie1992}
Axford, W.I., McKenzie, J.F., 1992, ``The origin of high speed solar wind
  streams'', in {\it Solar Wind Seven\/}, (Eds.) Marsch, E., Schwenn, R.,
  Proceedings of the 3rd COSPAR Colloquium held in Goslar, Germany, 16--20
  September 1991, vol.~3 of COSPAR Colloquia Series, pp. 1--5, Pergamon Press,
  Oxford; New York

\bibitem[Axford and McKenzie(1997)]{AxfordMcKenzie1997}
Axford, W.I., McKenzie, J.F., 1997, ``The Solar Wind'', in {\it Cosmic Winds
  and the Heliosphere\/}, (Eds.) Jokipii, J.R., Sonett, C.P., Giampapa, M.S.,
  Space Science Series, pp. 31--66, Arizona University Press, Tucson

\bibitem[Banaszkiewicz {\it et~al.\/}(1998)]{BanaszkiewiczEtal1998}
Banaszkiewicz, M., Axford, W.I., McKenzie, J.F., 1998, ``An analytic solar
  magnetic field model'', {\it Astron. Astrophys.\/}, {\bf 337}, 940--944.
  \newline ADS: \url{http://adsabs.harvard.edu/abs/1998A&A...337..940B}

\bibitem[Banerjee {\it et~al.\/}(1998)]{BanerjeeEtal1998}
Banerjee, D., Teriaca, L., Doyle, J.G., Wilhelm, K., 1998, ``Broadening of Si
  VIII lines observed in the solar polar coronal holes'', {\it Astron.
  Astrophys.\/}, {\bf 339}, 208--214. \newline ADS:
  \url{http://adsabs.harvard.edu/abs/1998A&A...339..208B}

\bibitem[Banerjee {\it et~al.\/}(2009)]{BanerjeeEtal2009}
Banerjee, D., P\'{e}rez-Su\'{a}rez, D., Doyle, J.G., 2009, ``Signatures of
  Alfv\'{e}n waves in the polar coronal holes as seen by EIS/Hinode'', {\it
  Astron. Astrophys.\/}, {\bf 501}, L15--L18. \newline ADS:
  \url{http://adsabs.harvard.edu/abs/2009A&A...501L..15B}

\bibitem[Barakat and Schunk(2006)]{BarakatSchunk2006}
Barakat, A.R., Schunk, R.W., 2006, ``A three-dimensional model of the
  generalized polar wind'', {\it J. Geophys. Res.\/}, {\bf 111}, A12\,314.
  \newline ADS: \url{http://adsabs.harvard.edu/abs/2006JGRA..11112314B}

\bibitem[Barnes and Hollweg(1974)]{BarnesHollweg1974}
Barnes, A., Hollweg, J.V., 1974, ``Large-amplitude hydromagnetic waves'', {\it
  J. Geophys. Res.\/}, {\bf 79}, 2302--2318. \newline ADS:
  \url{http://adsabs.harvard.edu/abs/1974JGR....79.2302B}

\bibitem[Barnes {\it et~al.\/}(2005)]{BarnesEtal2005}
Barnes, G., Longcope, D.W., Leka, K.D., 2005, ``Implementing a Magnetic Charge
  Topology Model for Solar Active Regions'', {\it Astrophys. J.\/}, {\bf 629},
  561--571. \newline ADS:
  \url{http://adsabs.harvard.edu/abs/2005ApJ...629..561B}

\bibitem[Bastian(2001)]{Bastian2001}
Bastian, T.S., 2001, ``Radio Wave Propagation in the Corona and the
  Interplanetary Medium'', {\it Astrophys. Space Sci.\/}, {\bf 277}, 107--116.
  \newline ADS: \url{http://adsabs.harvard.edu/abs/2001Ap&SS.277..107B}

\bibitem[Beckers(1968)]{Beckers1968}
Beckers, J.M., 1968, ``Solar Spicules (Invited Review Paper)'', {\it Solar
  Phys.\/}, {\bf 3}, 367--433. \newline ADS:
  \url{http://adsabs.harvard.edu/abs/1968SoPh....3..367B}

\bibitem[Bell and Noci(1973)]{BellNoci1973}
Bell, B., Noci, G., 1973, ``Are Coronal Holes M Regions?'', {\it Bull. Am.
  Astron. Soc.\/}, conference paper. \newline ADS:
  \url{http://adsabs.harvard.edu/abs/1973BAAS....5S.269B}

\bibitem[Bell and Noci(1976)]{BellNoci1976}
Bell, B., Noci, G., 1976, ``Intensity of the Fe XV emission line corona, the
  level of geomagnetic activity, and the velocity of the solar wind'', {\it J.
  Geophys. Res.\/}, {\bf 81}, 4508--4516. \newline ADS:
  \url{http://adsabs.harvard.edu/abs/1976JGR....81.4508B}

\bibitem[Bemporad {\it et~al.\/}(2008)]{BemporadEtal2008}
Bemporad, A., Matthaeus, W.H., Poletto, G., 2008, ``Low-frequency Ly$\alpha$
  Power Spectra Observed by UVCS in a Polar Coronal Hole'', {\it Astrophys. J.
  Lett.\/}, {\bf 677}, L137--L140. \newline ADS:
  \url{http://adsabs.harvard.edu/abs/2008ApJ...677L.137B}

\bibitem[Beresnyak and Lazarian(2009)]{BeresnyakLazarian2009}
Beresnyak, A., Lazarian, A., 2009, ``Structure of Stationary Strong Imbalanced
  Turbulence'', {\it Astrophys. J.\/}, {\bf 702}, 460--471. \newline ADS:
  \url{http://adsabs.harvard.edu/abs/2009ApJ...702..460B}

\bibitem[Berger and Title(2001)]{BergerTitle2001}
Berger, T.E., Title, A.M., 2001, ``On the Relation of G-Band Bright Points to
  the Photospheric Magnetic Field'', {\it Astrophys. J.\/}, {\bf 553},
  449--469. \newline ADS:
  \url{http://adsabs.harvard.edu/abs/2001ApJ...553..449B}

\bibitem[Beveridge {\it et~al.\/}(2003)]{BeveridgeEtal2003}
Beveridge, C., Longcope, D.W., Priest, E.R., 2003, ``A model for elemental
  coronal flux loops'', {\it Solar Phys.\/}, {\bf 216}, 27--40. \newline ADS:
  \url{http://adsabs.harvard.edu/abs/2003SoPh..216...27B}

\bibitem[Bieber {\it et~al.\/}(1996)]{BieberEtal1996}
Bieber, J.W., Wanner, W., Matthaeus, W.H., 1996, ``Dominant two-dimensional
  solar wind turbulence with implications for cosmic ray transport'', {\it J.
  Geophys. Res.\/}, {\bf 101}, 2511--2522. \newline ADS:
  \url{http://adsabs.harvard.edu/abs/1996JGR...101.2511B}

\bibitem[Blandford and Payne(1982)]{BlandfordPayne1982}
Blandford, R.D., Payne, D.G., 1982, ``Hydromagnetic flows from accretion discs
  and the production of radio jets'', {\it Mon. Not. Roy. Astron. Soc.\/}, {\bf
  199}, 883--903. \newline ADS:
  \url{http://adsabs.harvard.edu/abs/1982MNRAS.199..883B}

\bibitem[Bourouaine {\it et~al.\/}(2008)]{BourouaineEtal2008}
Bourouaine, S., Marsch, E., Vocks, C., 2008, ```On the efficiency of
  nonresonant ion heating by coronal Alfv\'{e}n waves'', {\it Astrophys. J.
  Lett.\/}, {\bf 684}, L119--L122. \newline ADS:
  \url{http://adsabs.harvard.edu/abs/2008ApJ...684L.119B}

\bibitem[Breech {\it et~al.\/}(2008)]{BreechEtal2008}
Breech, B., Matthaeus, W.H., Minnie, J., Bieber, J.W., Oughton, S., Smith,
  C.W., Isenberg, P.A., 2008, ``Turbulence transport throughout the
  heliosphere'', {\it J. Geophys. Res.\/}, {\bf 113}, A08\,105. \newline ADS:
  \url{http://adsabs.harvard.edu/abs/2008JGRA..11308105B}

\bibitem[Bryans {\it et~al.\/}(2009)]{BryansEtal2009}
Bryans, P., Landi, E., Savin, D.W., 2009, ``A New Approach to Analyzing Solar
  Coronal Spectra and Updated Collisional Ionization Equilibrium Calculations:
  II, Updated Ionization Rate Coefficients'', {\it Astrophys. J.\/}, {\bf 691},
  1540--1559. \newline ADS:
  \url{http://adsabs.harvard.edu/abs/2009ApJ...691.1540B}

\bibitem[Burlaga {\it et~al.\/}(1990)]{BurlagaEtal1990}
Burlaga, L.F., Mish, W.H., Whang, Y.C., 1990, ``Coalescence of recurrent
  streams of different sizes and amplitudes'', {\it J. Geophys. Res.\/}, {\bf
  95}, 4247--4255. \newline ADS:
  \url{http://adsabs.harvard.edu/abs/1990JGR....95.4247B}

\bibitem[Byhring {\it et~al.\/}(2008)]{ByhringEtal2008}
Byhring, H.S., Esser, R., Lie-Svendsen, {\O}., 2008, ``The Funnel Geometry of
  Open Flux Tubes in the Low Solar Corona Constrained by O VI and Ne VIII
  Outflow'', {\it Astrophys. J. Lett.\/}, {\bf 673}, L91--L94. \newline ADS:
  \url{http://adsabs.harvard.edu/abs/2008ApJ...673L..91B}

\bibitem[Centeno {\it et~al.\/}(2008)]{CentenoEtal2008}
Centeno, R., Trujillo~Bueno, J., Uitenbroek, H., Collados, M., 2008, ``The
  influence of coronal EUV irradiance on the emission in the He I 10830 {\AA}
  and D$_{3}$ multiplets'', {\it Astrophys. J.\/}, {\bf 677}, 742--750.
  \newline ADS: \url{http://adsabs.harvard.edu/abs/2008ApJ...677..742C}

\bibitem[Chae {\it et~al.\/}(1997)]{ChaeEtal1997}
Chae, J., Yun, H.S., Poland, A.I., 1997, ``Effects of Non-LTE Radiative Loss
  and Partial Ionization on the Structure of the Transition Region'', {\it
  Astrophys. J.\/}, {\bf 480}, 817--824. \newline ADS:
  \url{http://adsabs.harvard.edu/abs/1997ApJ...480..817C}

\bibitem[Chandran(2005)]{Chandran2005}
Chandran, B.D.G., 2005, ``Weak Compressible Magnetohydrodynamic Turbulence in
  the Solar Corona'', {\it Phys. Rev. Lett.\/}, {\bf 95}, 265\,004. \newline
  ADS: \url{http://adsabs.harvard.edu/abs/2005PhRvL..95z5004C}

\bibitem[Chandran {\it et~al.\/}(2009a)]{ChandranEtal2009a}
Chandran, B.D.G., Quataert, E., Howes, G.G., Hollweg, J.V., Dorland, W., 2009a,
  ``The Turbulent Heating Rate in Strong MHD Turbulence with Nonzero Cross
  Helicity'', {\it Astrophys. J.\/}, {\bf 701}, 652--657. \newline ADS:
  \url{http://adsabs.harvard.edu/abs/2009ApJ...701..652C}

\bibitem[Chandran {\it et~al.\/}(2009b)]{ChandranEtal2009b}
Chandran, B.D.G., Quataert, E., Howes, G.G., Xia, Q., Pongkitiwanichakul, P.,
  2009b, ``Constraining Low-Frequency Alfv\'{e}nic Turbulence in the Solar Wind
  Using Density Fluctuation Measurements'', {\it Astrophys. J.\/}, submitted

\bibitem[Chen {\it et~al.\/}(2001)]{ChenEtal2001}
Chen, L., Lin, Z., White, R., 2001, ``On resonant heating below the cyclotron
  frequency'', {\it Phys. Plasmas\/}, {\bf 8}, 4713--4716. \newline ADS:
  \url{http://adsabs.harvard.edu/abs/2001PhPl....8.4713C}

\bibitem[Chen {\it et~al.\/}(2009)]{ChenEtal2009}
Chen, Y., Li, X., Song, H.Q., Shi, Q.Q., Feng, S.W., Xia, L.D., 2009,
  ``Intrinsic Instability of Coronal Streamers'', {\it Astrophys. J.\/}, {\bf
  691}, 1936--1942. \newline ADS:
  \url{http://adsabs.harvard.edu/abs/2009ApJ...691.1936C}

\bibitem[Cho {\it et~al.\/}(2002)]{ChoEtal2002}
Cho, J., Lazarian, A., Vishniac, E.T., 2002, ``Simulations of
  Magnetohydrodynamic Turbulence in a Strongly Magnetized Medium'', {\it
  Astrophys. J.\/}, {\bf 564}, 291--301. \newline ADS:
  \url{http://adsabs.harvard.edu/abs/2002ApJ...564..291C}

\bibitem[Cirtain {\it et~al.\/}(2007)]{CirtainEtal2007}
Cirtain, J.W., Golub, L., Lundquist, L., van Ballegooijen, A., Savcheva, A.,
  Shimojo, M., DeLuca, E., Tsuneta, S., Sakao, T., Reeves, K., Weber, M., Kano,
  R., Narukage, N., Shibasaki, K., 2007, ``Evidence for Alfv\'{e}n waves in
  solar X-ray jets'', {\it Science\/}, {\bf 318}, 1580--1582. \newline ADS:
  \url{http://adsabs.harvard.edu/abs/2007Sci...318.1580C}

\bibitem[Close {\it et~al.\/}(2003)]{CloseEtal2003}
Close, R.M., Parnell, C.E., Mackay, D.H., Priest, E.R., 2003, ``Statistical
  Flux Tube Properties of 3D Magnetic Carpet Fields'', {\it Solar Phys.\/},
  {\bf 212}, 251--275. \newline ADS:
  \url{http://adsabs.harvard.edu/abs/2003SoPh..212..251C}

\bibitem[Cohen {\it et~al.\/}(2007)]{CohenEtal2007}
Cohen, O., Sokolov, I.V., Roussev, I.I., Arge, C.N., Manchester, W.B., Gombosi,
  T.I., Frazin, R.A., Park, H., Butala, M.D., Kamalabadi, F., Velli, M., 2007,
  ``A Semiempirical Magnetohydrodynamical Model of the Solar Wind'', {\it
  Astrophys. J. Lett.\/}, {\bf 654}, L163--L166. \newline ADS:
  \url{http://adsabs.harvard.edu/abs/2007ApJ...654L.163C}

\bibitem[Coleman(1968)]{Coleman1968}
Coleman, P.J., 1968, ``Turbulence, Viscosity, and Dissipation in the Solar-Wind
  Plasma'', {\it Astrophys. J.\/}, {\bf 153}, 371--388. \newline ADS:
  \url{http://adsabs.harvard.edu/abs/1968ApJ...153..371C}

\bibitem[Collier {\it et~al.\/}(1996)]{CollierEtal1996}
Collier, M.R., Hamilton, D.C., Gloeckler, G., Bochsler, P., Sheldon, R.B.,
  1996, ``Neon-20, Oxygen-16, and Helium-4 densities, temperatures, and
  suprathermal tails in the solar wind determined with WIND/MASS'', {\it
  Geophys. Res. Lett.\/}, {\bf 23}, 1191--1194. \newline ADS:
  \url{http://adsabs.harvard.edu/abs/1996GeoRL..23.1191C}

\bibitem[Cranmer(2000)]{Cranmer2000}
Cranmer, S.R., 2000, ``Ion Cyclotron Wave Dissipation in the Solar Corona: The
  Summed Effect of more than 2000 Ion Species'', {\it Astrophys. J.\/}, {\bf
  532}, 1197--1208. \newline ADS:
  \url{http://adsabs.harvard.edu/abs/2000ApJ...532.1197C}

\bibitem[Cranmer(2001)]{Cranmer2001}
Cranmer, S.R., 2001, ``Ion cyclotron diffusion of velocity distributions in the
  extended solar corona'', {\it J. Geophys. Res.\/}, {\bf 106}, 24,937--24,954.
  \newline ADS: \url{http://adsabs.harvard.edu/abs/2001JGR...10624937C}

\bibitem[Cranmer(2002a)]{Cranmer2002a}
Cranmer, S.R., 2002a, ``Coronal Holes and the High-Speed Solar Wind'', {\it
  Space Sci. Rev.\/}, {\bf 101}, 229--294. \newline ADS:
  \url{http://adsabs.harvard.edu/abs/2002SSRv..101..229C}

\bibitem[Cranmer(2002b)]{Cranmer2002b}
Cranmer, S.R., 2002b, ``Solar wind acceleration in coronal holes'', in {\it
  From Solar Min to Max: Half a Solar Cycle with SOHO\/}, (Ed.) Wilson, A.,
  Proceedings of the SOHO-11 Symposium, 11--15 March 2002, Davos, Switzerland,
  vol. SP-508 of ESA Conference Proceedings, pp. 361--366, ESA Publications
  Division, Noordwijk. \newline ADS:
  \url{http://adsabs.harvard.edu/abs/2002ESASP.508..361C}

\bibitem[Cranmer(2004a)]{Cranmer2004a}
Cranmer, S.R., 2004a, ``Observational Aspects of Wave Acceleration in Open
  Magnetic Regions'', in {\it Waves, Oscillations, and Small-Scale Transient
  Events in the Solar Atmosphere\/}, (Ed.) Lacoste, H., Proceedings of the
  SOHO--13 Workshop, 29 September--3 October 2003, Palma de Mallorca, Spain,
  vol. SP-547 of ESA Conference Proceedings, pp. 353--362, ESA Publications
  Division, Noordwijk. \newline ADS:
  \url{http://adsabs.harvard.edu/abs/2004ESASP.547..353C}

\bibitem[Cranmer(2004b)]{Cranmer2004b}
Cranmer, S.R., 2004b, ``Coronal Heating versus Solar Wind Acceleration'', in
  {\it Coronal Heating\/}, (Eds.) Walsh, R.W., Ireland, J., Danesy, D., Fleck,
  B., Proceedings of the SOHO--15 Workshop, 6--9 September 2004, St. Andrews,
  Scotland, UK, vol. SP-575 of ESA Conference Proceedings, pp. 154--163, ESA
  Publications Division, Noordwijk. \newline ADS:
  \url{http://adsabs.harvard.edu/abs/2004ESASP.575..154C}

\bibitem[Cranmer(2005)]{Cranmer2005}
Cranmer, S.R., 2005, ``Why is the Fast Solar Wind Fast and the Slow Solar Wind
  Slow? A Survey of Geometrical Models'', in {\it Connecting Sun and
  Heliosphere\/}, (Eds.) Fleck, B., Zurbuchen, T.H., Lacoste, H., Proceedings
  of Solar Wind 11/SOHO--16, 12--17 June 2005, Whistler, Canada, vol. SP-592 of
  ESA Conference Proceedings, pp. 159--164, ESA Publications Division,
  Noordwijk. \newline ADS:
  \url{http://adsabs.harvard.edu/abs/2005ESASP.592..159C}

\bibitem[Cranmer(2007)]{Cranmer2007}
Cranmer, S.R., 2007, ``Turbulence in the solar corona'', in {\it Turbulence and
  Nonlinear Processes in Astrophysical Plasmas\/}, (Eds.) Shaikh, D., Zank,
  G.P., Proceedings of the Sixth International Astrophysics Conference, Oahu,
  Hawaii, 16--22 March 2007, vol. 932 of AIP Conference Proceedings, pp.
  327--332, American Institute of Physics, Melville. \newline ADS:
  \url{http://adsabs.harvard.edu/abs/2007AIPC..932..327C}

\bibitem[Cranmer(2008a)]{Cranmer2008a}
Cranmer, S.R., 2008a, ``Winds of Main-Sequence Stars: Observational Limits and
  a Path to Theoretical Prediction'', in {\it Proceedings of the 14th Cambridge
  Workshop on Cool Stars, Stellar Systems, and the Sun\/}, (Ed.) van Belle, G.,
  5--10 November 2006, Pasadena, California, vol. 384 of ASP Conf. Proc., pp.
  317--326, Astronomical Society of the Pacific, San Francisco. \newline ADS:
  \url{http://adsabs.harvard.edu/abs/2008ASPC..384..317C}

\bibitem[Cranmer(2008b)]{Cranmer2008b}
Cranmer, S.R., 2008b, ``Turbulence-driven polar winds from T Tauri stars
  energized by magnetospheric accretion'', {\it Astrophys. J.\/}, {\bf 689},
  316--334. \newline ADS:
  \url{http://adsabs.harvard.edu/abs/2008ApJ...689..316C}

\bibitem[Cranmer and van Ballegooijen(2003)]{CranmerBallegooijen2003}
Cranmer, S.R., van Ballegooijen, A.A., 2003, ``Alfv{\'{e}}nic turbulence in the
  extended solar corona: kinetic effects and proton heating'', {\it Astrophys.
  J.\/}, {\bf 594}, 573--591. \newline ADS:
  \url{http://adsabs.harvard.edu/abs/2003ApJ...594..573C}

\bibitem[Cranmer and van Ballegooijen(2005)]{CranmerBallegooijen2005}
Cranmer, S.R., van Ballegooijen, A.A., 2005, ``On the generation, propagation,
  and reflection of Alfv{\'{e}}n waves from the solar photosphere to the
  distant heliosphere'', {\it Astrophys. J. Suppl. Ser.\/}, {\bf 156},
  265--293. \newline ADS:
  \url{http://adsabs.harvard.edu/abs/2005ApJS..156..265C}

\bibitem[Cranmer {\it et~al.\/}(1999a)]{CranmerEtal1999a}
Cranmer, S.R., Field, G.B., Kohl, J.L., 1999a, ``Spectroscopic constraints on
  models of ion cyclotron resonance heating in the polar solar corona and
  high-speed solar wind'', {\it Astrophys. J.\/}, {\bf 518}, 937--947. \newline
  ADS: \url{http://adsabs.harvard.edu/abs/1999ApJ...518..937C}

\bibitem[Cranmer {\it et~al.\/}(1999b)]{CranmerEtal1999b}
Cranmer, S.R., Kohl, J.L., Noci, G., Antonucci, E., Tondello, G., Huber,
  M.C.E., Strachan, L., Panasyuk, A.V., Gardner, L.D., Romoli, M., Fineschi,
  S., Dobrzycka, D., Raymond, J.C., Nicolosi, P., Siegmund, O.H.W., Spadaro,
  D., Benna, C., Ciaravella, A., Giordano, S., Habbal, S.R., Karovska, M., Li,
  X., Martin, R., Michels, J.G., Modigliani, A., Naletto, G., O'Neal, R.H.,
  Pernechele, C., Poletto, G., Smith, P.L., Suleiman, R.M., 1999b, ``An
  Empirical Model of a Polar Coronal Hole at Solar Minimum'', {\it Astrophys.
  J.\/}, {\bf 511}, 481--501. \newline ADS:
  \url{http://adsabs.harvard.edu/abs/1999ApJ...511..481C}

\bibitem[Cranmer {\it et~al.\/}(2007)]{CranmerEtal2007}
Cranmer, S.R., van Ballegooijen, A.A., Edgar, R.J., 2007, ``Self-consistent
  coronal heating and solar wind acceleration from anisotropic
  magnetohydrodynamic turbulence'', {\it Astrophys. J. Suppl. Ser.\/}, {\bf
  171}, 520--551. \newline ADS:
  \url{http://adsabs.harvard.edu/abs/2007ApJS..171..520C}

\bibitem[Cranmer {\it et~al.\/}(2008)]{CranmerEtal2008}
Cranmer, S.R., Panasyuk, A.V., Kohl, J.L., 2008, ``Improved Constraints on the
  Preferential Heating and Acceleration of Oxygen Ions in the Extended Solar
  Corona'', {\it Astrophys. J.\/}, {\bf 678}, 1480--1497. \newline ADS:
  \url{http://adsabs.harvard.edu/abs/2008ApJ...678.1480C}

\bibitem[Cranmer {\it et~al.\/}(2009)]{CranmerEtal2009}
Cranmer, S.R., Matthaeus, W.H., Breech, B.A., Kasper, J.C., 2009, ``Empirical
  Constraints on Proton and Electron Heating in the Fast Solar Wind'', {\it
  Astrophys. J.\/}, {\bf 702}, 1604--1614. \newline ADS:
  \url{http://adsabs.harvard.edu/abs/2009ApJ...702.1604C}

\bibitem[Crooker(2004)]{Crooker2004}
Crooker, N.U., 2004, ``What Is the International Heliophysical Year?'', {\it
  Eos Trans. AGU\/}, {\bf 85}, 351--353. \newline ADS:
  \url{http://adsabs.harvard.edu/abs/2004EOSTr..85R.351C}

\bibitem[Culhane {\it et~al.\/}(2007)]{CulhaneEtal2007}
Culhane, L., Harra, L.K., Baker, D., van Driel-Gesztelyi, L., Sun, J., Doschek,
  G.A., Brooks, D.H., Lundquist, L.L., Kamio, S., Young, P.R., Hansteen, V.H.,
  2007, ``Hinode EUV Study of Jets in the Sun's South Polar Corona'', {\it Pub.
  Astron. Soc. Japan\/}, {\bf 59}, S751--S756. \newline ADS:
  \url{http://adsabs.harvard.edu/abs/2007PASJ...59S.751C}

\bibitem[Dasso {\it et~al.\/}(2005)]{DassoEtal2005}
Dasso, S., Milano, L.J., Matthaeus, W.H., Smith, C.W., 2005, ``Anisotropy in
  Fast and Slow Solar Wind Fluctuations'', {\it Astrophys. J. Lett.\/}, {\bf
  635}, L181--L184. \newline ADS:
  \url{http://adsabs.harvard.edu/abs/2005ApJ...635L.181D}

\bibitem[David {\it et~al.\/}(1998)]{DavidEtal1998}
David, C., Gabriel, A.H., Bely-Dubau, F., Fludra, A., Lemaire, P., Wilhelm, K.,
  1998, ``Measurement of the electron temperature gradient in a solar coronal
  hole'', {\it Astron. Astrophys.\/}, {\bf 336}, L90--L94. \newline ADS:
  \url{http://adsabs.harvard.edu/abs/1998A&A...336L..90D}

\bibitem[Davila {\it et~al.\/}(2009)]{DavilaEtal2009}
Davila, J.M., Gopalswamy, N., Thompson, B.J., 2009, ``Universal processes in
  heliophysics'', in {\it Universal Heliophysical Processes\/}, (Eds.)
  Gopalswamy, N., Webb, D.F., Proceedings of IAU Symp. 257, 15--19 September
  2008, Ioannina, Greece, pp. 11--16, Cambridge University Press, Cambridge.
  \newline ADS: \url{http://adsabs.harvard.edu/abs/2009IAUS..257...11D}

\bibitem[De~Pontieu {\it et~al.\/}(2007)]{DePontieuEtal2007}
De~Pontieu, B., McIntosh, S.W., Carlsson, M., Hansteen, V.H., Tarbell, T.D.,
  Schrijver, C.J., Title, A.M., Shine, R.A., Tsuneta, S., Katsukawa, Y.,
  Ichimoto, K., Suematsu, Y., Shimizu, T., Nagata, S., 2007, ``Chromospheric
  Alfv\'{e}nic waves strong enough to power the solar wind'', {\it Science\/},
  {\bf 318}, 1574--1577. \newline ADS:
  \url{http://adsabs.harvard.edu/abs/2007Sci...318.1574D}

\bibitem[de~Toma and Arge(2005)]{deTomaArge2005}
de~Toma, G., Arge, C.N., 2005, ``Multi-wavelength Observations of Coronal
  Holes'', in {\it Large-scale Structures and their Role in Solar Activity\/},
  (Eds.) Sankarasubramanian, K., Penn, M., Pevtsov, A., Proceedings of the 22nd
  Sacramento Peak Workshop, 18--22 October 2004, Sunspot, New Mexico, vol. 346
  of ASP Conf. Proc., pp. 251--260, Astronomical Society of the Pacific, San
  Francisco. \newline ADS:
  \url{http://adsabs.harvard.edu/abs/2005ASPC..346..251T}

\bibitem[DeForest and Gurman(1998)]{DeForestGurman1998}
DeForest, C.E., Gurman, J.B., 1998, ``Observation of Quasi-periodic Compressive
  Waves in Solar Polar Plumes'', {\it Astrophys. J. Lett.\/}, {\bf 501},
  L217--L220. \newline ADS:
  \url{http://adsabs.harvard.edu/abs/1998ApJ...501L.217D}

\bibitem[DeForest {\it et~al.\/}(1997)]{DeForestEtal1997}
DeForest, C.E., Hoeksema, J.T., Gurman, J.B., Thompson, B.J., Plunkett, S.P.,
  Howard, R., Harrison, R.C., Hassler, D.M., 1997, ``Polar Plume Anatomy:
  Results of a Coordinated Observation'', {\it Solar Phys.\/}, {\bf 175},
  393--410. \newline ADS:
  \url{http://adsabs.harvard.edu/abs/1997SoPh..175..393D}

\bibitem[DeForest {\it et~al.\/}(2001)]{DeForestEtal2001}
DeForest, C.E., Lamy, P.L., Llebaria, A., 2001, ``Solar Polar Plume Lifetime
  and Coronal Hole Expansion: Determination from Long-Term Observations'', {\it
  Astrophys. J.\/}, {\bf 560}, 490--498. \newline ADS:
  \url{http://adsabs.harvard.edu/abs/2001ApJ...560..490D}

\bibitem[Del~Zanna and Bromage(1999)]{DelZannaBromage1999}
Del~Zanna, G., Bromage, B.J.I., 1999, ``The Elephant's Trunk: Spectroscopic
  diagnostics applied to SOHO/CDS observations of the August 1996 equatorial
  coronal hole'', {\it J. Geophys. Res.\/}, {\bf 104}, 9753--9766. \newline
  ADS: \url{http://adsabs.harvard.edu/abs/1999JGR...104.9753D}

\bibitem[Del~Zanna {\it et~al.\/}(1998)]{DelZannaEtal1998}
Del~Zanna, L., von Steiger, R., Velli, M., 1998, ``The Expansion of Coronal
  Plumes in the Fast Solar Wind'', {\it Space Sci. Rev.\/}, {\bf 85}, 349--356.
  \newline ADS: \url{http://adsabs.harvard.edu/abs/1998SSRv...85..349D}

\bibitem[Dmitruk and Matthaeus(2003)]{DmitrukMatthaeus2003}
Dmitruk, P., Matthaeus, W.H., 2003, ``Low-Frequency Waves and Turbulence in an
  Open Magnetic Region: Timescales and Heating Efficiency'', {\it Astrophys.
  J.\/}, {\bf 597}, 1097--1105. \newline ADS:
  \url{http://adsabs.harvard.edu/abs/2003ApJ...597.1097D}

\bibitem[Dmitruk {\it et~al.\/}(2001)]{DmitrukEtal2001}
Dmitruk, P., Milano, L.J., Matthaeus, W.H., 2001, ``Wave-driven Turbulent
  Coronal Heating in Open Field Line Regions: Nonlinear Phenomenological
  Model'', {\it Astrophys. J.\/}, {\bf 548}, 482--491. \newline ADS:
  \url{http://adsabs.harvard.edu/abs/2001ApJ...548..482D}

\bibitem[Dmitruk {\it et~al.\/}(2002)]{DmitrukEtal2002}
Dmitruk, P., Matthaeus, W.H., Milano, L.J., Oughton, S., Zank, G.P., Mullan,
  D.J., 2002, ``Coronal Heating Distribution Due to Low-Frequency, Wave-driven
  Turbulence'', {\it Astrophys. J.\/}, {\bf 575}, 571--577. \newline ADS:
  \url{http://adsabs.harvard.edu/abs/2002ApJ...575..571D}

\bibitem[Dmitruk {\it et~al.\/}(2004)]{DmitrukEtal2004}
Dmitruk, P., Matthaeus, W.H., Seenu, N., 2004, ``Test Particle Energization by
  Current Sheets and Nonuniform Fields in Magnetohydrodynamic Turbulence'',
  {\it Astrophys. J.\/}, {\bf 617}, 667--679. \newline ADS:
  \url{http://adsabs.harvard.edu/abs/2004ApJ...617..667D}

\bibitem[Dobrowolny {\it et~al.\/}(1980)]{DobrowolnyEtal1980}
Dobrowolny, M., Mangeney, A., Veltri, P., 1980, ``Fully developed anisotropic
  hydromagnetic turbulence in interplanetary space'', {\it Phys. Rev. Lett.\/},
  {\bf 45}, 144--147. \newline ADS:
  \url{http://adsabs.harvard.edu/abs/1980PhRvL..45..144D}

\bibitem[Dobrzycka {\it et~al.\/}(2000)]{DobrzyckaEtal2000}
Dobrzycka, D., Raymond, J.C., Cranmer, S.R., 2000, ``Ultraviolet Spectroscopy
  of Polar Coronal Jets'', {\it Astrophys. J.\/}, {\bf 538}, 922--931. \newline
  ADS: \url{http://adsabs.harvard.edu/abs/2000ApJ...538..922D}

\bibitem[Dobrzycka {\it et~al.\/}(2006)]{DobrzyckaEtal2006}
Dobrzycka, D., Raymond, J., DeLuca, E., Gurman, J., Fludra, A., Biesecker, D.,
  2006, ``SOHO Observations of Polar Coronal Jets over the Last Solar Cycle'',
  in {\it 10 Years of SOHO and Beyond\/}, (Eds.) Lacoste, H., Ouwehand, L.,
  Proceedings of the SOHO-17 Conference, 7--12 May 2006, Giardini Naxos, Italy,
  vol. SP-617 of ESA Conference Proceedings, pp. 87.1--87.4, ESA Publications
  Division, Noordwijk. \newline ADS:
  \url{http://adsabs.harvard.edu/abs/2006ESASP.617E..87D}

\bibitem[Dolla and Solomon(2008)]{DollaSolomon2008}
Dolla, L., Solomon, J., 2008, ``Solar off-limb line widths: Alfv\'{e}n waves,
  ion-cyclotron waves, and preferential heating'', {\it Astron. Astrophys.\/},
  {\bf 483}, 271--283. \newline ADS:
  \url{http://adsabs.harvard.edu/abs/2008A&A...483..271D}

\bibitem[Doschek {\it et~al.\/}(2001)]{DoschekEtal2001}
Doschek, G.A., Feldman, U., Laming, J.M., Sch\"{u}hle, U., Wilhelm, K., 2001,
  ``Properties of Solar Polar Coronal Hole Plasmas Observed above the Limb'',
  {\it Astrophys. J.\/}, {\bf 546}, 559--568. \newline ADS:
  \url{http://adsabs.harvard.edu/abs/2001ApJ...546..559D}

\bibitem[Dowdy {\it et~al.\/}(1986)]{DowdyEtal1986}
Dowdy, J.F., Rabin, D., Moore, R.L., 1986, ``On the magnetic structure of the
  quiet transition region'', {\it Solar Phys.\/}, {\bf 105}, 35--45. \newline
  ADS: \url{http://adsabs.harvard.edu/abs/1986SoPh..105...35D}

\bibitem[Doyle {\it et~al.\/}(1999)]{DoyleEtal1999}
Doyle, J.G., Teriaca, L., Banerjee, D., 1999, ``Coronal hole diagnostics out to
  8 $R_{\odot}$'', {\it Astron. Astrophys.\/}, {\bf 349}, 956--960. \newline
  ADS: \url{http://adsabs.harvard.edu/abs/1999A&A...349..956D}

\bibitem[Dunn and Zirker(1973)]{DunnZirker1973}
Dunn, R.B., Zirker, J.B., 1973, ``The Solar Filigree'', {\it Solar Phys.\/},
  {\bf 33}, 281--304. \newline ADS:
  \url{http://adsabs.harvard.edu/abs/1973SoPh...33..281D}

\bibitem[Dupree {\it et~al.\/}(1996)]{DupreeEtal1996}
Dupree, A.K., Penn, M.J., Jones, H.P., 1996, ``He I 10830 {\AA} Wing Asymmetry
  in Polar Coronal Holes: Evidence for Radial Outflows'', {\it Astrophys. J.
  Lett.\/}, {\bf 467}, L121--L124. \newline ADS:
  \url{http://adsabs.harvard.edu/abs/1996ApJ...467L.121D}

\bibitem[Dupree {\it et~al.\/}(2005)]{DupreeEtal2005}
Dupree, A.K., Brickhouse, N.S., Smith, G.H., Strader, J., 2005, ``A Hot Wind
  from the Classical T Tauri Stars: TW Hydrae and T Tauri'', {\it Astrophys. J.
  Lett.\/}, {\bf 625}, L131--L134. \newline ADS:
  \url{http://adsabs.harvard.edu/abs/2005ApJ...625L.131D}

\bibitem[Eastwood(2008)]{Eastwood2008}
Eastwood, J.P., 2008, ``The science of space weather'', {\it Phil. Trans. Roy.
  Soc. A\/}, {\bf 366}, 4489--4500. \newline ADS:
  \url{http://adsabs.harvard.edu/abs/2008RSPTA.366.4489E}

\bibitem[Ergun {\it et~al.\/}(1999)]{ErgunEtal1999}
Ergun, R.E., Carlson, C.W., Muschietti, L., Roth, I., McFadden, J.P., 1999,
  ``Properties of fast solitary structures'', {\it Nonlinear Proc. Geophys.\/},
  {\bf 6}, 187--194. \newline ADS:
  \url{http://adsabs.harvard.edu/abs/1999NPGeo...6..187E}

\bibitem[Esser and Edgar(2000)]{EsserEdgar2000}
Esser, R., Edgar, R.J., 2000, ``Reconciling Spectroscopic Electron Temperature
  Measurements in the Solar Corona with In Situ Charge State Observations'',
  {\it Astrophys. J.\/}, {\bf 532}, L71--L74. \newline ADS:
  \url{http://adsabs.harvard.edu/abs/2000ApJ...532L..71E}

\bibitem[Esser {\it et~al.\/}(1999)]{EsserEtal1999}
Esser, R., Fineschi, S., Dobrzycka, D., Habbal, S.R., Edgar, R.J., Raymond,
  J.C., Kohl, J.L., Guhathakurta, M., 1999, ``Plasma Properties in Coronal
  Holes Derived from Measurements of Minor Ion Spectral Lines and Polarized
  White Light Intensity'', {\it Astrophys. J. Lett.\/}, {\bf 510}, L63--L67.
  \newline ADS: \url{http://adsabs.harvard.edu/abs/1999ApJ...510L..63E}

\bibitem[Feigelson and Montmerle(1999)]{FeigelsonMontmerle1999}
Feigelson, E.D., Montmerle, T., 1999, ``High-energy processes in young stellar
  objects'', {\it Ann. Rev. Astron. Astrophys.\/}, {\bf 37}, 363--408. \newline
  ADS: \url{http://adsabs.harvard.edu/abs/1999ARA&A..37..363F}

\bibitem[Feldman(1998)]{Feldman1998}
Feldman, U., 1998, ``FIP Effect in the Solar Upper Atmosphere: Spectroscopic
  Results'', {\it Space Sci. Rev.\/}, {\bf 85}, 227--240. \newline ADS:
  \url{http://adsabs.harvard.edu/abs/1998SSRv...85..227F}

\bibitem[Feldman and Widing(2003)]{FeldmanWiding2003}
Feldman, U., Widing, K.G., 2003, ``Elemental Abundances in the Solar Upper
  Atmosphere Derived by Spectroscopic Means'', {\it Space Sci. Rev.\/}, {\bf
  107}, 665--720. \newline ADS:
  \url{http://adsabs.harvard.edu/abs/2003SSRv..107..665F}

\bibitem[Feldman {\it et~al.\/}(1999)]{FeldmanEtal1999}
Feldman, U., Widing, K.G., Warren, H.P., 1999, ``Morphology of the quiet solar
  upper atmosphere in the $4 \times 10^{4} < T_{e} < 1.4 \times 10^{6}$ K
  temperature regime'', {\it Astrophys. J.\/}, {\bf 522}, 1133--1147. \newline
  ADS: \url{http://adsabs.harvard.edu/abs/1999ApJ...522.1133F}

\bibitem[Feldman {\it et~al.\/}(2005)]{FeldmanEtal2005}
Feldman, U., Landi, E., Schwadron, N.A., 2005, ``On the sources of fast and
  slow solar wind'', {\it J. Geophys. Res.\/}, {\bf 110}, A07\,109. \newline
  ADS: \url{http://adsabs.harvard.edu/abs/2005JGRA..11007109F}

\bibitem[Feldman {\it et~al.\/}(2007)]{FeldmanEtal2007}
Feldman, U., Landi, E., Doschek, G.A., 2007, ``Diagnostics of Suprathermal
  Electrons in Active-Region Plasmas Using He-like UV Lines'', {\it Astrophys.
  J.\/}, {\bf 660}, 1674--1682. \newline ADS:
  \url{http://adsabs.harvard.edu/abs/2007ApJ...660.1674F}

\bibitem[Feng {\it et~al.\/}(2007)]{FengEtal2007}
Feng, X., Zhou, Y., Wu, S.T., 2007, ``A Novel Numerical Implementation for
  Solar Wind Modeling by the Modified Conservation Element/Solution Element
  Method'', {\it Astrophys. J.\/}, {\bf 655}, 1110--1126. \newline ADS:
  \url{http://adsabs.harvard.edu/abs/2007ApJ...655.1110F}

\bibitem[Ferreira {\it et~al.\/}(2006)]{FerreiraEtal2006}
Ferreira, J., Dougados, C., Cabrit, S., 2006, ``Which jet launching
  mechanism(s) in T Tauri stars?'', {\it Astron. Astrophys.\/}, {\bf 453},
  785--796. \newline ADS:
  \url{http://adsabs.harvard.edu/abs/2008A&A...453..785F}

\bibitem[Feynman and Gabriel(2000)]{FeynmanGabriel2000}
Feynman, J., Gabriel, S.B., 2000, ``On space weather consequences and
  predictions'', {\it J. Geophys. Res.\/}, {\bf 105}, 10\,543--10\,564.
  \newline ADS: \url{http://adsabs.harvard.edu/abs/2000JGR...10510543F}

\bibitem[Filippov {\it et~al.\/}(2009)]{FilippovEtal2009}
Filippov, B., Golub, L., Koutchmy, S., 2009, ``X-Ray Jet Dynamics in a Polar
  Coronal Hole Region'', {\it Solar Phys.\/}, {\bf 254}, 259--269. \newline
  ADS: \url{http://adsabs.harvard.edu/abs/2009SoPh..254..259F}

\bibitem[Fisher and Guhathakurta(1995)]{FisherGuhathakurta1995}
Fisher, R., Guhathakurta, M., 1995, ``Physical Properties of Polar Coronal Rays
  and Holes as Observed with the Spartan 201-01 Coronagraph'', {\it Astrophys.
  J. Lett.\/}, {\bf 447}, L139--L142. \newline ADS:
  \url{http://adsabs.harvard.edu/abs/1995ApJ...447L.139F}

\bibitem[Fisher and Sime(1984)]{FisherSime1984}
Fisher, R., Sime, D.G., 1984, ``Solar activity cycle variation of the K
  corona'', {\it Astrophys. J.\/}, {\bf 285}, 354--358. \newline ADS:
  \url{http://adsabs.harvard.edu/abs/1984ApJ...285..354F}

\bibitem[Fisk(2003)]{Fisk2003}
Fisk, L.A., 2003, ``Acceleration of the solar wind as a result of the
  reconnection of open magnetic flux with coronal loops'', {\it J. Geophys.
  Res.\/}, {\bf 108}, 1157. \newline ADS:
  \url{http://adsabs.harvard.edu/abs/2003JGRA..108.1157F}

\bibitem[Fisk(2005)]{Fisk2005}
Fisk, L.A., 2005, ``The open magnetic flux of the Sun: I: Transport by
  reconnections with coronal loops'', {\it Astrophys. J.\/}, {\bf 626},
  563--573. \newline ADS:
  \url{http://adsabs.harvard.edu/abs/2005ApJ...626..563F}

\bibitem[Fisk and Schwadron(2001)]{FiskSchwadron2001}
Fisk, L.A., Schwadron, N.A., 2001, ``The Behavior of the Open Magnetic Field of
  the Sun'', {\it Astrophys. J.\/}, {\bf 560}, 425--438. \newline ADS:
  \url{http://adsabs.harvard.edu/abs/2001ApJ...560..425F}

\bibitem[Fisk and Zhao(2009)]{FiskZhao2009}
Fisk, L.A., Zhao, L., 2009, ``The heliospheric magnetic field and the solar
  wind during the solar cycle'', in {\it Universal Heliophysical Processes\/},
  (Eds.) Gopalswamy, N., Webb, D.F., Proceedings of IAU Symp. 257, 15--19
  September 2008, Ioannina, Greece, pp. 109--120, Cambridge University Press,
  Cambridge. \newline ADS:
  \url{http://adsabs.harvard.edu/abs/2009IAUS..257..109F}

\bibitem[Fisk and Zurbuchen(2006)]{FiskZurbuchen2006}
Fisk, L.A., Zurbuchen, T.H., 2006, ``Distribution and properties of open
  magnetic flux outside of coronal holes'', {\it J. Geophys. Res.\/}, {\bf
  111}, A09\,115. \newline ADS:
  \url{http://adsabs.harvard.edu/abs/2006JGRA..11109115F}

\bibitem[Fisk {\it et~al.\/}(1999)]{FiskEtal1999}
Fisk, L.A., Schwadron, N.A., Zurbuchen, T.H., 1999, ``Acceleration of the fast
  solar wind by the emergence of new magnetic flux'', {\it J. Geophys. Res.\/},
  {\bf 104}, 19\,765--19\,772. \newline ADS:
  \url{http://adsabs.harvard.edu/abs/1999JGR...10419765F}

\bibitem[Frazin {\it et~al.\/}(2007)]{FrazinEtal2007}
Frazin, R.A., V\'{a}squez, A.M., Kamalabadi, F., Park, H., 2007,
  ``Three-dimensional Tomographic Analysis of a High-Cadence LASCO-C2 Polarized
  Brightness Sequence'', {\it Astrophys. J. Lett.\/}, {\bf 671}, L201--L204.
  \newline ADS: \url{http://adsabs.harvard.edu/abs/2007ApJ...671L.201F}

\bibitem[Gabriel(1976)]{Gabriel1976}
Gabriel, A.H., 1976, ``A magnetic model of the solar transition region'', {\it
  Phil. Trans. Roy. Soc. A\/}, {\bf 281}, 339--352. \newline ADS:
  \url{http://adsabs.harvard.edu/abs/1976RSPTA.281..339G}

\bibitem[Gabriel {\it et~al.\/}(2003)]{GabrielEtal2003}
Gabriel, A.H., Bely-Dubau, F., Lemaire, P., 2003, ``The contribution of polar
  plumes to the fast solar wind'', {\it Astrophys. J.\/}, {\bf 589}, 623--634.
  \newline ADS: \url{http://adsabs.harvard.edu/abs/2003ApJ...589..623G}

\bibitem[Galinsky and Shevchenko(2000)]{GalinskyShevchenko2000}
Galinsky, V.L., Shevchenko, V.I., 2000, ``Nonlinear Cyclotron Resonant
  Wave-Particle Interaction in a Nonuniform Magnetic Field'', {\it Phys. Rev.
  Lett.\/}, {\bf 85}, 90--93. \newline ADS:
  \url{http://adsabs.harvard.edu/abs/2000PhRvL..85...90G}

\bibitem[Gardner {\it et~al.\/}(2003)]{GardnerEtal2003}
Gardner, L.D., Kohl, J.L., Daigneau, P.S., Smith, P.L., Strachan, L., Howard,
  R.A., Socker, D.G., Davila, J.M., Noci, G., Romoli, M., Fineschi, S., 2003,
  ``Advanced spectroscopic and coronographic explorer: Science payload design
  concept'', {\it Proc. SPIE\/}, {\bf 4843}, 1--7. \newline ADS:
  \url{http://adsabs.harvard.edu/abs/2003SPIE.4843....1G}

\bibitem[Gary and Borovsky(2008)]{GaryBorovsky2008}
Gary, S.P., Borovsky, J.E., 2008, ``Damping of long-wavelength kinetic
  Alfv\'{e}n fluctuations: Linear theory'', {\it J. Geophys. Res.\/}, {\bf
  113}, A12\,104. \newline ADS:
  \url{http://adsabs.harvard.edu/abs/2008JGRA..11312104G}

\bibitem[Gibson {\it et~al.\/}(1999)]{GibsonEtal1999}
Gibson, S.E., Fludra, A., Bagenal, F., Biesecker, D., Del~Zanna, G., Bromage,
  B., 1999, ``Solar minimum streamer densities and temperatures using Whole Sun
  Month coordinated data sets'', {\it J. Geophys. Res.\/}, {\bf 104},
  9691--9700. \newline ADS:
  \url{http://adsabs.harvard.edu/abs/1999JGR...104.9691G}

\bibitem[Giordano {\it et~al.\/}(2000)]{GiordanoEtal2000}
Giordano, S., Antonucci, E., Noci, G., Romoli, M., Kohl, J.L., 2000,
  ``Identification of the Coronal Sources of the Fast Solar Wind'', {\it
  Astrophys. J. Lett.\/}, {\bf 531}, L79--L82. \newline ADS:
  \url{http://adsabs.harvard.edu/abs/2000ApJ...531L..79G}

\bibitem[Gloeckler {\it et~al.\/}(2003)]{GloecklerEtal2003}
Gloeckler, G., Zurbuchen, T.H., Geiss, J., 2003, ``Implications of the observed
  anticorrelation between solar wind speed and coronal electron temperature'',
  {\it J. Geophys. Res.\/}, {\bf 108}, 1158. \newline ADS:
  \url{http://adsabs.harvard.edu/abs/2003JGRA..108.1158G}

\bibitem[Gold(1955)]{Gold1955}
Gold, T., 1955, ``The symmetry of the corona of 1954 June 30'', {\it Mon. Not.
  Roy. Astron. Soc.\/}, {\bf 115}, 340--342. \newline ADS:
  \url{http://adsabs.harvard.edu/abs/1955MNRAS.115..340G}

\bibitem[Goldreich and Sridhar(1995)]{GoldreichSridhar1995}
Goldreich, P., Sridhar, S., 1995, ``Toward a Theory of Interstellar Turbulence:
  2, Strong Alfv\'{e}nic Turbulence'', {\it Astrophys. J.\/}, {\bf 438},
  763--775. \newline ADS:
  \url{http://adsabs.harvard.edu/abs/1995ApJ...438..763G}

\bibitem[Goldstein {\it et~al.\/}(1996)]{GoldsteinEtal1996}
Goldstein, B.E., Neugebauer, M., Phillips, J.L., Bame, S., Gosling, J.T.,
  McComas, D., Wang, Y.-M., Sheeley~Jr, N.R., Suess, S.T., 1996, ``Ulysses
  plasma parameters: Latitudinal, radial, and temporal variations'', {\it
  Astron. Astrophys.\/}, {\bf 316}, 296--303. \newline ADS:
  \url{http://adsabs.harvard.edu/abs/1996A&A...316..296G}

\bibitem[Goldstein {\it et~al.\/}(1997)]{GoldsteinEtal1997}
Goldstein, M.L., Roberts, D.A., Matthaeus, W.H., 1997, ``Magnetohydrodynamic
  Turbulence in Cosmic Winds'', in {\it Cosmic Winds and the Heliosphere\/},
  (Eds.) Jokipii, J.R., Sonett, C.P., Giampapa, M.S., Space Science Series, pp.
  521--580, University of Arizona Press, Tucson

\bibitem[G\'{o}mez {\it et~al.\/}(2000)]{GomezEtal2000}
G\'{o}mez, D.O., Dmitruk, P.A., Milano, L.J., 2000, ``Recent theoretical
  results on coronal heating'', {\it Solar Phys.\/}, {\bf 195}, 299--318.
  \newline ADS: \url{http://adsabs.harvard.edu/abs/2000SoPh..195..299G}

\bibitem[Gosling and Szabo(2008)]{GoslingSzabo2008}
Gosling, J.T., Szabo, A., 2008, ``Bifurcated current sheets produced by
  magnetic reconnection in the solar wind'', {\it J. Geophys. Res.\/}, {\bf
  113}, A10\,103. \newline ADS:
  \url{http://adsabs.harvard.edu/abs/2008JGRA..11310103G}

\bibitem[Gosling {\it et~al.\/}(2005)]{GoslingEtal2005}
Gosling, J.T., Skoug, R.M., McComas, D.J., Smith, C.W., 2005, ``Magnetic
  disconnection from the Sun: Observations of a reconnection exhaust in the
  solar wind at the heliospheric current sheet'', {\it Geophys. Res. Lett.\/},
  {\bf 32}, L05\,105. \newline ADS:
  \url{http://adsabs.harvard.edu/abs/2005GeoRL..3205105G}

\bibitem[G\"{u}del(2007)]{Gudel2007}
G\"{u}del, M., 2007, ``The Sun in Time: Activity and Environment'', {\it Living
  Rev. Solar Phys.\/}, {\bf 4}, lrsp-2007-3. URL (cited on 1 July 2009):
  \newline\url{http://www.livingreviews.org/lrsp-2007-3}

\bibitem[G\"{u}del {\it et~al.\/}(2001)]{GudelEtal2001}
G\"{u}del, M., Audard, M., Briggs, K., Haberl, F., Magee, H., Maggio, A., Mewe,
  R., Pallavicini, R., Pye, J., 2001, ``The XMM-Newton view of stellar coronae:
  X-ray spectroscopy of the corona of AB Doradus'', {\it Astron. Astrophys.\/},
  {\bf 365}, L336--L343. \newline ADS:
  \url{http://adsabs.harvard.edu/abs/2001A&A...365L.336G}

\bibitem[G\"{u}del {\it et~al.\/}(2003)]{GudelEtal2003}
G\"{u}del, M., Arzner, K., Audard, M., Mewe, R., 2003, ``Tomography of a
  stellar X-ray corona: alpha Coronae Borealis'', {\it Astron. Astrophys.\/},
  {\bf 403}, 155--171. \newline ADS:
  \url{http://adsabs.harvard.edu/abs/2003A&A...403..155G}

\bibitem[Gudiksen(2005)]{Gudiksen2005}
Gudiksen, B.V., 2005, ``DC Heating: Is it Enough?'', in {\it Connecting Sun and
  Heliosphere\/}, (Eds.) Fleck, B., Zurbuchen, T.H., Lacoste, H., Proceedings
  of Solar Wind 11/SOHO--16, 12--17 June 2005, Whistler, Canada, vol. SP-592 of
  ESA Conference Proceedings, pp. 165--169, ESA Publications Division,
  Noordwijk. \newline ADS:
  \url{http://adsabs.harvard.edu/abs/2005ESASP.592..165G}

\bibitem[Guhathakurta and Holzer(1994)]{GuhathakurtaHolzer1994}
Guhathakurta, M., Holzer, T.E., 1994, ``Density structure inside a polar
  coronal hole'', {\it Astrophys. J.\/}, {\bf 426}, 782--786. \newline ADS:
  \url{http://adsabs.harvard.edu/abs/1994ApJ...426..782G}

\bibitem[Guhathakurta {\it et~al.\/}(1999a)]{GuhathakurtaEtal1999a}
Guhathakurta, M., Sittler, E., Fisher, R., McComas, D., Thompson, B., 1999a,
  ``Coronal magnetic field topology and source of fast solar wind'', {\it
  Geophys. Res. Lett.\/}, {\bf 26}, 2901--2904. \newline ADS:
  \url{http://adsabs.harvard.edu/abs/1999GeoRL..26.2901G}

\bibitem[Guhathakurta {\it et~al.\/}(1999b)]{GuhathakurtaEtal1999b}
Guhathakurta, M., Fludra, A., Gibson, S.E., Biesecker, D., Fisher, R., 1999b,
  ``Physical properties of a coronal hole from a coronal diagnostic
  spectrometer, Mauna Loa Coronagraph, and LASCO observations during the Whole
  Sun Month'', {\it J. Geophys. Res.\/}, {\bf 104}, 9801--9808. \newline ADS:
  \url{http://adsabs.harvard.edu/abs/1999JGR...104.9801G}

\bibitem[Guo {\it et~al.\/}(2008)]{GuoEtal2008}
Guo, Z., Crabtree, C., Chen, L., 2008, ``Theory of charged particle heating by
  low-frequency Alfv\'{e}n waves'', {\it Phys. Plasmas\/}, {\bf 15}, 032311.
  \newline ADS: \url{http://adsabs.harvard.edu/abs/2008PhPl...15c2311G}

\bibitem[Habbal {\it et~al.\/}(1993)]{HabbalEtal1993}
Habbal, S.R., Esser, R., Arndt, M.B., 1993, ``How reliable are coronal hole
  temperatures deduced from observations?'', {\it Astrophys. J.\/}, {\bf 413},
  435--444. \newline ADS:
  \url{http://adsabs.harvard.edu/abs/1993ApJ...413..435H}

\bibitem[Habbal {\it et~al.\/}(2001)]{HabbalEtal2001}
Habbal, S.R., Woo, R., Arnaud, J., 2001, ``On the Predominance of the Radial
  Component of the Magnetic Field in the Solar Corona'', {\it Astrophys. J.\/},
  {\bf 558}, 852--858. \newline ADS:
  \url{http://adsabs.harvard.edu/abs/2001ApJ...558..852H}

\bibitem[Habbal {\it et~al.\/}(2008)]{HabbalEtal2008}
Habbal, S.R., Scholl, I.F., McIntosh, S.W., 2008, ``Impact of Active Regions on
  Coronal Hole Outflows'', {\it Astrophys. J. Lett.\/}, {\bf 683}, L75--L78.
  \newline ADS: \url{http://adsabs.harvard.edu/abs/2008ApJ...683L..75H}

\bibitem[Hagenaar {\it et~al.\/}(1999)]{HagenaarEtal1999}
Hagenaar, H.J., Schrijver, C.J., Title, A.M., Shine, R.A., 1999, ``Dispersal of
  Magnetic Flux in the Quiet Solar Photosphere'', {\it Astrophys. J.\/}, {\bf
  511}, 932--944. \newline ADS:
  \url{http://adsabs.harvard.edu/abs/1999ApJ...511..932H}

\bibitem[Hagenaar {\it et~al.\/}(2008)]{HagenaarEtal2008}
Hagenaar, H.J., DeRosa, M.L., Schrijver, C.J., 2008, ``The Dependence of
  Ephemeral Region Emergence on Local Flux Imbalance'', {\it Astrophys. J.\/},
  {\bf 678}, 541--548. \newline ADS:
  \url{http://adsabs.harvard.edu/abs/2008ApJ...678..541H}

\bibitem[Hansteen and Leer(1995)]{HansteenLeer1995}
Hansteen, V.H., Leer, E., 1995, ``Coronal heating, densities, and temperatures
  and solar wind acceleration'', {\it J. Geophys. Res.\/}, {\bf 100},
  21,577--21,594. \newline ADS:
  \url{http://adsabs.harvard.edu/abs/1995JGR...10021577H}

\bibitem[Hansteen {\it et~al.\/}(1997)]{HansteenEtal1997}
Hansteen, V.H., Leer, E., Holzer, T.E., 1997, ``The Role of Helium in the Outer
  Solar Atmosphere'', {\it Astrophys. J.\/}, {\bf 482}, 498--509. \newline ADS:
  \url{http://adsabs.harvard.edu/abs/1997ApJ...482..498H}

\bibitem[Harmon and Coles(2005)]{HarmonColes2005}
Harmon, J.K., Coles, W.A., 2005, ``Modeling radio scattering and scintillation
  observations of the inner solar wind using oblique Alfv\'{e}n/ion cyclotron
  waves'', {\it J. Geophys. Res.\/}, {\bf 110}, A03\,101. \newline ADS:
  \url{http://adsabs.harvard.edu/abs/2005JGRA..11003101H}

\bibitem[Harra {\it et~al.\/}(2007)]{HarraEtal2007}
Harra, L.K., Hara, H., Imada, S., Young, P.R., Williams, D.R., Sterling, A.C.,
  Korendyke, C., Attrill, G.D.R., 2007, ``Coronal Dimming Observed with Hinode:
  Outflows Related to a Coronal Mass Ejection'', {\it Pub. Astron. Soc.
  Japan\/}, {\bf 59}, S801--S806. \newline ADS:
  \url{http://adsabs.harvard.edu/abs/2007PASJ...59S.801H}

\bibitem[Hartigan {\it et~al.\/}(1995)]{HartiganEtal1995}
Hartigan, P., Edwards, S., Ghandour, L., 1995, ``Disk Accretion and Mass Loss
  from Young Stars'', {\it Astrophys. J.\/}, {\bf 452}, 736--768. \newline ADS:
  \url{http://adsabs.harvard.edu/abs/1995ApJ...452..736H}

\bibitem[Hartmann(2000)]{Hartmann2000}
Hartmann, L., 2000, {\it Accretion Processes in Star Formation\/}, Cambridge
  University Press, Cambridge

\bibitem[Harvey {\it et~al.\/}(1975)]{HarveyEtal1975}
Harvey, J., Krieger, A.S., Timothy, A.F., Vaiana, G.S., 1975, ``Comparison of
  Skylab X-ray and Ground-based Helium Observations'', {\it Osserv. Mem. Oss.
  Astrofis. Arcetri\/}, {\bf 104}, 50--58. \newline ADS:
  \url{http://adsabs.harvard.edu/abs/1975MmArc.104...50H}

\bibitem[Harvey(1965)]{Harvey1965}
Harvey, J.W., 1965, ``Coronal Polar Rays and Polar Magnetic Fields'', {\it
  Astrophys. J.\/}, {\bf 141}, 832--833. \newline ADS:
  \url{http://adsabs.harvard.edu/abs/1965ApJ...141..832H}

\bibitem[Harvey and Sheeley~Jr(1979)]{HarveySheeley1979}
Harvey, J.W., Sheeley~Jr, N.R., 1979, ``Coronal holes and solar magnetic
  fields'', {\it Space Sci. Rev.\/}, {\bf 23}, 139--158. \newline ADS:
  \url{http://adsabs.harvard.edu/abs/1979SSRv...23..139H}

\bibitem[Harvey and Recely(2002)]{HarveyRecely2002}
Harvey, K.L., Recely, F., 2002, ``Polar Coronal Holes During Cycles 22 and
  23'', {\it Solar Phys.\/}, {\bf 211}, 31--52. \newline ADS:
  \url{http://adsabs.harvard.edu/abs/2002SoPh..211...31H}

\bibitem[Hassler {\it et~al.\/}(1999)]{HasslerEtal1999}
Hassler, D.M., Dammasch, I.E., Lemaire, P., Brekke, P., Curdt, W., Mason, H.E.,
  Vial, J.-C., Wilhelm, K., 1999, ``Solar Wind Outflow and the Chromospheric
  Magnetic Network'', {\it Science\/}, {\bf 283}, 810--813. \newline ADS:
  \url{http://adsabs.harvard.edu/abs/1999Sci...283..810H}

\bibitem[He {\it et~al.\/}(2007)]{HeEtal2007}
He, J.-S., Tu, C.-Y., Marsch, E., 2007, ``Can the solar wind originate from a
  quiet Sun region?'', {\it Astron. Astrophys.\/}, {\bf 468}, 307--312.
  \newline ADS: \url{http://adsabs.harvard.edu/abs/2007A&A...468..307H}

\bibitem[Hefti {\it et~al.\/}(1998)]{HeftiEtal1998}
Hefti, S., Gr{\"{u}}nwaldt, H., Ipavich, F.M., Bochsler, P., Hovestadt, D.,
  Aellig, M.R., Hilchenbach, M., Kallenbach, R., Galvin, A.B., Geiss, J.,
  Gliem, F., Gloeckler, G., Klecker, B., Marsch, E., M{\"{o}}bius, E.,
  Neugebauer, M., Wurz, P., 1998, ``Kinetic properties of solar wind minor ions
  and protons measured with SOHO/CELIAS'', {\it J. Geophys. Res.\/}, {\bf 103},
  29,697--29,704. \newline ADS:
  \url{http://adsabs.harvard.edu/abs/1998JGR...10329697H}

\bibitem[Hick {\it et~al.\/}(1995)]{HickEtal1995}
Hick, P., Jackson, B.V., Rappoport, S., Woan, G., Slater, G., Strong, K.,
  Uchida, Y., 1995, ``Synoptic IPS and Yohkoh soft X-ray observations'', {\it
  Geophys. Res. Lett.\/}, {\bf 22}, 643--646. \newline ADS:
  \url{http://adsabs.harvard.edu/abs/1995GeoRL..22..643H}

\bibitem[Hoeksema and Scherrer(1986)]{HoeksemaScherrer1986}
Hoeksema, J.T., Scherrer, P.H., 1986, ``An atlas of photospheric magnetic field
  observations and computed coronal magnetic fields: 1976--1985'', {\it Solar
  Phys.\/}, {\bf 105}, 205--211. \newline ADS:
  \url{http://adsabs.harvard.edu/abs/1986SoPh..105..205H}

\bibitem[Hollweg(1986)]{Hollweg1986}
Hollweg, J.V., 1986, ``Transition region, corona, and solar wind in coronal
  holes'', {\it J. Geophys. Res.\/}, {\bf 91}, 4111--4125. \newline ADS:
  \url{http://adsabs.harvard.edu/abs/1986JGR....91.4111H}

\bibitem[Hollweg(1990)]{Hollweg1990}
Hollweg, J.V., 1990, ``Heating of the solar corona'', {\it Comput. Phys.
  Rep.\/}, {\bf 12}, 205--232. \newline ADS:
  \url{http://adsabs.harvard.edu/abs/1990CoPhR..12..205H}

\bibitem[Hollweg(2000)]{Hollweg2000}
Hollweg, J.V., 2000, ``Compressibility of ion cyclotron and whistler waves: Can
  radio measurements detect high-frequency waves of solar origin in the
  corona?'', {\it J. Geophys. Res.\/}, {\bf 105}, 7573--7582. \newline ADS:
  \url{http://adsabs.harvard.edu/abs/2000JGR...105.7573H}

\bibitem[Hollweg(2006)]{Hollweg2006}
Hollweg, J.V., 2006, ``Drivers of the solar wind: Then and now'', {\it Phil.
  Trans. Roy. Soc. A\/}, {\bf 364}, 505--527. \newline ADS:
  \url{http://adsabs.harvard.edu/abs/2006RSPTA.364..505H}

\bibitem[Hollweg and Isenberg(2002)]{HollwegIsenberg2002}
Hollweg, J.V., Isenberg, P.A., 2002, ``Generation of the fast solar wind: A
  review with emphasis on the resonant cyclotron interaction'', {\it J.
  Geophys. Res.\/}, {\bf 107}, 1147. \newline ADS:
  \url{http://adsabs.harvard.edu/abs/2002JGRA.107gSSH12H}

\bibitem[Hollweg and Johnson(1988)]{HollwegJohnson1988}
Hollweg, J.V., Johnson, W., 1988, ``Transition region, corona, and solar wind
  in coronal holes: Some two-fluid models'', {\it J. Geophys. Res.\/}, {\bf
  93}, 9547--9554. \newline ADS:
  \url{http://adsabs.harvard.edu/abs/1988JGR....93.9547H}

\bibitem[Hollweg and Markovskii(2002)]{HollwegMarkovskii2002}
Hollweg, J.V., Markovskii, S.A., 2002, ``Cyclotron resonances of ions with
  obliquely propagating waves in coronal holes and the fast solar wind'', {\it
  J. Geophys. Res.\/}, {\bf 107}, 1080. \newline ADS:
  \url{http://adsabs.harvard.edu/abs/2002JGRA.107f.SSH1H}

\bibitem[Hossain {\it et~al.\/}(1995)]{HossainEtal1995}
Hossain, M., Gray, P.C., Pontius, D.H., Matthaeus, W.H., Oughton, S., 1995,
  ``Phenomenology for the decay of energy-containing eddies in homogeneous MHD
  turbulence'', {\it Phys. Fluids\/}, {\bf 7}, 2886--2904. \newline ADS:
  \url{http://adsabs.harvard.edu/abs/1995PhFl....7.2886H}

\bibitem[Huber {\it et~al.\/}(1974)]{HuberEtal1974}
Huber, M.C.E., Foukal, P.V., Noyes, R.W., Reeves, E.M., Schmahl, E.J., Timothy,
  J.G., Vernazza, J.E., Withbroe, G.~L., 1974, ``Extreme-ultraviolet
  observations of coronal holes: Initial results from Skylab'', {\it Astrophys.
  J. Lett.\/}, {\bf 194}, L115--L118. \newline ADS:
  \url{http://adsabs.harvard.edu/abs/1974ApJ...194L.115H}

\bibitem[Hudson(2002)]{Hudson2002}
Hudson, H.S., 2002, ``Coronal holes as seen in soft X-rays by Yohkoh'', in {\it
  From Solar Min to Max: Half a Solar Cycle with SOHO\/}, (Ed.) Wilson, A.,
  Proceedings of the SOHO-11 Symposium, 11--15 March 2002, Davos, Switzerland,
  vol. SP-508 of ESA Conference Proceedings, pp. 341--349, ESA Publications
  Division, Noordwijk. \newline ADS:
  \url{http://adsabs.harvard.edu/abs/2002ESASP.508..341H}

\bibitem[Hundhausen(1972)]{Hundhausen1972}
Hundhausen, A.J., 1972, {\it Coronal Expansion and the Solar Wind\/}, vol.~5 of
  Physics and Chemistry in Space, Springer, Berlin; New York

\bibitem[Isenberg(1990)]{Isenberg1990}
Isenberg, P.A., 1990, ``Investigations of a turbulence-driven solar wind
  model'', {\it J. Geophys. Res.\/}, {\bf 95}, 6437--6442. \newline ADS:
  \url{http://adsabs.harvard.edu/abs/1990JGR....95.6437I}

\bibitem[Isenberg(2001)]{Isenberg2001}
Isenberg, P.A., 2001, ``Heating of coronal holes and generation of the solar
  wind by ion-cyclotron resonance'', {\it Space Sci. Rev.\/}, {\bf 95},
  119--131. \newline ADS:
  \url{http://adsabs.harvard.edu/abs/2001SSRv...95..119I}

\bibitem[Isenberg and Vasquez(2009)]{IsenbergVasquez2009}
Isenberg, P.A., Vasquez, B.J., 2009, ``Preferential acceleration and
  perpendicular heating of minor ions in a collisionless coronal hole'', {\it
  Astrophys. J.\/}, {\bf 696}, 591--600. \newline ADS:
  \url{http://adsabs.harvard.edu/abs/2009ApJ...696..591I}

\bibitem[Isenberg {\it et~al.\/}(2001)]{IsenbergEtal2001}
Isenberg, P.A., Lee, M.A., Hollweg, J.V., 2001, ``The kinetic shell model of
  coronal heating and acceleration by ion-cyclotron waves: 1. Outward
  propagating waves'', {\it J. Geophys. Res.\/}, {\bf 106}, 5649--5660.
  \newline ADS: \url{http://adsabs.harvard.edu/abs/2001JGR...106.5649I}

\bibitem[Janardhan {\it et~al.\/}(2008)]{JanardhanEtal2008}
Janardhan, P., Tripathi, D., Mason, H.E., 2008, ``The solar wind disappearance
  event of 11 May 1999: Source region evolution'', {\it Astron. Astrophys.\/},
  {\bf 488}, L1--L4. \newline ADS:
  \url{http://adsabs.harvard.edu/abs/2001A&A...488L...1J}

\bibitem[Janse {\it et~al.\/}(2007)]{JanseEtal2007}
Janse, {\AA.M.}, Lie-Svendsen, {\O}., Leer, E., 2007, ``Solar wind originating
  in funnels: fast or slow?'', {\it Astron. Astrophys.\/}, {\bf 474},
  997--1013. \newline ADS:
  \url{http://adsabs.harvard.edu/abs/2007A&A...474..997J}

\bibitem[Jardine and van Ballegooijen(2005)]{JardineBallegooijen2005}
Jardine, M., van Ballegooijen, A.A., 2005, ``Slingshot prominences above
  stellar X-ray coronae'', {\it Mon. Not. Roy. Astron. Soc.\/}, {\bf 361},
  1173--1179. \newline ADS:
  \url{http://adsabs.harvard.edu/abs/2005MNRAS.361.1173J}

\bibitem[Jiang {\it et~al.\/}(2009)]{JiangEtal2009}
Jiang, Y.W., Liu, S., Petrosian, V., 2009, ``Cascade and Damping of
  Alfv\'{e}n-Cyclotron Fluctuations: Application to Solar Wind Turbulence'',
  {\it Astrophys. J.\/}, {\bf 698}, 163--183. \newline ADS:
  \url{http://adsabs.harvard.edu/abs/2009ApJ...698..163J}

\bibitem[Jones(2005)]{Jones2005}
Jones, H.P., 2005, ``Magnetic Fields and Flows in Open Magnetic Structures'',
  in {\it Large-scale Structures and their Role in Solar Activity\/}, (Eds.)
  Sankarasubramanian, K., Penn, M., Pevtsov, A., Proceedings of the 22nd
  Sacramento Peak Workshop, 18-22 October 2004, Sunspot, New Mexico, vol. 346
  of ASP Conf. Proc., pp. 229--244, Astronomical Society of the Pacific, San
  Francisco. \newline ADS:
  \url{http://adsabs.harvard.edu/abs/2005ASPC..346..229J}

\bibitem[Kahler(2000)]{Kahler2000}
Kahler, S., 2000, ``Skylab'', in {\it Encyclopedia of Astronomy and
  Astrophysics\/}, (Ed.) Murdin, P., p. 2238, Institute of Physics Publishing,
  Bristol. \newline ADS:
  \url{http://adsabs.harvard.edu/abs/2000eaa..bookE2238K}

\bibitem[Khazanov and Singh(2007)]{KhazanovSingh2007}
Khazanov, I., Singh, N., 2007, ``Ion and electron accelerations by large-scale
  shear Alfv\'{e}n waves via cross-field instabilities'', {\it Geophys. Res.
  Lett.\/}, {\bf 34}, L20\,111. \newline ADS:
  \url{http://adsabs.harvard.edu/abs/2007GeoRL..3420111K}

\bibitem[Klimchuk(2006)]{Klimchuk2006}
Klimchuk, J.A., 2006, ``On Solving the Coronal Heating Problem'', {\it Solar
  Phys.\/}, {\bf 234}, 41--77. \newline ADS:
  \url{http://adsabs.harvard.edu/abs/2006SoPh..234...41K}

\bibitem[Ko {\it et~al.\/}(1996)]{KoEtal1996}
Ko, Y.-K., Fisk, L.A., Gloeckler, G., Geiss, J., 1996, ``Limitations on
  suprathermal tails of electrons in the lower solar corona'', {\it Geophys.
  Res. Lett.\/}, {\bf 23}, 2785--2788. \newline ADS:
  \url{http://adsabs.harvard.edu/abs/1996GeoRL..23.2785K}

\bibitem[Ko {\it et~al.\/}(1997)]{KoEtal1997}
Ko, Y.-K., Fisk, L.A., Geiss, J., Gloeckler, G., Guhathakurta, M., 1997, ``An
  Empirical Study of the Electron Temperature and Heavy Ion Velocities in the
  South Polar Coronal Hole'', {\it Solar Phys.\/}, {\bf 171}, 345--361.
  \newline ADS: \url{http://adsabs.harvard.edu/abs/1997SoPh..171..345K}

\bibitem[Kohl and Cranmer(1999)]{KohlCranmer1999}
Kohl, J.L., Cranmer, S.R. (Eds.), 1999, {\it Coronal Holes and Solar Wind
  Acceleration\/}, Kluwer Academic Publishers, Dordrecht; Boston; London

\bibitem[Kohl and Withbroe(1982)]{KohlWithbroe1982}
Kohl, J.L., Withbroe, G.L., 1982, ``EUV spectroscopic plasma diagnostics for
  the solar wind acceleration region'', {\it Astrophys. J.\/}, {\bf 256},
  263--270. \newline ADS:
  \url{http://adsabs.harvard.edu/abs/1982ApJ...256..263K}

\bibitem[Kohl {\it et~al.\/}(1978)]{KohlEtal1978}
Kohl, J.L., Reeves, E.M., Kirkham, B., 1978, ``The Lyman Alpha Coronagraph'',
  in {\it New Instrumentation for Space Astronomy\/}, (Eds.) van~der Hucht,
  K.A., Viana, G., Proceedings of the 20th COSPAR Meeting, Tel Aviv, Israel,
  7--18 June 1977, pp. 91--94, Pergamon Press, Oxford. \newline ADS:
  \url{http://adsabs.harvard.edu/abs/1978nisa.conf...91K}

\bibitem[Kohl {\it et~al.\/}(1980)]{KohlEtal1980}
Kohl, J.L., Weiser, H., Withbroe, G.L., Noyes, R.W., Parkinson, W.H., Reeves,
  E.M., Munro, R.H., MacQueen, R.M., 1980, ``Measurements of coronal kinetic
  temperatures from 1.5 to 3 solar radii'', {\it Astrophys. J. Lett.\/}, {\bf
  241}, L117--L121. \newline ADS:
  \url{http://adsabs.harvard.edu/abs/1980ApJ...241L.117K}

\bibitem[Kohl {\it et~al.\/}(1995)]{KohlEtal1995}
Kohl, J.L., Esser, R., Gardner, L.D., Habbal, S., Daigneau, P.S., Dennis, E.F.,
  Nystrom, G.U., Panasyuk, A., Raymond, J.C., Smith, P.L., Strachan, L., van
  Ballegooijen, A.A., Noci, G., Fineschi, S., Romoli, M., Ciaravella, A.,
  Modigliani, A., Huber, M.C.E., Antonucci, E., Benna, C., Giordano, S.,
  Tondello, G., Nicolosi, P., Naletto, G., Pernechele, C., Spadaro, D.,
  Poletto, G., Livi, S., von~der L\"{u}he, O., Geiss, J., Timothy, J.G.,
  Gloeckler, G., Allegra, A., Basile, G., Brusa, R., Wood, B., Siegmund,
  O.H.W., Fowler, W., Fisher, R., Jhabvala, M., 1995, ``The Ultraviolet
  Coronagraph Spectrometer for the Solar and Heliospheric Observatory'', {\it
  Solar Phys.\/}, {\bf 162}, 313--356. \newline ADS:
  \url{http://adsabs.harvard.edu/abs/1995SoPh..162..313K}

\bibitem[Kohl {\it et~al.\/}(1997)]{KohlEtal1997}
Kohl, J.L., Noci, G., Antonucci, E., Tondello, G., Huber, M.C.E., Gardner,
  L.D., Nicolosi, P., Strachan, L., Fineschi, S., Raymond, J.C., Romoli, M.,
  Spadaro, D., Panasyuk, A., Siegmund, O.H.W., Benna, C., Ciaravella, A.,
  Cranmer, S.R., Giordano, S., Karovska, M., Martin, R., Michels, J.,
  Modigliani, A., Naletto, G., Pernechele, C., Poletto, G., Smith, P.L., 1997,
  ``First Results from the SOHO Ultraviolet Coronagraph Spectrometer'', {\it
  Solar Phys.\/}, {\bf 175}, 613--644. \newline ADS:
  \url{http://adsabs.harvard.edu/abs/1997SoPh..175..613K}

\bibitem[Kohl {\it et~al.\/}(1998)]{KohlEtal1998}
Kohl, J.L., Noci, G., Antonucci, E., Tondello, G., Huber, M.C.E., Cranmer,
  S.R., Strachan, L., Panasyuk, A.V., Gardner, L.D., Romoli, M., Fineschi, S.,
  Dobrzycka, D., Raymond, J.C., Nicolosi, P., Siegmund, O.H.W., Spadaro, D.,
  Benna, C., Ciaravella, A., Giordano, S., Habbal, S.R., Karovska, M., Li, X.,
  Martin, R., Michels, J.G., Modigliani, A., Naletto, G., O'Neal, R.H.,
  Pernechele, C., Poletto, G., Smith, P.L., Suleiman, R.M., 1998, ``UVCS/SOHO
  empirical determinations of anisotropic velocity distributions in the solar
  corona'', {\it Astrophys. J. Lett.\/}, {\bf 501}, L127--L131. \newline ADS:
  \url{http://adsabs.harvard.edu/abs/1998ApJ...501L.127K}

\bibitem[Kohl {\it et~al.\/}(1999)]{KohlEtal1999}
Kohl, J.L., Esser, R., Cranmer, S.R., Fineschi, S., Gardner, L.D., Panasyuk,
  A.V., Strachan, L., Suleiman, R.M., Frazin, R.A., Noci, G., 1999, ``EUV
  Spectral Line Profiles in Polar Coronal Holes from 1.3 to 3.0 $R_{\odot}$'',
  {\it Astrophys. J. Lett.\/}, {\bf 510}, L59--L62. \newline ADS:
  \url{http://adsabs.harvard.edu/abs/1999ApJ...510L..59K}

\bibitem[Kohl {\it et~al.\/}(2006)]{KohlEtal2006}
Kohl, J.L., Noci, G., Cranmer, S.R., Raymond, J.C., 2006, ``Ultraviolet
  spectroscopy of the extended solar corona'', {\it Astron. Astrophys.
  Review\/}, {\bf 13}, 31--157. \newline ADS:
  \url{http://adsabs.harvard.edu/abs/2006A&ARv..13...31K}

\bibitem[Kojima {\it et~al.\/}(2007)]{KojimaEtal2007}
Kojima, M., Tokumaru, M., Fujiki, K., Itoh, H., Murakami, T., Hakamada, K.,
  2007, ``What Coronal Parameters Determine Solar Wind Speed?'', in {\it New
  Solar Physics with Solar-B Mission\/}, (Eds.) Shibata, K., Nagata, S.,
  Sakurai, T., Proceedings of the 6th Solar-B Science Meeting, 8--11 November
  2005, Kyoto, Japan, vol. 369 of ASP Conf. Proc., pp. 549--555, Astronomical
  Society of the Pacific, San Francisco. \newline ADS:
  \url{http://adsabs.harvard.edu/abs/2007ASPC..369..549K}

\bibitem[Krieger {\it et~al.\/}(1973)]{KriegerEtal1973}
Krieger, A.S., Timothy, A.F., Roelof, E.C., 1973, ``A Coronal Hole and Its
  Identification as the Source of a High Velocity Solar Wind Stream'', {\it
  Solar Phys.\/}, {\bf 29}, 505--525. \newline ADS:
  \url{http://adsabs.harvard.edu/abs/1973SoPh...29..505K}

\bibitem[Krista and Gallagher(2009)]{KristaGallagher2009}
Krista, L.D., Gallagher, P.T., 2009, ``Automated Coronal Hole Detection Using
  Local Intensity Thresholding Techniques'', {\it Solar Phys.\/}, {\bf 256},
  87--100. \newline ADS:
  \url{http://adsabs.harvard.edu/abs/2009SoPh..256...87K}

\bibitem[Kwan {\it et~al.\/}(2007)]{KwanEtal2007}
Kwan, J., Edwards, S., Fischer, W., 2007, ``Modeling T Tauri Winds from He I
  10830 {\AA} Profiles'', {\it Astrophys. J.\/}, {\bf 657}, 897--915. \newline
  ADS: \url{http://adsabs.harvard.edu/abs/2007ApJ...657..897K}

\bibitem[Lada(1985)]{Lada1985}
Lada, C.J., 1985, ``Cold outflows, energetic winds, and enigmatic jets around
  young stellar objects'', {\it Ann. Rev. Astron. Astrophys.\/}, {\bf 23},
  267--317. \newline ADS:
  \url{http://adsabs.harvard.edu/abs/1985ARA&A..23..267L}

\bibitem[Lamers and Cassinelli(1999)]{LamersCassinelli1999}
Lamers, H.J.G.L.M., Cassinelli, J.P., 1999, {\it Introduction to Stellar
  Winds\/}, Cambridge University Press, Cambridge

\bibitem[Laming(2004)]{Laming2004}
Laming, J.M., 2004, ``A Unified Picture of the First Ionization Potential and
  Inverse First Ionization Potential Effects'', {\it Astrophys. J.\/}, {\bf
  614}, 1063--1072. \newline ADS:
  \url{http://adsabs.harvard.edu/abs/2004ApJ...614.1063L}

\bibitem[Laming(2009)]{Laming2009}
Laming, J.M., 2009, ``Non-WKB Models of the FIP Effect: Implications for Solar
  Coronal Heating and the Coronal Helium and Neon Abundances'', {\it Astrophys.
  J.\/}, {\bf 695}, 954--969. \newline ADS:
  \url{http://adsabs.harvard.edu/abs/2009ApJ...695..954L}

\bibitem[Laming and Lepri(2007)]{LamingLepri2007}
Laming, J.M., Lepri, S.T., 2007, ``Ion Charge States in the Fast Solar Wind:
  New Data Analysis and Theoretical Refinements'', {\it Astrophys. J.\/}, {\bf
  660}, 1642--1652. \newline ADS:
  \url{http://adsabs.harvard.edu/abs/2007ApJ...660.1642L}

\bibitem[Landi(2008)]{Landi2008}
Landi, E., 2008, ``The off-disk thermal structure of a polar coronal hole'',
  {\it Astrophys. J.\/}, {\bf 685}, 1270--1276. \newline ADS:
  \url{http://adsabs.harvard.edu/abs/2008ApJ...685.1270L}

\bibitem[Landi and Cranmer(2009)]{LandiCranmer2009}
Landi, E., Cranmer, S.R., 2009, ``Ion temperatures in the low solar corona:
  Polar coronal holes at solar minimum'', {\it Astrophys. J.\/}, {\bf 691},
  794--805. \newline ADS:
  \url{http://adsabs.harvard.edu/abs/2009ApJ...691..794L}

\bibitem[Leamon and McIntosh(2007)]{LeamonMcIntosh2007}
Leamon, R.J., McIntosh, S.W., 2007, ``Empirical Solar Wind Forecasting from the
  Chromosphere'', {\it Astrophys. J.\/}, {\bf 659}, 738--742. \newline ADS:
  \url{http://adsabs.harvard.edu/abs/2007ApJ...659..738L}

\bibitem[Leamon {\it et~al.\/}(1999)]{LeamonEtal1999}
Leamon, R.J., Smith, C.W., Ness, N.F., Wong, H.K., 1999, ``Dissipation range
  dynamics: Kinetic Alfv\'{e}n waves and the importance of $\beta_e$'', {\it J.
  Geophys. Res.\/}, {\bf 104}, 22\,331--22\,344. \newline ADS:
  \url{http://adsabs.harvard.edu/abs/1999JGR...10422331L}

\bibitem[Lee and Wu(2000)]{LeeWu2000}
Lee, L.C., Wu, B.H., 2000, ``Heating and Acceleration of Protons and Minor Ions
  by Fast Shocks in the Solar Corona'', {\it Astrophys. J.\/}, {\bf 535},
  1014--1026. \newline ADS:
  \url{http://adsabs.harvard.edu/abs/2000ApJ...535.1014L}

\bibitem[Leer and Holzer(1980)]{LeerHolzer1980}
Leer, E., Holzer, T.E., 1980, ``Energy addition in the solar wind'', {\it J.
  Geophys. Res.\/}, {\bf 85}, 4681--4688. \newline ADS:
  \url{http://adsabs.harvard.edu/abs/1980JGR....85.4681L}

\bibitem[Leer {\it et~al.\/}(1982)]{LeerEtal1982}
Leer, E., Holzer, T.E., Fl{\aa}, T., 1982, ``Acceleration of the solar wind'',
  {\it Space Sci. Rev.\/}, {\bf 33}, 161--200. \newline ADS:
  \url{http://adsabs.harvard.edu/abs/1982SSRv...33..161L}

\bibitem[Lemaire and Pierrard(2001)]{LemairePierrard2001}
Lemaire, J., Pierrard, V., 2001, ``Kinetic models of solar and polar winds'',
  {\it Astrophys. Space Sci.\/}, {\bf 277}, 169--180. \newline ADS:
  \url{http://adsabs.harvard.edu/abs/2001Ap&SS.277..169L}

\bibitem[Levine(1974)]{Levine1974}
Levine, R.H., 1974, ``A new theory of coronal heating'', {\it Astrophys. J.\/},
  {\bf 190}, 457--466. \newline ADS:
  \url{http://adsabs.harvard.edu/abs/1974ApJ...190..457L}

\bibitem[Levine(1982)]{Levine1982}
Levine, R.H., 1982, ``Open magnetic fields and the solar cycle, I, Photospheric
  sources of open magnetic flux'', {\it Solar Phys.\/}, {\bf 79}, 203--230.
  \newline ADS: \url{http://adsabs.harvard.edu/abs/1982SoPh...79..203L}

\bibitem[Li and Habbal(2001)]{LiHabbal2001}
Li, X., Habbal, S.R., 2001, ``Damping of fast and ion cyclotron oblique waves
  in the multi-ion fast solar wind'', {\it J. Geophys. Res.\/}, {\bf 106},
  10\,669--10\,680. \newline ADS:
  \url{http://adsabs.harvard.edu/abs/2001JGR...10610669L}

\bibitem[Li {\it et~al.\/}(1998)]{LiEtal1998}
Li, X., Habbal, S.R., Kohl, J., Noci, G., 1998, ``The Effect of Temperature
  Anisotropy on Observations of Doppler Dimming and Pumping in the Inner
  Corona'', {\it Astrophys. J. Lett.\/}, {\bf 501}, L133--L137. \newline ADS:
  \url{http://adsabs.harvard.edu/abs/1998ApJ...501L.133L}

\bibitem[Li {\it et~al.\/}(1999)]{LiEtal1999}
Li, X., Habbal, S.R., Hollweg, J.V., Esser, R., 1999, ``Heating and cooling of
  protons by turbulence-driven ion-cyclotron waves in the fast solar wind'',
  {\it J. Geophys. Res.\/}, {\bf 104}, 2521--2535. \newline ADS:
  \url{http://adsabs.harvard.edu/abs/1999JGR...104.2521L}

\bibitem[Liewer {\it et~al.\/}(2004)]{LiewerEtal2004}
Liewer, P.C., Neugebauer, M., Zurbuchen, T., 2004, ``Characteristics of
  active-region sources of solar wind near solar maximum'', {\it Solar
  Phys.\/}, {\bf 223}, 209--229. \newline ADS:
  \url{http://adsabs.harvard.edu/abs/2004SoPh..223..209L}

\bibitem[Lippincott(1957)]{Lippincott1957}
Lippincott, S.L., 1957, ``Chromospheric Spicules'', {\it Smithsonian Contrib.
  Astrophys.\/}, {\bf 2}, 15--23. \newline ADS:
  \url{http://adsabs.harvard.edu/abs/1957SCoA....2...15L}

\bibitem[Longcope(2004)]{Longcope2004}
Longcope, D., 2004, ``Quantifying Magnetic Reconnection and the Heat it
  Generates'', in {\it Coronal Heating\/}, (Eds.) Walsh, R.W., Ireland, J.,
  Danesy, D., Fleck, B., Proceedings of the SOHO--15 Workshop, 6--9 September
  2004, St. Andrews, Scotland, UK, vol. SP-575 of ESA Conference Proceedings,
  pp. 198--209, ESA Publications Division, Noordwijk. \newline ADS:
  \url{http://adsabs.harvard.edu/abs/2004ESASP.575..198L}

\bibitem[Longcope(1996)]{Longcope1996}
Longcope, D.W., 1996, ``Topology and Current Ribbons: A Model for Current,
  Reconnection, and Flaring in a Complex, Evolving Corona'', {\it Solar
  Phys.\/}, {\bf 169}, 91--121. \newline ADS:
  \url{http://adsabs.harvard.edu/abs/1996SoPh..169...91L}

\bibitem[Longcope and Kankelborg(1999)]{LongcopeKankelborg1999}
Longcope, D.W., Kankelborg, C.C., 1999, ``Coronal Heating by Collision and
  Cancellation of Magnetic Elements'', {\it Astrophys. J.\/}, {\bf 524},
  483--495. \newline ADS:
  \url{http://adsabs.harvard.edu/abs/1999ApJ...524..483L}

\bibitem[Lu and Li(2007)]{LuLi2007}
Lu, Q., Li, X., 2007, ``Heating of ions by low-frequency Alfv\'{e}n waves'',
  {\it Phys. Plasmas\/}, {\bf 14}, 042303. \newline ADS:
  \url{http://adsabs.harvard.edu/abs/2007PhPl...14d2303L}

\bibitem[Luhmann {\it et~al.\/}(2002)]{LuhmannEtal2002}
Luhmann, J.G., Li, Y., Arge, C.N., Gazis, P.R., Ulrich, R., 2002, ``Solar cycle
  changes in coronal holes and space weather cycles'', {\it J. Geophys.
  Res.\/}, {\bf 107}, 1154. \newline ADS:
  \url{http://adsabs.harvard.edu/abs/2002JGRA..107.1154L}

\bibitem[Luo and Melrose(2006)]{LuoMelrose2006}
Luo, Q., Melrose, D., 2006, ``Anisotropic weak turbulence of Alfv\'{e}n waves
  in collisionless astrophysical plasmas'', {\it Mon. Not. Roy. Astron.
  Soc.\/}, {\bf 368}, 1151--1158. \newline ADS:
  \url{http://adsabs.harvard.edu/abs/2006MNRAS.368.1151L}

\bibitem[MacBride {\it et~al.\/}(2008)]{MacBrideEtal2008}
MacBride, B.T., Smith, C.W., Forman, M.A., 2008, ``The Turbulent Cascade at 1
  AU: Energy Transfer and the Third-Order Scaling for MHD'', {\it Astrophys.
  J.\/}, {\bf 679}, 1644--1660. \newline ADS:
  \url{http://adsabs.harvard.edu/abs/2008ApJ...679.1644M}

\bibitem[Malanushenko and Jones(2004)]{MalanushenkoJones2004}
Malanushenko, O.V., Jones, H.P., 2004, ``Analysis of He I 1083 nm Imaging
  Spectroscopy Using a Spectral Standard'', {\it Solar Phys.\/}, {\bf 222},
  43--60. \newline ADS: \url{http://adsabs.harvard.edu/abs/2004SoPh..222...43M}

\bibitem[Mancuso {\it et~al.\/}(2002)]{MancusoEtal2002}
Mancuso, S., Raymond, J.C., Kohl, J., Ko, Y.-K., Uzzo, M., Wu, R., 2002,
  ``UVCS/SOHO observations of a CME-driven shock: Consequences on ion heating
  mechanisms behind a coronal shock'', {\it Astron. Astrophys.\/}, {\bf 383},
  267--274. \newline ADS:
  \url{http://adsabs.harvard.edu/abs/2002A&A...383..267M}

\bibitem[Markovskii(2001)]{Markovskii2001}
Markovskii, S.A., 2001, ``Generation of ion-cyclotron waves in coronal holes by
  a global resonant magnetohydrodynamic mode'', {\it Astrophys. J.\/}, {\bf
  557}, 337--342. \newline ADS:
  \url{http://adsabs.harvard.edu/abs/2001ApJ...557..337M}

\bibitem[Markovskii and Hollweg(2002)]{MarkovskiiHollweg2002}
Markovskii, S.A., Hollweg, J.V., 2002, ``Electron heat flux instabilities in
  coronal holes: Implications for ion heating'', {\it Geophys. Res. Lett.\/},
  {\bf 29}, 1843. \newline ADS:
  \url{http://adsabs.harvard.edu/abs/2002GeoRL..29q..24M}

\bibitem[Markovskii and Hollweg(2004)]{MarkovskiiHollweg2004}
Markovskii, S.A., Hollweg, J.V., 2004, ``Intermittent Heating of the Solar
  Corona by Heat Flux-generated Ion Cyclotron Waves'', {\it Astrophys. J.\/},
  {\bf 609}, 1112--1122. \newline ADS:
  \url{http://adsabs.harvard.edu/abs/2004ApJ...609.1112M}

\bibitem[Markovskii {\it et~al.\/}(2006)]{MarkovskiiEtal2006}
Markovskii, S.A., Vasquez, B.J., Smith, C.W., Hollweg, J.V., 2006,
  ``Dissipation of the Perpendicular Turbulent Cascade in the Solar Wind'',
  {\it Astrophys. J.\/}, {\bf 639}, 1177--1185. \newline ADS:
  \url{http://adsabs.harvard.edu/abs/2006ApJ...639.1177M}

\bibitem[Markovskii {\it et~al.\/}(2009)]{MarkovskiiEtal2009}
Markovskii, S.A., Vasquez, B.J., Hollweg, J.V., 2009, ``Proton heating by
  nonlinear field-aligned Alfv\'{e}n waves in solar coronal holes'', {\it
  Astrophys. J.\/}, {\bf 695}, 1413--1420. \newline ADS:
  \url{http://adsabs.harvard.edu/abs/2009ApJ...695.1413M}

\bibitem[Marsch(1999)]{Marsch1999}
Marsch, E., 1999, ``Solar wind models from the sun to 1 AU: Constraints by in
  situ and remote sensing measurements'', {\it Space Sci. Rev.\/}, {\bf 87},
  1--24. \newline ADS: \url{http://adsabs.harvard.edu/abs/1999SSRv...87....1M}

\bibitem[Marsch(2005)]{Marsch2005}
Marsch, E., 2005, ``Importance of Kinetic Effects in Heating the Open and
  Closed Corona'', in {\it Connecting Sun and Heliosphere\/}, (Eds.) Fleck, B.,
  Zurbuchen, T.H., Lacoste, H., Proceedings of Solar Wind 11/SOHO--16, 12--17
  June 2005, Whistler, Canada, vol. SP-592 of ESA Conference Proceedings, pp.
  191--198, ESA Publications Division, Noordwijk. \newline ADS:
  \url{http://adsabs.harvard.edu/abs/2005ESASP.592..191M}

\bibitem[Marsch(2006)]{Marsch2006}
Marsch, E., 2006, ``Kinetic Physics of the Solar Corona and Solar Wind'', {\it
  Living Rev. Solar Phys.\/}, {\bf 3}, lrsp-2006-1. URL (cited on 1 July 2009):
  \newline\url{http://www.livingreviews.org/lrsp-2006-1}

\bibitem[Marsch {\it et~al.\/}(1982)]{MarschEtal1982}
Marsch, E., M{\"{u}}hlh{\"{a}}user, K.-H., Schwenn, R., Rosenbauer, H., Pilipp,
  W.G., Neubauer, F.M., 1982, ``Solar wind protons: Three-dimensional velocity
  distributions and derived plasma parameters measured between 0.3 and 1 AU'',
  {\it J. Geophys. Res.\/}, {\bf 87}, 52--72. \newline ADS:
  \url{http://adsabs.harvard.edu/abs/1982JGR....87...52M}

\bibitem[Marsch {\it et~al.\/}(2008)]{MarschEtal2008}
Marsch, E., Tian, H., Sun, J., Curdt, W., Wiegelmann, T., 2008, ``Plasma flows
  guided by strong magnetic fields in the solar corona'', {\it Astrophys.
  J.\/}, {\bf 685}, 1262--1269. \newline ADS:
  \url{http://adsabs.harvard.edu/abs/2008ApJ...685.1262M}

\bibitem[Marsden and Fleck(2007)]{MarsdenFleck2007}
Marsden, R.G., Fleck, B., 2007, ``Solar Orbiter: A Mission Update'', in {\it
  The Physics of Chromospheric Plasmas\/}, (Eds.) Heinzel, P., Dorotovi\v{c},
  I., Proceedings of the conference held 9--13 October 2006, Coimbra, Portugal,
  vol. 368 of ASP Conf. Proc., p. 645, Astronomical Society of the Pacific, San
  Francisco. \newline ADS:
  \url{http://adsabs.harvard.edu/abs/2007ASPC..368..645M}

\bibitem[Matthaeus {\it et~al.\/}(1999)]{MatthaeusEtal1999}
Matthaeus, W.H., Zank, G.P., Oughton, S., Mullan, D.J., Dmitruk, P., 1999,
  ``Coronal Heating by Magnetohydrodynamic Turbulence Driven by Reflected
  Low-Frequency Waves'', {\it Astrophys. J. Lett.\/}, {\bf 523}, L93--L96.
  \newline ADS: \url{http://adsabs.harvard.edu/abs/1999ApJ...523L..93M}

\bibitem[Matthaeus {\it et~al.\/}(2003)]{MatthaeusEtal2003}
Matthaeus, W.H., Dmitruk, P., Oughton, S., Mullan, D., 2003, ``Turbulent
  dissipation in the solar wind and corona'', in {\it Solar Wind Ten\/}, (Eds.)
  Velli, M., Bruno, R., Malara, F., Proceedings of the Tenth International
  Solar Wind Conference, Pisa, Italy, 17--21 June 2002, vol. 679 of AIP
  Conference Proceedings, pp. 427--432, American Institute of Physics,
  Melville. \newline ADS:
  \url{http://adsabs.harvard.edu/abs/2003AIPC..679..427M}

\bibitem[McComas {\it et~al.\/}(1998)]{McComasEtal1998}
McComas, D.J., Bame, S.J., Barker, P., Feldman, W.C., Phillips, J.L., Riley,
  P., Griffee, J.W., 1998, ``Solar Wind Electron Proton Alpha Monitor (SWEPAM)
  for the Advanced Composition Explorer'', {\it Space Sci. Rev.\/}, {\bf 86},
  563--612. \newline ADS:
  \url{http://adsabs.harvard.edu/abs/1998SSRv...86..563M}

\bibitem[McComas {\it et~al.\/}(2002)]{McComasEtal2002}
McComas, D.J., Elliott, H.A., Gosling, J.T., Reisenfeld, D.B., Skoug, R.M.,
  Goldstein, B.E., Neugebauer, M., Balogh, A., 2002, ``Ulysses' second
  fast-latitude scan: Complexity near solar maximum and the reformation of
  polar coronal holes'', {\it Geophys. Res. Lett.\/}, {\bf 29}, 1290. \newline
  ADS: \url{http://adsabs.harvard.edu/abs/2002GeoRL..29i...4M}

\bibitem[McComas {\it et~al.\/}(2007)]{McComasEtal2007}
McComas, D.J., Velli, M., Lewis, W.S., Acton, L.W., Balat-Pichelin, M.,
  Bothmer, V., Dirling, R.B., Feldman, W.C., Gloeckler, G., Habbal, S.R.,
  Hassler, D.M., Mann, I., Matthaeus, W.H., McNutt, R.L., Mewaldt, R.A.,
  Murphy, N., Ofman, L., Sittler, E.C., Smith, C.W., Zurbuchen, T.H., 2007,
  ``Understanding coronal heating and solar wind acceleration: Case for in situ
  near-Sun measurements'', {\it Rev. Geophys.\/}, {\bf 45}, RG1004. \newline
  ADS: \url{http://adsabs.harvard.edu/abs/2007RvGeo..45.1004M}

\bibitem[McIntosh(2009)]{McIntosh2009}
McIntosh, S.W., 2009, ``The Inconvenient Truth About Coronal Dimmings'', {\it
  Astrophys. J.\/}, {\bf 693}, 1306--1309. \newline ADS:
  \url{http://adsabs.harvard.edu/abs/2009ApJ...693.1306M}

\bibitem[McIntosh {\it et~al.\/}(2007)]{McIntoshEtal2007}
McIntosh, S.W., Davey, A.R., Hassler, D.M., Armstrong, J.D., Curdt, W.,
  Wilhelm, K., Lin, G., 2007, ``Observations Supporting the Role of
  Magnetoconvection in Energy Supply to the Quiescent Solar Atmosphere'', {\it
  Astrophys. J.\/}, {\bf 654}, 650--664. \newline ADS:
  \url{http://adsabs.harvard.edu/abs/2007ApJ...654..650M}

\bibitem[Mecheri and Marsch(2008)]{MecheriMarsch2008}
Mecheri, R., Marsch, E., 2008, ``Drift instabilities in the solar corona within
  the multi-fluid description'', {\it Astron. Astrophys.\/}, {\bf 481},
  853--860. \newline ADS:
  \url{http://adsabs.harvard.edu/abs/2008A&A...481..853M}

\bibitem[Medvedev(2000)]{Medvedev2000}
Medvedev, M.V., 2000, ``Particle Heating by Nonlinear Alfv\'{e}nic Turbulence
  in Advection-dominated Accretion Flows'', {\it Astrophys. J.\/}, {\bf 541},
  811--820. \newline ADS:
  \url{http://adsabs.harvard.edu/abs/2000ApJ...541..811M}

\bibitem[Mikhailenko {\it et~al.\/}(2008)]{MikhailenkoEtal2008}
Mikhailenko, V.S., Mikhailenko, V.V., Stepanov, K.N., 2008, ``Ion cyclotron
  instabilities of parallel shear flow of collisional plasma'', {\it Phys.
  Plasmas\/}, {\bf 15}, 092901. \newline ADS:
  \url{http://adsabs.harvard.edu/abs/2008PhPl...15i2901M}

\bibitem[Miller(1908)]{Miller1908}
Miller, J.A., 1908, ``The Determination of the Heliocentric Position of a
  Certain Class of Coronal Streamers'', {\it Astrophys. J.\/}, {\bf 27},
  286--295. \newline ADS:
  \url{http://adsabs.harvard.edu/abs/1908ApJ....27..286M}

\bibitem[Miralles {\it et~al.\/}(2001a)]{MirallesEtal2001a}
Miralles, M.P., Cranmer, S.R., Panasyuk, A.V., Romoli, M., Kohl, J.L., 2001a,
  ``Comparison of empirical models for polar and equatorial coronal holes'',
  {\it Astrophys. J. Lett.\/}, {\bf 549}, L257--L260. \newline ADS:
  \url{http://adsabs.harvard.edu/abs/2001ApJ...549L.257M}

\bibitem[Miralles {\it et~al.\/}(2001b)]{MirallesEtal2001b}
Miralles, M.P., Cranmer, S.R., Kohl, J.L., 2001b, ``Ultraviolet Coronagraph
  Spectrometer Observations of a High-Latitude Coronal Hole with High Oxygen
  Temperatures and the Next Solar Cycle Polarity'', {\it Astrophys. J.
  Lett.\/}, {\bf 560}, L193--L196. \newline ADS:
  \url{http://adsabs.harvard.edu/abs/2001ApJ...560L.193M}

\bibitem[Miralles {\it et~al.\/}(2004)]{MirallesEtal2004}
Miralles, M.P., Cranmer, S.R., Kohl, J.L., 2004, ``Low-latitude coronal holes
  during solar maximum'', {\it Adv. Space Res.\/}, {\bf 33}, 696--700. \newline
  ADS: \url{http://adsabs.harvard.edu/abs/2004AdSpR..33..696M}

\bibitem[Miralles {\it et~al.\/}(2006)]{MirallesEtal2006}
Miralles, M.P., Cranmer, S.R., Kohl, J.L., 2006, ``Coronal Hole Properties
  During the First Decade of UVCS/SOHO'', in {\it 10 Years of SOHO and
  Beyond\/}, (Eds.) Lacoste, H., Ouwehand, L., Proceedings of the SOHO-17
  Conference, 7--12 May 2006, Giardini Naxos, Italy, vol. SP-617 of ESA
  Conference Proceedings, pp. 15.1--15.4, ESA Publications Division, Noordwijk.
  \newline ADS: \url{http://adsabs.harvard.edu/abs/2006ESASP.617E..15M}

\bibitem[Miralles {\it et~al.\/}(2007)]{MirallesEtal2007}
Miralles, M.P., Cranmer, S.R., Raymond, J.C., Kohl, J.L., 2007,
  ``Multi-Instrument Searches for Polar Jets: Characterizing Jet Heating and
  Cooling'', conference paper. \newline ADS:
  \url{http://adsabs.harvard.edu/abs/2007IUGG...24..691M}

\bibitem[Mitchell(1932)]{Mitchell1932}
Mitchell, S.A., 1932, ``The Spectrum of the Corona'', {\it Astrophys. J.\/},
  {\bf 75}, 1--33. \newline ADS:
  \url{http://adsabs.harvard.edu/abs/1932ApJ....75....1M}

\bibitem[Moreno-Insertis {\it et~al.\/}(2008)]{MorenoInsertisEtal2008}
Moreno-Insertis, F., Galsgaard, K., Ugarte-Urra, I., 2008, ``Jets in Coronal
  Holes: Hinode Observations and Three-dimensional Computer Modeling'', {\it
  Astrophys. J. Lett.\/}, {\bf 673}, L211--L214. \newline ADS:
  \url{http://adsabs.harvard.edu/abs/2008ApJ...673L.211M}

\bibitem[Morgan {\it et~al.\/}(2004)]{MorganEtal2004}
Morgan, H., Habbal, S.R., Rifai, S., Li, X., 2004, ``Hydrogen Ly$\alpha$
  Intensity Oscillations Observed by the Solar and Heliospheric Observatory
  Ultraviolet Coronagraph Spectrometer'', {\it Astrophys. J.\/}, {\bf 605},
  521--527. \newline ADS:
  \url{http://adsabs.harvard.edu/abs/2004ApJ...605..521M}

\bibitem[Mullan and Yakovlev(1995)]{MullanYakovlev1995}
Mullan, D.J., Yakovlev, O.I., 1995, ``Remote Sensing of the Solar Wind Using
  Radio Waves: Part 1'', {\it Irish Astron. J.\/}, {\bf 22}, 119--136. \newline
  ADS: \url{http://adsabs.harvard.edu/abs/1995IrAJ...22..119M}

\bibitem[Munro and Jackson(1977)]{MunroJackson1977}
Munro, R.H., Jackson, B.V., 1977, ``Physical properties of a polar coronal hole
  from 2 to 5 solar radii'', {\it Astrophys. J.\/}, {\bf 213}, 874--886.
  \newline ADS: \url{http://adsabs.harvard.edu/abs/1977ApJ...213..874M}

\bibitem[Munro and Withbroe(1972)]{MunroWithbroe1972}
Munro, R.H., Withbroe, G.L., 1972, ``Properties of a Coronal `hole' Derived
  from Extreme-Ultraviolet Observations'', {\it Astrophys. J.\/}, {\bf 176},
  511--520. \newline ADS:
  \url{http://adsabs.harvard.edu/abs/1972ApJ...176..511M}

\bibitem[Neugebauer {\it et~al.\/}(1998)]{NeugebauerEtal1998}
Neugebauer, M., Forsyth, R.J., Galvin, A.B., Harvey, K.L., Hoeksema, J.T.,
  Lazarus, A.J., Lepping, R.P., Linker, J.A., Mikic, Z., Steinberg, J.T., von
  Steiger, R., Wang, Y.-M., Wimmer-Schweingruber, R.F., 1998, ``Spatial
  structure of the solar wind and comparisons with solar data and models'',
  {\it J. Geophys. Res.\/}, {\bf 103}, 14\,587--14\,600. \newline ADS:
  \url{http://adsabs.harvard.edu/abs/1998JGR...10314587N}

\bibitem[Neupert and Pizzo(1974)]{NeupertPizzo1974}
Neupert, W.M., Pizzo, V., 1974, ``Solar coronal holes as sources of recurrent
  geomagnetic disturbances'', {\it J. Geophys. Res.\/}, {\bf 79}, 3701--3709.
  \newline ADS: \url{http://adsabs.harvard.edu/abs/1974JGR....79.3701N}

\bibitem[Newkirk(1967)]{Newkirk1967}
Newkirk, G., 1967, ``Structure of the Solar Corona'', {\it Ann. Rev. Astron.
  Astrophys.\/}, {\bf 5}, 213--266. \newline ADS:
  \url{http://adsabs.harvard.edu/abs/1967ARA&A...5..213N}

\bibitem[Newkirk and Harvey(1968)]{NewkirkHarvey1968}
Newkirk, G., Harvey, J., 1968, ``Coronal Polar Plumes'', {\it Solar Phys.\/},
  {\bf 3}, 321--343. \newline ADS:
  \url{http://adsabs.harvard.edu/abs/1968SoPh....3..321N}

\bibitem[Ng and Bhattacharjee(2008)]{NgBhattacharjee2008}
Ng, C.S., Bhattacharjee, A., 2008, ``A Constrained Tectonics Model for Coronal
  Heating'', {\it Astrophys. J.\/}, {\bf 675}, 899--905. \newline ADS:
  \url{http://adsabs.harvard.edu/abs/2008ApJ...675..899N}

\bibitem[Noci(1973)]{Noci1973}
Noci, G., 1973, ``Energy Budget in Coronal Holes'', {\it Solar Phys.\/}, {\bf
  28}, 403--407. \newline ADS:
  \url{http://adsabs.harvard.edu/abs/1973SoPh...28..403N}

\bibitem[Noci {\it et~al.\/}(1987)]{NociEtal1987}
Noci, G., Kohl, J.L., Withbroe, G.L., 1987, ``Solar wind diagnostics from
  Doppler-enhanced scattering'', {\it Astrophys. J.\/}, {\bf 315}, 706--715.
  \newline ADS: \url{http://adsabs.harvard.edu/abs/1987ApJ...315..706N}

\bibitem[Noci {\it et~al.\/}(1997)]{NociEtal1997}
Noci, G., Kohl, J.L., Antonucci, E., Tondello, G., Huber, M.C.E., Fineschi, S.,
  Gardner, L.D., Naletto, G., Nicolosi, P., Raymond, J.C., Romoli, M., Spadaro,
  D., Siegmund, O.H.W., Benna, C., Ciaravella, A., Giordano, S., Michels, J.,
  Modigliani, A., Panasyuk, A., Pernechele, C., Poletto, G., Smith, P.L.,
  Strachan, L., 1997, ``First results from UVCS/SOHO'', {\it Adv. Space
  Res.\/}, {\bf 20}, 2219--2230. \newline ADS:
  \url{http://adsabs.harvard.edu/abs/1997AdSpR..20.2219N}

\bibitem[Nolte {\it et~al.\/}(1976)]{NolteEtal1976}
Nolte, J.T., Krieger, A.S., Timothy, A.F., Gold, R.E., Roelof, E.C., Vaiana,
  G., Lazarus, A.J., Sullivan, J.D., McIntosh, P.S., 1976, ``Coronal holes as
  sources of solar wind'', {\it Solar Phys.\/}, {\bf 46}, 303--322. \newline
  ADS: \url{http://adsabs.harvard.edu/abs/1976SoPh...46..303N}

\bibitem[Ofman(2005)]{Ofman2005}
Ofman, L., 2005, ``MHD Waves and Heating in Coronal Holes'', {\it Space Sci.
  Rev.\/}, {\bf 120}, 67--94. \newline ADS:
  \url{http://adsabs.harvard.edu/abs/2005SSRv..120...67O}

\bibitem[Ofman {\it et~al.\/}(1999)]{OfmanEtal1999}
Ofman, L., Nakariakov, V.M., DeForest, C.E., 1999, ``Slow Magnetosonic Waves in
  Coronal Plumes'', {\it Astrophys. J.\/}, {\bf 514}, 441--447. \newline ADS:
  \url{http://adsabs.harvard.edu/abs/1999ApJ...514..441O}

\bibitem[Ofman {\it et~al.\/}(2000)]{OfmanEtal2000}
Ofman, L., Romoli, M., Poletto, G., Noci, G., Kohl, J.L., 2000, ``UVCS WLC
  Observations of Compressional Waves in the South Polar Coronal Hole'', {\it
  Astrophys. J.\/}, {\bf 529}, 592--598. \newline ADS:
  \url{http://adsabs.harvard.edu/abs/2000ApJ...529..592O}

\bibitem[Oughton {\it et~al.\/}(2004)]{OughtonEtal2004}
Oughton, S., Dmitruk, P., Matthaeus, W.H., 2004, ``Reduced magnetohydrodynamics
  and parallel spectral transfer'', {\it Phys. Plasmas\/}, {\bf 11},
  2214--2225. \newline ADS:
  \url{http://adsabs.harvard.edu/abs/2004PhPl...11.2214O}

\bibitem[Oughton {\it et~al.\/}(2006)]{OughtonEtal2006}
Oughton, S., Dmitruk, P., Matthaeus, W.H., 2006, ``A two-component
  phenomenology for homogeneous magnetohydrodynamic turbulence'', {\it Phys.
  Rev. Lett.\/}, {\bf 13}, 042306. \newline ADS:
  \url{http://adsabs.harvard.edu/abs/2006PhPl...13d2306O}

\bibitem[Pagel {\it et~al.\/}(2004)]{PagelEtal2004}
Pagel, A.C., Crooker, N.U., Zurbuchen, T.H., Gosling, J.T., 2004, ``Correlation
  of solar wind entropy and oxygen ion charge state ratio'', {\it J. Geophys.
  Res.\/}, {\bf 109}, A01\,113. \newline ADS:
  \url{http://adsabs.harvard.edu/abs/2004JGRA..10901113P}

\bibitem[Parashar {\it et~al.\/}(2009)]{ParasharEtal2009}
Parashar, T.N., Shay, M.A., Cassak, P.A., Matthaeus, W.H., 2009, ``Kinetic
  dissipation and anisotropic heating in a turbulent collisionless plasma'',
  {\it Phys. Plasmas\/}, {\bf 16}, 032310. \newline ADS:
  \url{http://adsabs.harvard.edu/abs/2009PhPl...16c2310P}

\bibitem[Parhi {\it et~al.\/}(1999)]{ParhiEtal1999}
Parhi, S., Suess, S.T., Sulkanen, M., 1999, ``Can Kelvin-Helmholtz
  instabilities of jet-like structures and plumes cause solar wind fluctuations
  at 1 AU?'', {\it J. Geophys. Res.\/}, {\bf 104}, 14\,781--14\,788. \newline
  ADS: \url{http://adsabs.harvard.edu/abs/1999JGR...10414781P}

\bibitem[Pariat {\it et~al.\/}(2009)]{PariatEtal2009}
Pariat, E., Antiochos, S.K., DeVore, C.R., 2009, ``A model for solar polar
  jets'', {\it Astrophys. J.\/}, {\bf 691}, 61--74. \newline ADS:
  \url{http://adsabs.harvard.edu/abs/2009ApJ...691...61P}

\bibitem[Parker(1958a)]{Parker1958a}
Parker, E.N., 1958a, ``Dynamics of the interplanetary gas and magnetic
  fields'', {\it Astrophys. J.\/}, {\bf 128}, 664--676. \newline ADS:
  \url{http://adsabs.harvard.edu/abs/1958ApJ...128..664P}

\bibitem[Parker(1958b)]{Parker1958b}
Parker, E.N., 1958b, ``Suprathermal particle generation in the solar corona'',
  {\it Astrophys. J.\/}, {\bf 128}, 677--685. \newline ADS:
  \url{http://adsabs.harvard.edu/abs/1958ApJ...128..677P}

\bibitem[Parker(1991)]{Parker1991}
Parker, E.N., 1991, ``Heating solar coronal holes'', {\it Astrophys. J.\/},
  {\bf 372}, 719--727. \newline ADS:
  \url{http://adsabs.harvard.edu/abs/1991ApJ...372..719P}

\bibitem[Parnell(2007)]{Parnell2007}
Parnell, C.E., 2007, ``3D magnetic reconnection, flares and coronal heating'',
  {\it Mem. Soc. Astron. Italiana\/}, {\bf 78}, 229--235. \newline ADS:
  \url{http://adsabs.harvard.edu/abs/2007MmSAI..78..229P}

\bibitem[Pasachoff {\it et~al.\/}(2007)]{PasachoffEtal2007}
Pasachoff, J.M., Ru\v{s}in, V., Druckm\"{u}ller, M., Saniga, M., 2007, ``Fine
  Structures in the White-Light Solar Corona at the 2006 Eclipse'', {\it
  Astrophys. J.\/}, {\bf 665}, 824--829. \newline ADS:
  \url{http://adsabs.harvard.edu/abs/2007ApJ...665..824P}

\bibitem[Peter and Judge(1999)]{PeterJudge1999}
Peter, H., Judge, P.G., 1999, ``On the Doppler Shifts of Solar Ultraviolet
  Emission Lines'', {\it Astrophys. J.\/}, {\bf 522}, 1148--1166. \newline ADS:
  \url{http://adsabs.harvard.edu/abs/1999ApJ...522.1148P}

\bibitem[Pevtsov {\it et~al.\/}(2003)]{PevtsovEtal2003}
Pevtsov, A.A., Fisher, G.H., Acton, L.W., Longcope, D.W., Johns-Krull, C.M.,
  Kankelborg, C.C., Metcalf, T.R., 2003, ``The Relationship Between X-Ray
  Radiance and Magnetic Flux'', {\it Astrophys. J.\/}, {\bf 598}, 1387--1391.
  \newline ADS: \url{http://adsabs.harvard.edu/abs/2003ApJ...598.1387P}

\bibitem[Piddington(1972)]{Piddington1972}
Piddington, J.H., 1972, ``A model of the quiet solar atmosphere'', {\it Solar
  Phys.\/}, {\bf 27}, 402--419. \newline ADS:
  \url{http://adsabs.harvard.edu/abs/1972SoPh...27..402P}

\bibitem[Pierrard and Lamy(2003)]{PierrardLamy2003}
Pierrard, V., Lamy, H., 2003, ``The effects of the velocity filtration
  mechanism on the minor ions of the corona'', {\it Solar Phys.\/}, {\bf 216},
  47--58. \newline ADS: \url{http://adsabs.harvard.edu/abs/2003SoPh..216...47P}

\bibitem[Pierrard {\it et~al.\/}(2004)]{PierrardEtal2004}
Pierrard, V., Lamy, H., Lemaire, J., 2004, ``Exospheric distributions of minor
  ions in the solar wind'', {\it J. Geophys. Res.\/}, {\bf 109}, A02\,118.
  \newline ADS: \url{http://adsabs.harvard.edu/abs/2004JGRA..10902118P}

\bibitem[Pinto {\it et~al.\/}(2009)]{PintoEtal2009}
Pinto, R., Grappin, R., Wang, Y.-M., L\'{e}orat, J., 2009, ``Time-dependent
  hydrodynamical simulations of slow solar wind, coronal inflows, and polar
  plumes'', {\it Astron. Astrophys.\/}, {\bf 497}, 537--543. \newline ADS:
  \url{http://adsabs.harvard.edu/abs/2009A&A...497..537P}

\bibitem[Pneuman(1968)]{Pneuman1968}
Pneuman, G.W., 1968, ``Some General Properties of Helmeted Coronal
  Structures'', {\it Solar Phys.\/}, {\bf 3}, 578--597. \newline ADS:
  \url{http://adsabs.harvard.edu/abs/1968SoPh....3..578P}

\bibitem[Pneuman(1973)]{Pneuman1973}
Pneuman, G.W., 1973, ``The Solar Wind and the Temperature-Density Structure of
  the Solar Corona'', {\it Solar Phys.\/}, {\bf 28}, 247--262. \newline ADS:
  \url{http://adsabs.harvard.edu/abs/1973SoPh...28..247P}

\bibitem[Pneuman(1980)]{Pneuman1980}
Pneuman, G.W., 1980, ``The physical structure of coronal holes: Influence of
  magnetic fields and coronal heating'', {\it Astron. Astrophys.\/}, {\bf 81},
  161--166. \newline ADS:
  \url{http://adsabs.harvard.edu/abs/1980A&A....81..161P}

\bibitem[Podesta and Bhattacharjee(2009)]{PodestaBhattacharjee2009}
Podesta, J.J., Bhattacharjee, A., 2009, ``Theory of incompressible MHD
  turbulence with cross-helicity'', {\it Phys. Rev. Lett.\/}, submitted

\bibitem[Poduval and Zhao(2004)]{PoduvalZhao2004}
Poduval, B., Zhao, X.P., 2004, ``Discrepancies in the prediction of solar wind
  using potential field source surface model: An investigation of possible
  sources'', {\it J. Geophys. Res.\/}, {\bf 109}, A08\,102. \newline ADS:
  \url{http://adsabs.harvard.edu/abs/2004JGRA..10908102P}

\bibitem[Poletto {\it et~al.\/}(2002)]{PolettoEtal2002}
Poletto, G., Suess, S.T., Biesecker, D.A., Esser, R., Gloeckler, G., Ko, Y.-K.,
  Zurbuchen, T.H., 2002, ``Low-latitude solar wind during the Fall 1998
  SOHO-Ulysses quadrature'', {\it J. Geophys. Res.\/}, {\bf 107}, 1300.
  \newline ADS: \url{http://adsabs.harvard.edu/abs/2002JGRA..107.1300P}

\bibitem[Priest and Forbes(2000)]{PriestForbes2000}
Priest, E.R., Forbes, T.G., 2000, {\it Magnetic Reconnection: MHD Theory and
  Applications\/}, Cambridge University Press, New York

\bibitem[Priest {\it et~al.\/}(2002)]{PriestEtal2002}
Priest, E.R., Heyvaerts, J.F., Title, A.M., 2002, ``A Flux-Tube Tectonics Model
  for Solar Coronal Heating Driven by the Magnetic Carpet'', {\it Astrophys.
  J.\/}, {\bf 576}, 533--551. \newline ADS:
  \url{http://adsabs.harvard.edu/abs/2002ApJ...576..533P}

\bibitem[Ralchenko {\it et~al.\/}(2007)]{RalchenkoEtal2007}
Ralchenko, Y., Feldman, U., Doschek, G.A., 2007, ``Is There a High-Energy
  Particle Population in the Quiet Solar Corona?'', {\it Astrophys. J.\/}, {\bf
  659}, 1682--1692. \newline ADS:
  \url{http://adsabs.harvard.edu/abs/2007ApJ...659.1682R}

\bibitem[Raouafi and Solanki(2004)]{RaouafiSolanki2004}
Raouafi, N.-E., Solanki, S.K., 2004, ``Effect of the electron density
  stratification on off-limb O VI line profiles: How large is the velocity
  distribution anisotropy in the solar corona?'', {\it Astron. Astrophys.\/},
  {\bf 427}, 725--733. \newline ADS:
  \url{http://adsabs.harvard.edu/abs/2004A&A...427..725R}

\bibitem[Raouafi and Solanki(2006)]{RaouafiSolanki2006}
Raouafi, N.-E., Solanki, S.K., 2006, ``Sensitivity of solar off-limb line
  profiles to electron density stratification and the velocity distribution
  anisotropy'', {\it Astron. Astrophys.\/}, {\bf 445}, 735--745. \newline ADS:
  \url{http://adsabs.harvard.edu/abs/2006A&A...445..735R}

\bibitem[Raouafi {\it et~al.\/}(2007)]{RaouafiEtal2007}
Raouafi, N.-E., Harvey, J.W., Solanki, S.K., 2007, ``Properties of Solar Polar
  Coronal Plumes Constrained by Ultraviolet Coronagraph Spectrometer Data'',
  {\it Astrophys. J.\/}, {\bf 658}, 643--656. \newline ADS:
  \url{http://adsabs.harvard.edu/abs/2007ApJ...658..643R}

\bibitem[Raouafi {\it et~al.\/}(2008)]{RaouafiEtal2008}
Raouafi, N.-E., Petrie, G.J.D., Norton, A.A., Henney, C.J., Solanki, S.K.,
  2008, ``Evidence for Polar Jets as Precursors of Polar Plume Formation'',
  {\it Astrophys. J. Lett.\/}, {\bf 682}, L137--L140. \newline ADS:
  \url{http://adsabs.harvard.edu/abs/2008ApJ...682L.137R}

\bibitem[Rappazzo {\it et~al.\/}(2008)]{RappazzoEtal2008}
Rappazzo, A.F., Velli, M., Einaudi, G., Dahlburg, R.~B., 2008, ``Nonlinear
  Dynamics of the Parker Scenario for Coronal Heating'', {\it Astrophys. J.\/},
  {\bf 677}, 1348--1366. \newline ADS:
  \url{http://adsabs.harvard.edu/abs/2008ApJ...677.1348R}

\bibitem[Reisenfeld {\it et~al.\/}(1999)]{ReisenfeldEtal1999}
Reisenfeld, D.B., McComas, D.J., Steinberg, J.T., 1999, ``Evidence of a solar
  origin for pressure balance structures in the high-latitude solar wind'',
  {\it Geophys. Res. Lett.\/}, {\bf 26}, 1805--1808. \newline ADS:
  \url{http://adsabs.harvard.edu/abs/1999GeoRL..26.1805R}

\bibitem[Reisenfeld {\it et~al.\/}(2001)]{ReisenfeldEtal2001}
Reisenfeld, D.B., Gary, S.P., Gosling, J.T., Steinberg, J.T., McComas, D.J.,
  Goldstein, B.E., Neugebauer, M., 2001, ``Helium energetics in the
  high-latitude solar wind: Ulysses observations'', {\it J. Geophys. Res.\/},
  {\bf 106}, 5693--5708. \newline ADS:
  \url{http://adsabs.harvard.edu/abs/2001JGR...106.5693R}

\bibitem[Riley {\it et~al.\/}(2001)]{RileyEtal2001}
Riley, P., Linker, J.A., Miki\'{c}, Z., 2001, ``An empirically-driven global
  MHD model of the solar corona and inner heliosphere'', {\it J. Geophys.
  Res.\/}, {\bf 106}, 15\,889--15\,902. \newline ADS:
  \url{http://adsabs.harvard.edu/abs/2001JGR...10615889R}

\bibitem[Roberts(2000)]{Roberts2000}
Roberts, B., 2000, ``Waves and Oscillations in the Corona'', {\it Solar
  Phys.\/}, {\bf 193}, 139--152. \newline ADS:
  \url{http://adsabs.harvard.edu/abs/2000SoPh..193..139R}

\bibitem[Romoli {\it et~al.\/}(2002)]{RomoliEtal2002}
Romoli, M., Frazin, R.A., Kohl, J.L., Gardner, L.D., Cranmer, S.R., Reardon,
  K., Fineschi, S., 2002, ``In-flight Calibration of the UVCS White Light
  Channel'', in {\it The Radiometric Calibration of SOHO\/}, (Eds.) Pauluhn,
  A., Huber, M.C.E., von Steiger, R., vol. SR-002 of ISSI Scientific Report,
  pp. 181--201, ESA Publications Division, Noordwijk

\bibitem[Rosner {\it et~al.\/}(1978)]{RosnerEtal1978}
Rosner, R., Tucker, W.H., Vaiana, G.S., 1978, ``Dynamics of the quiescent solar
  corona'', {\it Astrophys. J.\/}, {\bf 220}, 643--665. \newline ADS:
  \url{http://adsabs.harvard.edu/abs/1978ApJ...220..643R}

\bibitem[Roussev {\it et~al.\/}(2003)]{RoussevEtal2003}
Roussev, I.I., Gombosi, T.I., Sokolov, I.V., Velli, M., Manchester, W.,
  DeZeeuw, D.L., Liewer, P., T\'{o}th, G., Luhmann, J., 2003, ``A
  Three-dimensional Model of the Solar Wind Incorporating Solar Magnetogram
  Observations'', {\it Astrophys. J. Lett.\/}, {\bf 595}, L57--L61. \newline
  ADS: \url{http://adsabs.harvard.edu/abs/2003ApJ...595L..57R}

\bibitem[Rowlands {\it et~al.\/}(1966)]{RowlandsEtal1966}
Rowlands, J., Shapiro, V.D., Shevchenko, V.I., 1966, ``Quasilinear theory of
  plasma cyclotron instability'', {\it Sov. Phys. JETP\/}, {\bf 23}, 651--660

\bibitem[Rust(1983)]{Rust1983}
Rust, D.M., 1983, ``Coronal disturbances and their terrestrial effects'', {\it
  Space Sci. Rev.\/}, {\bf 34}, 21--36. \newline ADS:
  \url{http://adsabs.harvard.edu/abs/1983SSRv...34...21R}

\bibitem[Ryutova {\it et~al.\/}(2001)]{RyutovaEtal2001}
Ryutova, M., Habbal, S., Woo, R., Tarbell, T., 2001, ``Photospheric Network as
  the Energy Source for the quiet-Sun corona'', {\it Solar Phys.\/}, {\bf 200},
  213--234. \newline ADS:
  \url{http://adsabs.harvard.edu/abs/2001SoPh..200..213R}

\bibitem[Saito(1958)]{Saito1958}
Saito, K., 1958, ``Polar Rays of the Solar Corona'', {\it Pub. Astron. Soc.
  Japan\/}, {\bf 10}, 49--78. \newline ADS:
  \url{http://adsabs.harvard.edu/abs/1958PASJ...10...49S}

\bibitem[Sakao {\it et~al.\/}(2007)]{SakaoEtal2007}
Sakao, T., Kano, R., Narukage, N., Kotoku, J., Bando, T., DeLuca, E.E.,
  Lundquist, L.L., Tsuneta, S., Harra, L.K., Katsukawa, Y., Kubo, M., Hara, H.,
  Matsuzaki, K., Shimojo, M., Bookbinder, J.A., Golub, L., Korreck, K.E., Su,
  Y., Shibasaki, K., Shimizu, T., Nakatani, I., 2007, ``Continuous Plasma
  Outflows from the Edge of a Solar Active Region as a Possible Source of Solar
  Wind'', {\it Science\/}, {\bf 318}, 1585--1588. \newline ADS:
  \url{http://adsabs.harvard.edu/abs/2007Sci...318.1585S}

\bibitem[Sanz-Forcada and Dupree(2008)]{SanzForcadaDupree2008}
Sanz-Forcada, J., Dupree, A.K., 2008, ``Active cool stars and He I 10830 {\AA}:
  The coronal connection'', {\it Astron. Astrophys.\/}, {\bf 488}, 715--721.
  \newline ADS: \url{http://adsabs.harvard.edu/abs/2008A&A...488..715S}

\bibitem[Schatten {\it et~al.\/}(1969)]{SchattenEtal1969}
Schatten, K.H., Wilcox, J.M., Ness, N.F., 1969, ``A model of interplanetary and
  coronal magnetic fields'', {\it Solar Phys.\/}, {\bf 6}, 442--455. \newline
  ADS: \url{http://adsabs.harvard.edu/abs/1969SoPh....6..442S}

\bibitem[Scholl and Habbal(2008)]{SchollHabbal2008}
Scholl, I.F., Habbal, S.R., 2008, ``Automatic Detection and Classification of
  Coronal Holes and Filaments Based on EUV and Magnetogram Observations of the
  Solar Disk'', {\it Solar Phys.\/}, {\bf 248}, 425--439. \newline ADS:
  \url{http://adsabs.harvard.edu/abs/2008SoPh..248..425S}

\bibitem[Schrijver and Title(2003)]{SchrijverTitle2003}
Schrijver, C.J., Title, A.M., 2003, ``The Magnetic Connection between the Solar
  Photosphere and the Corona'', {\it Astrophys. J. Lett.\/}, {\bf 597},
  L165--L168. \newline ADS:
  \url{http://adsabs.harvard.edu/abs/2003ApJ...597L.165S}

\bibitem[Schrijver {\it et~al.\/}(1997)]{SchrijverEtal1997}
Schrijver, C.J., Title, A.M., van Ballegooijen, A.A., Hagenaar, H.J., Shine,
  R.A., 1997, ``Sustaining the Quiet Photospheric Network: The Balance of Flux
  Emergence, Fragmentation, Merging, and Cancellation'', {\it Astrophys. J.\/},
  {\bf 487}, 424--436. \newline ADS:
  \url{http://adsabs.harvard.edu/abs/1997ApJ...487..424S}

\bibitem[Schwadron and McComas(2003)]{SchwadronMcComas2003}
Schwadron, N.A., McComas, D.J., 2003, ``Solar wind scaling law'', {\it
  Astrophys. J.\/}, {\bf 599}, 1395--1403. \newline ADS:
  \url{http://adsabs.harvard.edu/abs/2003ApJ...599.1395S}

\bibitem[Schwadron and McComas(2008)]{SchwadronMcComas2008}
Schwadron, N.A., McComas, D.J., 2008, ``The Solar Wind Power from Magnetic
  Flux'', {\it Astrophys. J. Lett.\/}, {\bf 686}, L33--L36. \newline ADS:
  \url{http://adsabs.harvard.edu/abs/2008ApJ...686L..33S}

\bibitem[Schwadron {\it et~al.\/}(2006)]{SchwadronEtal2006}
Schwadron, N.A., McComas, D.J., DeForest, C., 2006, ``Relationship between
  Solar Wind and Coronal Heating: Scaling Laws from Solar X-Rays'', {\it
  Astrophys. J.\/}, {\bf 642}, 1173--1176. \newline ADS:
  \url{http://adsabs.harvard.edu/abs/2006ApJ...642.1173S}

\bibitem[Schwartz {\it et~al.\/}(1981)]{SchwartzEtal1981}
Schwartz, S.J., Feldman, W.C., Gary, S.P., 1981, ``The source of proton
  anisotropy in the high-speed solar wind'', {\it J. Geophys. Res.\/}, {\bf
  86}, 541--546. \newline ADS:
  \url{http://adsabs.harvard.edu/abs/1981JGR....86..541S}

\bibitem[Schwenn(2006)]{Schwenn2006}
Schwenn, R., 2006, ``Solar Wind Sources and Their Variations Over the Solar
  Cycle'', {\it Space Sci. Rev.\/}, {\bf 124}, 51--76. \newline ADS:
  \url{http://adsabs.harvard.edu/abs/2006SSRv..124...51S}

\bibitem[Scudder(1992{\natexlab{a}})]{Scudder1992a}
Scudder, J.D., 1992{\natexlab{a}}, ``Why all stars should posses circumstellar
  temperature inversions'', {\it Astrophys. J.\/}, {\bf 398}, 319--349.
  \newline ADS: \url{http://adsabs.harvard.edu/abs/1992ApJ...398..319S}

\bibitem[Scudder(1992{\natexlab{b}})]{Scudder1992b}
Scudder, J.D., 1992{\natexlab{b}}, ``On the causes of temperature change in
  inhomogenous low-density astrophysical plasmas'', {\it Astrophys. J.\/}, {\bf
  398}, 299--318. \newline ADS:
  \url{http://adsabs.harvard.edu/abs/1992ApJ...398..299S}

\bibitem[Scudder(1994)]{Scudder1994}
Scudder, J.D., 1994, ``Ion and electron suprathermal tail strengths in the
  transition region: Support for the velocity filtration model of the corona'',
  {\it Astrophys. J.\/}, {\bf 427}, 446--452. \newline ADS:
  \url{http://adsabs.harvard.edu/abs/1994ApJ...427..446S}

\bibitem[Seely {\it et~al.\/}(1997)]{SeelyEtal1997}
Seely, J.F., Feldman, U., Sch\"{u}hle, U., Wilhelm, K., Curdt, W., Lemaire, P.,
  1997, ``Turbulent Velocities and Ion Temperatures in the Solar Corona
  Obtained from SUMER Line Widths'', {\it Astrophys. J. Lett.\/}, {\bf 484},
  L87--L90. \newline ADS:
  \url{http://adsabs.harvard.edu/abs/1997ApJ...484L..87S}

\bibitem[Serviss(1909)]{Serviss1909}
Serviss, G.P., 1909, {\it Curiosities of the Sky: A Popular Presentation of the
  Great Riddles and Mysteries of Astronomy\/}, Harper \& Brothers, New York

\bibitem[Shebalin {\it et~al.\/}(1983)]{ShebalinEtal1983}
Shebalin, J.V., Matthaeus, W.H., Montgomery, D., 1983, ``Anisotropy in MHD
  turbulence due to a mean magnetic field'', {\it J. Plasma Phys.\/}, {\bf 29},
  525--547. \newline ADS:
  \url{http://adsabs.harvard.edu/abs/1983JPlPh..29..525S}

\bibitem[Sheeley~Jr {\it et~al.\/}(1997)]{SheeleyEtal1997}
Sheeley~Jr, N.R., Wang, Y.-M., Hawley, S~H., Brueckner, G.E., Dere, K.P.,
  Howard, R.A., Koomen, M.J., Korendyke, C.M., Michels, D.J., Paswaters, S.E.,
  Socker, D.G., St.~Cyr, O.C., Wang, D., Lamy, P.L., Llebaria, A., Schwenn, R.,
  Simnett, G.M., Plunkett, S., Biesecker, D.A., 1997, ``Measurements of Flow
  Speeds in the Corona between 2 and 30 $R_{\odot}$'', {\it Astrophys. J.\/},
  {\bf 484}, 472--478. \newline ADS:
  \url{http://adsabs.harvard.edu/abs/1997ApJ...484..472S}

\bibitem[Shibata {\it et~al.\/}(1994)]{ShibataEtal1994}
Shibata, K., Nitta, N., Strong, K.T., Matsumoto, R., Yokoyama, T., Hirayama,
  T., Hudson, H., Ogawara, Y., 1994, ``A gigantic coronal jet ejected from a
  compact active region in a coronal hole'', {\it Astrophys. J. Lett.\/}, {\bf
  431}, L51--L53. \newline ADS:
  \url{http://adsabs.harvard.edu/abs/1994ApJ...431L..51S}

\bibitem[Shimojo {\it et~al.\/}(2007)]{ShimojoEtal2007}
Shimojo, M., Narukage, N., Kano, R., Sakao, T., Tsuneta, S., Shibasaki, K.,
  Cirtain, J.W., Lundquist, L.L., Reeves, K., Savcheva, A., 2007, ``Fine
  Structures of Solar X-Ray Jets Observed with the X-Ray Telescope aboard
  Hinode'', {\it Pub. Astron. Soc. Japan\/}, {\bf 59}, S745--S750. \newline
  ADS: \url{http://adsabs.harvard.edu/abs/2007PASJ...59S.745S}

\bibitem[Singh and Khazanov(2004)]{SinghKhazanov2004}
Singh, N., Khazanov, G., 2004, ``Numerical simulation of waves driven by plasma
  currents generated by low-frequency Alfv\'{e}n waves in a multi-ion plasma'',
  {\it J. Geophys. Res.\/}, {\bf 109}, A05\,210. \newline ADS:
  \url{http://adsabs.harvard.edu/abs/2004JGRA..10905210S}

\bibitem[Singh {\it et~al.\/}(2007)]{SinghEtal2007}
Singh, N., Khazanov, G., Mukhter, A., 2007, ``Electrostatic wave generation and
  transverse ion acceleration by Alfv\'{e}nic wave components of broadband
  extremely low frequency turbulence'', {\it J. Geophys. Res.\/}, {\bf 112},
  A06\,210. \newline ADS:
  \url{http://adsabs.harvard.edu/abs/2007JGRA..11206210S}

\bibitem[Sittler and Guhathakurta(1999)]{SittlerGuhathakurta1999}
Sittler, E.C., Guhathakurta, M., 1999, ``Semiempirical Two-dimensional
  MagnetoHydrodynamic Model of the Solar Corona and Interplanetary Medium'',
  {\it Astrophys. J.\/}, {\bf 523}, 812--826. \newline ADS:
  \url{http://adsabs.harvard.edu/abs/1999ApJ...523..812S}

\bibitem[Smith {\it et~al.\/}(2006)]{SmithEtal2006}
Smith, C.W., Hamilton, K., Vasquez, B.J., Leamon, R.J., 2006, ``Dependence of
  the Dissipation Range Spectrum of Interplanetary Magnetic Fluctuationson the
  Rate of Energy Cascade'', {\it Astrophys. J. Lett.\/}, {\bf 645}, L85--L88.
  \newline ADS: \url{http://adsabs.harvard.edu/abs/2006ApJ...645L..85S}

\bibitem[Spangler(2002)]{Spangler2002}
Spangler, S.R., 2002, ``The Amplitude of Magnetohydrodynamic Turbulence in the
  Inner Solar Wind'', {\it Astrophys. J.\/}, {\bf 576}, 997--1004. \newline
  ADS: \url{http://adsabs.harvard.edu/abs/2002ApJ...576..997S}

\bibitem[Spruit(1981)]{Spruit1981}
Spruit, H.C., 1981, ``Motion of magnetic flux tubes in the solar convection
  zone and chromosphere'', {\it Astron. Astrophys.\/}, {\bf 98}, 155--160.
  \newline ADS: \url{http://adsabs.harvard.edu/abs/1981A&A....98..155S}

\bibitem[Spruit(1984)]{Spruit1984}
Spruit, H.C., 1984, ``Interaction of Fluxtubes With Convection'', in {\it
  Small-Scale Dynamical Processes in Quiet Stellar Atmospheres\/}, (Ed.) Keil,
  S.L., 25--29 July 1983, Sunspot, New Mexico, p. 249, National Solar
  Observatory, Sunspot, NM. \newline ADS:
  \url{http://adsabs.harvard.edu/abs/1984ssdp.conf..249S}

\bibitem[Stoddard {\it et~al.\/}(1966)]{StoddardEtal1966}
Stoddard, L.G., Carson, D.G., Saito, K., 1966, ``Polar Rays of the Solar Corona
  Observed at the 1963 Eclipse'', {\it Astrophys. J.\/}, {\bf 145}, 796--799.
  \newline ADS: \url{http://adsabs.harvard.edu/abs/1966ApJ...145..796S}

\bibitem[Strachan {\it et~al.\/}(1993)]{StrachanEtal1993}
Strachan, L., Kohl, J.L., Weiser, H., Withbroe, G.L., Munro, R.H., 1993, ``A
  Doppler dimming determination of coronal outflow velocity'', {\it Astrophys.
  J.\/}, {\bf 412}, 410--420. \newline ADS:
  \url{http://adsabs.harvard.edu/abs/1993ApJ...412..410S}

\bibitem[Strachan {\it et~al.\/}(2002)]{StrachanEtal2002}
Strachan, L., Suleiman, R., Panasyuk, A.V., Biesecker, D.A., Kohl, J.L., 2002,
  ``Empirical densities, kinetic temperatures, and outflow velocities in the
  equatorial streamer belt at solar minimum'', {\it Astrophys. J.\/}, {\bf
  571}, 1008--1014. \newline ADS:
  \url{http://adsabs.harvard.edu/abs/2002ApJ...571.1008S}

\bibitem[Strauss(1976)]{Strauss1976}
Strauss, H.R., 1976, ``Nonlinear, three-dimensional magnetohydrodynamics of
  noncircular tokamaks'', {\it Phys. Fluids\/}, {\bf 19}, 134--140. \newline
  ADS: \url{http://adsabs.harvard.edu/abs/1976PhFl...19..134S}

\bibitem[Sturrock {\it et~al.\/}(1999)]{SturrockEtal1999}
Sturrock, P.A., Roald, C.B., Wolfson, R., 1999, ``Chromospheric Magnetic
  Reconnection and its Implication for Coronal Heating'', {\it Astrophys. J.
  Lett.\/}, {\bf 524}, L75--L78. \newline ADS:
  \url{http://adsabs.harvard.edu/abs/1999ApJ...524L..75S}

\bibitem[Suess(1979)]{Suess1979}
Suess, S.~T., 1979, ``Models of coronal hole flows'', {\it Space Sci. Rev.\/},
  {\bf 23}, 159--200. \newline ADS:
  \url{http://adsabs.harvard.edu/abs/1979SSRv...23..159S}

\bibitem[Susino {\it et~al.\/}(2008)]{SusinoEtal2008}
Susino, R., Ventura, R., Spadaro, D., Vourlidas, A., Landi, E., 2008,
  ``Physical parameters along the boundaries of a mid-latitude streamer and in
  its adjacent regions'', {\it Astron. Astrophys.\/}, {\bf 488}, 303--310.
  \newline ADS: \url{http://adsabs.harvard.edu/abs/2008A&A...488..303S}

\bibitem[Suzuki(2006)]{Suzuki2006}
Suzuki, T.K., 2006, ``Forecasting Solar Wind Speeds'', {\it Astrophys. J.
  Lett.\/}, {\bf 640}, L75--L78. \newline ADS:
  \url{http://adsabs.harvard.edu/abs/2006ApJ...640L..75S}

\bibitem[Suzuki and Inutsuka(2006)]{SuzukiInutsuka2006}
Suzuki, T.K., Inutsuka, S.-I., 2006, ``Solar winds driven by nonlinear
  low-frequency Alfv\'{e}n waves from the photosphere: Parametric study for
  fast/slow winds and disappearance of solar winds'', {\it J. Geophys. Res.\/},
  {\bf 111}, A06\,101. \newline ADS:
  \url{http://adsabs.harvard.edu/abs/2006JGRA..11106101S}

\bibitem[Suzuki {\it et~al.\/}(2007)]{SuzukiEtal2007}
Suzuki, T.K., Lazarian, A., Beresnyak, A., 2007, ``Cascading of Fast-Mode
  Balanced and Imbalanced Turbulence'', {\it Astrophys. J.\/}, {\bf 662},
  1033--1042. \newline ADS:
  \url{http://adsabs.harvard.edu/abs/2007ApJ...662.1033S}

\bibitem[Tanskanen {\it et~al.\/}(2005)]{TanskanenEtal2005}
Tanskanen, E.I., Slavin, J.A., Tanskanen, A.J, Viljanen, A., Pulkkinen, T.I.,
  Koskinen, H.E.J., Pulkkinen, A., Eastwood, J., 2005, ``Magnetospheric
  substorms are strongly modulated by interplanetary high-speed streams'', {\it
  Geophys. Res. Lett.\/}, {\bf 32}, L16\,104. \newline ADS:
  \url{http://adsabs.harvard.edu/abs/2005GeoRL..3216104T}

\bibitem[Tappin {\it et~al.\/}(1999)]{TappinEtal1999}
Tappin, S.J., Simnett, G.M., Lyons, M.A., 1999, ``A determination of the
  outflow speeds in the lower solar wind'', {\it Astron. Astrophys.\/}, {\bf
  350}, 302--309. \newline ADS:
  \url{http://adsabs.harvard.edu/abs/1999A&A...350..302T}

\bibitem[Telloni {\it et~al.\/}(2009)]{TelloniEtal2009}
Telloni, D., Antonucci, E., Bruno, R., D'Amicis, R., 2009, ``Persistent and
  Self-Similar Large-Scale Density Fluctuations in the Solar Corona'', {\it
  Astrophys. J.\/}, {\bf 693}, 1022--1028. \newline ADS:
  \url{http://adsabs.harvard.edu/abs/2009ApJ...693.1022T}

\bibitem[Teplitskaya {\it et~al.\/}(2007)]{TeplitskayaEtal2007}
Teplitskaya, R.B., Turova, I.P., Ozhogina, O.A., 2007, ``The Lower Chromosphere
  in a Coronal Hole'', {\it Solar Phys.\/}, {\bf 243}, 143--161. \newline ADS:
  \url{http://adsabs.harvard.edu/abs/2007SoPh..243..143T}

\bibitem[Teriaca {\it et~al.\/}(2003)]{TeriacaEtal2003}
Teriaca, L., Poletto, G., Romoli, M., Biesecker, D.A., 2003, ``The nascent
  solar wind: Origin and acceleration'', {\it Astrophys. J.\/}, {\bf 588},
  566--577. \newline ADS:
  \url{http://adsabs.harvard.edu/abs/2003ApJ...588..566T}

\bibitem[Thieme {\it et~al.\/}(1990)]{ThiemeEtal1990}
Thieme, K.M., Marsch, E., Schwenn, R., 1990, ``Spatial structures in high-speed
  streams as signatures of fine structures in coronal holes'', {\it Ann.
  Geophys.\/}, {\bf 8}, 713--723. \newline ADS:
  \url{http://adsabs.harvard.edu/abs/1990AnGeo...8..713T}

\bibitem[Thompson {\it et~al.\/}(2000)]{ThompsonEtal2000}
Thompson, B.J., Cliver, E.W., Nitta, N., Delann\'{e}e, C., Delaboudini\`{e}re,
  J.-P., 2000, ``Coronal dimmings and energetic CMEs in April-May 1998'', {\it
  Geophys. Res. Lett.\/}, {\bf 27}, 1431--1434. \newline ADS:
  \url{http://adsabs.harvard.edu/abs/2000GeoRL..27.1431T}

\bibitem[Timothy {\it et~al.\/}(1975)]{TimothyEtal1975}
Timothy, A.F., Krieger, A.S., Vaiana, G.S., 1975, ``The structure and evolution
  of coronal holes'', {\it Solar Phys.\/}, {\bf 42}, 135--156. \newline ADS:
  \url{http://adsabs.harvard.edu/abs/1975SoPh...42..135T}

\bibitem[Title and Schrijver(1998)]{TitleSchrijver1998}
Title, A.M., Schrijver, C.J., 1998, ``The Sun's Magnetic Carpet'', in {\it
  Proceedings of the 10th Cambridge Workshop on Cool Stars, Stellar Systems,
  and the Sun\/}, (Eds.) Donahue, R.A., Bookbinder, J.A., 15--19 July 1997,
  Cambridge, Massachusetts, vol. 154 of ASP Conf. Proc., pp. 345--358,
  Astronomical Society of the Pacific, San Francisco. \newline ADS:
  \url{http://adsabs.harvard.edu/abs/1998ASPC..154..345T}

\bibitem[Tomczyk and McIntosh(2009)]{TomczykMcIntosh2009}
Tomczyk, S., McIntosh, S.W., 2009, ``Time-Distance Seismology of the Solar
  Corona with CoMP'', {\it Astrophys. J.\/}, {\bf 697}, 1384--1391. \newline
  ADS: \url{http://adsabs.harvard.edu/abs/2009ApJ...697.1384T}

\bibitem[Tomczyk {\it et~al.\/}(2007)]{TomczykEtal2007}
Tomczyk, S., McIntosh, S.W., Keil, S.L., Judge, P.G., Schad, T., Seeley, D.H.,
  Edmondson, J., 2007, ``Alfv\'{e}n waves in the solar corona'', {\it
  Science\/}, {\bf 317}, 1192--1196. \newline ADS:
  \url{http://adsabs.harvard.edu/abs/2007Sci...317.1192T}

\bibitem[T\'{o}th {\it et~al.\/}(2005)]{TothEtal2005}
T\'{o}th, G., Sokolov, I.V., Gombosi, T.I., Chesney, D.R., Clauer, C.R.,
  De~Zeeuw, D.L., Hansen, K.C., Kane, K.J., Manchester, W.B., Oehmke, R.C.,
  Powell, K.G., Ridley, A.J., Roussev, I.I., Stout, Q.F., Volberg, O., Wolf,
  R.A., Sazykin, S., Chan, A., Yu, B., K\'{o}ta, J., 2005, ``Space Weather
  Modeling Framework: A new tool for the space science community'', {\it J.
  Geophys. Res.\/}, {\bf 110}, A12\,226. \newline ADS:
  \url{http://adsabs.harvard.edu/abs/2005JGRA..11012226T}

\bibitem[Tousey {\it et~al.\/}(1968)]{TouseyEtal1968}
Tousey, R., Sandlin, G.D., Purcell, J.D., 1968, ``On Some Aspects of XUV
  Spectroheliograms'', in {\it Structure and Development of Solar Active
  Regions\/}, (Ed.) Kiepenheuer, K.O., Proceedings of IAU Symp. 35, 4--8
  September 1967, Budapest, Hungary, pp. 411--419, D. Reidel, Dordrecht.
  \newline ADS: \url{http://adsabs.harvard.edu/abs/1968IAUS...35..411T}

\bibitem[Tsuneta {\it et~al.\/}(2008)]{TsunetaEtal2008}
Tsuneta, S., Ichimoto, K., Katsukawa, Y., Lites, B.W., Matsuzaki, K., Nagata,
  S., Orozco~Su\'{a}rez, D., Shimizu, T., Shimojo, M., Shine, R.A., Suematsu,
  Y., Suzuki, T.K., Tarbell, T.D., Title, A.M., 2008, ``The Magnetic Landscape
  of the Sun's Polar Region'', {\it Astrophys. J.\/}, {\bf 688}, 1374--1381.
  \newline ADS: \url{http://adsabs.harvard.edu/abs/2008ApJ...688.1374T}

\bibitem[Tu and Marsch(1994)]{TuMarsch1994}
Tu, C.-Y., Marsch, E., 1994, ``On the nature of compressive fluctuations in the
  solar wind'', {\it J. Geophys. Res.\/}, {\bf 99}, 21\,481--21\,509. \newline
  ADS: \url{http://adsabs.harvard.edu/abs/1994JGR....9921481T}

\bibitem[Tu and Marsch(1995)]{TuMarsch1995}
Tu, C.-Y., Marsch, E., 1995, ``MHD Structures, Waves and Turbulence in the
  Solar Wind: Observations and Theories'', {\it Space Sci. Rev.\/}, {\bf 73},
  1--210. \newline ADS: \url{http://adsabs.harvard.edu/abs/1995SSRv...73....1T}

\bibitem[Tu and Marsch(1997)]{TuMarsch1997}
Tu, C.-Y., Marsch, E., 1997, ``Two-fluid model for heating of the solar corona
  and acceleration of the solar wind by high-frequency Alfv{\'{e}}n waves'',
  {\it Solar Phys.\/}, {\bf 171}, 363--391. \newline ADS:
  \url{http://adsabs.harvard.edu/abs/1997SoPh..171..363T}

\bibitem[Tu and Marsch(2001)]{TuMarsch2001}
Tu, C.-Y., Marsch, E., 2001, ``On cyclotron wave heating and acceleration of
  solar wind ions in the outer corona'', {\it J. Geophys. Res.\/}, {\bf 106},
  8233--8252. \newline ADS:
  \url{http://adsabs.harvard.edu/abs/2001JGR...106.8233T}

\bibitem[Tu {\it et~al.\/}(1998)]{TuEtal1998}
Tu, C.-Y., Marsch, E., Wilhelm, K., Curdt, W., 1998, ``Ion Temperatures in a
  Solar Polar Coronal Hole Observed by SUMER on SOHO'', {\it Astrophys. J.\/},
  {\bf 503}, 475--482. \newline ADS:
  \url{http://adsabs.harvard.edu/abs/1998ApJ...503..475T}

\bibitem[Tu {\it et~al.\/}(2005)]{TuEtal2005}
Tu, C.-Y., Zhou, C., Marsch, E., Xia, L.-D., Zhao, L., Wang, J.-X., Wilhelm,
  K., 2005, ``Solar wind origin in coronal funnels'', {\it Science\/}, {\bf
  308}, 519--523. \newline ADS:
  \url{http://adsabs.harvard.edu/abs/2005Sci...308..519T}

\bibitem[Usmanov and Goldstein(2006)]{UsmanovGoldstein2006}
Usmanov, A.V., Goldstein, M.L., 2006, ``A three-dimensional MHD solar wind
  model with pickup protons'', {\it J. Geophys. Res.\/}, {\bf 111}, A07\,101.
  \newline ADS: \url{http://adsabs.harvard.edu/abs/2006JGRA..11107101U}

\bibitem[van Ballegooijen(1994)]{vanBallegooijen1994}
van Ballegooijen, A.A., 1994, ``Energy release in stellar magnetospheres'',
  {\it Space Sci. Rev.\/}, {\bf 68}, 299--307. \newline ADS:
  \url{http://adsabs.harvard.edu/abs/1994SSRv...68..299V}

\bibitem[van~de Hulst(1950)]{vandeHulst1950}
van~de Hulst, H.C., 1950, ``The electron density of the solar corona'', {\it
  Bull. Astron. Inst. Netherlands\/}, {\bf 11}, 135--150. \newline ADS:
  \url{http://adsabs.harvard.edu/abs/1950BAN....11..135V}

\bibitem[Vaquero(2003)]{Vaquero2003}
Vaquero, J.M., 2003, ``The Solar Corona in the Eclipse of 24 June 1778'', {\it
  Solar Phys.\/}, {\bf 216}, 41--45. \newline ADS:
  \url{http://adsabs.harvard.edu/abs/2003SoPh..216...41V}

\bibitem[Vasquez and Hollweg(1996)]{VasquezHollweg1996}
Vasquez, B.J., Hollweg, J.V., 1996, ``Formation of arc-shaped Alfv\'{e}n waves
  and rotational discontinuities from oblique linearly polarized wave trains'',
  {\it J. Geophys. Res.\/}, {\bf 101}, 13,527--13,540. \newline ADS:
  \url{http://adsabs.harvard.edu/abs/1996JGR...10113527V}

\bibitem[Vasquez and Hollweg(1998)]{VasquezHollweg1998}
Vasquez, B.J., Hollweg, J.V., 1998, ``Formation of spherically polarized
  Alfv\'{e}n waves and imbedded rotational discontinuities from a small number
  of entirely oblique waves'', {\it J. Geophys. Res.\/}, {\bf 103}, 335--347.
  \newline ADS: \url{http://adsabs.harvard.edu/abs/1998JGR...103..335V}

\bibitem[Verdini and Velli(2007)]{VerdiniVelli2007}
Verdini, A., Velli, M., 2007, ``Alfv\'{e}n waves and turbulence in the solar
  atmosphere and solar wind'', {\it Astrophys. J.\/}, {\bf 662}, 669--676.
  \newline ADS: \url{http://adsabs.harvard.edu/abs/2007ApJ...662..669V}

\bibitem[Verdini {\it et~al.\/}(2009)]{VerdiniEtal2009}
Verdini, A., Velli, M., Buchlin, E., 2009, ``Reflection Driven MHD Turbulence
  in the Solar Atmosphere and Wind'', {\it Earth, Moon, \& Planets\/}, {\bf
  104}, 121--125. \newline ADS:
  \url{http://adsabs.harvard.edu/abs/2009EM&P..104..121V}

\bibitem[Voitenko and Goossens(2002)]{VoitenkoGoossens2002}
Voitenko, Y.M., Goossens, M., 2002, ``Excitation of high-frequency Alfv\'{e}n
  waves by plasma outflows from coronal reconnection events'', {\it Solar
  Phys.\/}, {\bf 206}, 285--313. \newline ADS:
  \url{http://adsabs.harvard.edu/abs/2002SoPh..206..285V}

\bibitem[Voitenko and Goossens(2003)]{VoitenkoGoossens2003}
Voitenko, Y.M., Goossens, M., 2003, ``Kinetic excitation mechanisms for
  ion-cyclotron kinetic Alfv\'{e}n waves in Sun-Earth connection'', {\it Space
  Sci. Rev.\/}, {\bf 107}, 387--401. \newline ADS:
  \url{http://adsabs.harvard.edu/abs/2003SSRv..107..387V}

\bibitem[Voitenko and Goossens(2004)]{VoitenkoGoossens2004}
Voitenko, Y.M., Goossens, M., 2004, ``Cross-Field Heating of Coronal Ions by
  Low-Frequency Kinetic Alfv{\'{e}}n Waves'', {\it Astrophys. J. Lett.\/}, {\bf
  605}, L149--L152. \newline ADS:
  \url{http://adsabs.harvard.edu/abs/2004ApJ...605L.149V}

\bibitem[von Steiger {\it et~al.\/}(1995)]{vonSteigerEtal1995}
von Steiger, R., Wimmer-Schweingruber, R.F., Geiss, J., Gloeckler, G., 1995,
  ``Abundance variations in the solar wind'', {\it Adv. Space Res.\/}, {\bf
  15}, 3--12. \newline ADS:
  \url{http://adsabs.harvard.edu/abs/1995AdSpR..15Q...3V}

\bibitem[Vranjes and Poedts(2008)]{VranjesPoedts2008}
Vranjes, J., Poedts, S., 2008, ``Growing drift-cyclotron modes in the hot solar
  atmosphere'', {\it Astron. Astrophys.\/}, {\bf 482}, 653--656. \newline ADS:
  \url{http://adsabs.harvard.edu/abs/2008A&A...482..653V}

\bibitem[Vr\v{s}nak {\it et~al.\/}(2007)]{VrsnakEtal2007}
Vr\v{s}nak, B., Temmer, M., Veronig, A.M., 2007, ``Coronal Holes and Solar Wind
  High-Speed Streams: I. Forecasting the Solar Wind Parameters'', {\it Solar
  Phys.\/}, {\bf 240}, 315--330. \newline ADS:
  \url{http://adsabs.harvard.edu/abs/2007SoPh..240..315V}

\bibitem[Waldmeier(1955)]{Waldmeier1955}
Waldmeier, M., 1955, ``Ergenbnisse der Z\"{u}rcher Sonnenfinsternisexpedition
  1954: 1. Vorl\"{a}ufige Photometrie der Korona, Mit 7 Textabbildungen'', {\it
  Zeit. Astrophysik\/}, {\bf 36}, 275--292. \newline ADS:
  \url{http://adsabs.harvard.edu/abs/1955ZA.....36..275W}

\bibitem[Waldmeier(1956)]{Waldmeier1956}
Waldmeier, M., 1956, ``Synoptische Karten der Sonnenkorona'', {\it Zeit.
  Astrophysik\/}, {\bf 38}, 219--236. \newline ADS:
  \url{http://adsabs.harvard.edu/abs/1956ZA.....38..219W}

\bibitem[Waldmeier(1957)]{Waldmeier1957}
Waldmeier, M., 1957, {\it Die Sonnenkorona\/}, vol.~2, Birkh\"{a}user, Basel

\bibitem[Waldmeier(1975)]{Waldmeier1975}
Waldmeier, M., 1975, ``The coronal hole at the 7 March 1970 solar eclipse'',
  {\it Solar Phys.\/}, {\bf 40}, 351--358. \newline ADS:
  \url{http://adsabs.harvard.edu/abs/1975SoPh...40..351W}

\bibitem[Waldmeier(1981)]{Waldmeier1981}
Waldmeier, M., 1981, ``Cyclic variations of the polar coronal hole'', {\it
  Solar Phys.\/}, {\bf 70}, 251--258. \newline ADS:
  \url{http://adsabs.harvard.edu/abs/1981SoPh...70..251W}

\bibitem[Wang and Wu(2009)]{WangWu2009}
Wang, C.B., Wu, C.S., 2009, ``Pseudoheating of protons in the presence of
  Alfv\'{e}nic turbulence'', {\it Phys. Plasmas\/}, {\bf 16}, 020\,703.
  \newline ADS: \url{http://adsabs.harvard.edu/abs/2009PhPl...16b0703W}

\bibitem[Wang and Siscoe(1980)]{WangSiscoe1980}
Wang, P.K., Siscoe, G.L., 1980, ``Ancient Chinese Observations of Physical
  Phenomena Attending Solar Eclipses'', {\it Solar Phys.\/}, {\bf 66},
  187--193. \newline ADS:
  \url{http://adsabs.harvard.edu/abs/1980SoPh...66..187W}

\bibitem[Wang(1994)]{Wang1994}
Wang, Y.-M., 1994, ``Polar plumes and the solar wind'', {\it Astrophys. J.
  Lett.\/}, {\bf 435}, L153--L156. \newline ADS:
  \url{http://adsabs.harvard.edu/abs/1994ApJ...435L.153W}

\bibitem[Wang(1998)]{Wang1998}
Wang, Y.-M., 1998, ``Network Activity and the Evaporative Formation of Polar
  Plumes'', {\it Astrophys. J. Lett.\/}, {\bf 501}, L145--L148. \newline ADS:
  \url{http://adsabs.harvard.edu/abs/1998ApJ...501L.145W}

\bibitem[Wang(2009)]{Wang2009}
Wang, Y.-M., 2009, ``Coronal holes and open magnetic flux'', {\it Space Sci.
  Rev.\/}, {\bf 144}, 383--399. \newline ADS:
  \url{http://adsabs.harvard.edu/abs/2009SSRv..144..383W}

\bibitem[Wang and Sheeley~Jr(1990)]{WangSheeley1990}
Wang, Y.-M., Sheeley~Jr, N.R., 1990, ``Solar wind speed and coronal flux-tube
  expansion'', {\it Astrophys. J.\/}, {\bf 355}, 726--732. \newline ADS:
  \url{http://adsabs.harvard.edu/abs/1990ApJ...355..726W}

\bibitem[Wang and Sheeley~Jr(1991)]{WangSheeley1991}
Wang, Y.-M., Sheeley~Jr, N.R., 1991, ``Why fast solar wind originates from
  slowly expanding coronal flux tubes'', {\it Astrophys. J. Lett.\/}, {\bf
  372}, L45--L48. \newline ADS:
  \url{http://adsabs.harvard.edu/abs/1991ApJ...372L..45W}

\bibitem[Wang and Sheeley~Jr(2006)]{WangSheeley2006}
Wang, Y.-M., Sheeley~Jr, N.R., 2006, ``Sources of the Solar Wind at Ulysses
  during 1990--2006'', {\it Astrophys. J.\/}, {\bf 653}, 708--718. \newline
  ADS: \url{http://adsabs.harvard.edu/abs/2006ApJ...653..708W}

\bibitem[Wang {\it et~al.\/}(1996)]{WangEtal1996}
Wang, Y.-M., Hawley, S.H., Sheeley~Jr, N.R., 1996, ``The magnetic nature of
  coronal holes'', {\it Science\/}, {\bf 271}, 464--469. \newline ADS:
  \url{http://adsabs.harvard.edu/abs/1996Sci...271..464W}

\bibitem[Wang {\it et~al.\/}(1998)]{WangEtal1998}
Wang, Y.-M., Sheeley~Jr, N.R., Socker, D.G., Howard, R.A., Brueckner, G.E.,
  Michels, D.J., Moses, D., St.~Cyr, O.C., Llebaria, A., Delaboudini\`{e}re,
  J.-P., 1998, ``Observations of Correlated White-Light and Extreme-Ultraviolet
  Jets from Polar Coronal Holes'', {\it Astrophys. J.\/}, {\bf 508}, 899--907.
  \newline ADS: \url{http://adsabs.harvard.edu/abs/1998ApJ...508..899W}

\bibitem[Wang {\it et~al.\/}(2000)]{WangEtal2000}
Wang, Y.-M., Sheeley~Jr, N.R., Socker, D.G., Howard, R.A., Rich, N.B., 2000,
  ``The dynamical nature of coronal streamers'', {\it J. Geophys. Res.\/}, {\bf
  105}, 25\,133--25\,142. \newline ADS:
  \url{http://adsabs.harvard.edu/abs/2000JGR...10525133W}

\bibitem[Wang {\it et~al.\/}(2007)]{WangEtal2007}
Wang, Y.-M., Biersteker, J.B., Sheeley~Jr, N.R., Koutchmy, S., Mouette, J.,
  Druckm\"{u}ller, M., 2007, ``The Solar Eclipse of 2006 and the Origin of
  Raylike Features in the White-Light Corona'', {\it Astrophys. J.\/}, {\bf
  660}, 882--892. \newline ADS:
  \url{http://adsabs.harvard.edu/abs/2007ApJ...660..882W}

\bibitem[Wang {\it et~al.\/}(2009)]{WangEtal2009}
Wang, Y.-M., Ko, Y.-K., Grappin, R., 2009, ``Slow Solar Wind from Open Regions
  with Strong Low-Coronal Heating'', {\it Astrophys. J.\/}, {\bf 691},
  760--769. \newline ADS:
  \url{http://adsabs.harvard.edu/abs/2009ApJ...691..760W}

\bibitem[Warren {\it et~al.\/}(1997)]{WarrenEtal1997}
Warren, H.P., Mariska, J.T., Wilhelm, K., Lemaire, P., 1997, ``Doppler Shifts
  and Nonthermal Broadening in the Quiet Solar Transition Region: O VI'', {\it
  Astrophys. J. Lett.\/}, {\bf 484}, L91--L94. \newline ADS:
  \url{http://adsabs.harvard.edu/abs/1997ApJ...484L..91W}

\bibitem[Wilcox(1968)]{Wilcox1968}
Wilcox, J.M., 1968, ``The Interplanetary Magnetic Field. Solar Origin and
  Terrestrial Effects'', {\it Space Sci. Rev.\/}, {\bf 8}, 258--328. \newline
  ADS: \url{http://adsabs.harvard.edu/abs/1968SSRv....8..258W}

\bibitem[Wilhelm(2006)]{Wilhelm2006}
Wilhelm, K., 2006, ``Solar coronal-hole plasma densities and temperatures'',
  {\it Astron. Astrophys.\/}, {\bf 455}, 697--708. \newline ADS:
  \url{http://adsabs.harvard.edu/abs/2006A&A...455..697W}

\bibitem[Wilhelm {\it et~al.\/}(1995)]{WilhelmEtal1995}
Wilhelm, K., Curdt, W., Marsch, E., Sch\"{u}hle, U., Lemaire, P., Gabriel, A.,
  Vial, J.-C., Grewing, M., Huber, M.C.E., Jordan, S.D., Poland, A.I., Thomas,
  R.J., K\"{u}hne, M., Timothy, J.G., Hassler, D.M., Siegmund, O.H.W., 1995,
  ``SUMER: Solar Ultraviolet Measurements of Emitted Radiation'', {\it Solar
  Phys.\/}, {\bf 162}, 189--231. \newline ADS:
  \url{http://adsabs.harvard.edu/abs/1995SoPh..162..189W}

\bibitem[Wilhelm {\it et~al.\/}(2000)]{WilhelmEtal2000}
Wilhelm, K., Dammasch, I.E., Marsch, E., Hassler, D.M., 2000, ``On the source
  regions of the fast solar wind in polar coronal holes'', {\it Astron.
  Astrophys.\/}, {\bf 353}, 749--756. \newline ADS:
  \url{http://adsabs.harvard.edu/abs/2000A&A...353..749W}

\bibitem[Wilhelm {\it et~al.\/}(2004)]{WilhelmEtal2004}
Wilhelm, K., Dwivedi, B.N., Marsch, E., Feldman, U., 2004, ``Observations of
  the Sun at Vacuum-Ultraviolet Wavelengths from Space, I: Concepts and
  Instrumentation'', {\it Space Sci. Rev.\/}, {\bf 111}, 415--480. \newline
  ADS: \url{http://adsabs.harvard.edu/abs/2004SSRv..111..415W}

\bibitem[Wilhelm {\it et~al.\/}(2007)]{WilhelmEtal2007}
Wilhelm, K., Marsch, E., Dwivedi, B.N., Feldman, U., 2007, ``Observations of
  the Sun at Vacuum-Ultraviolet Wavelengths from Space, II: Results and
  Interpretations'', {\it Space Sci. Rev.\/}, {\bf 133}, 103--179. \newline
  ADS: \url{http://adsabs.harvard.edu/abs/2007SSRv..133..103W}

\bibitem[Withbroe(1988)]{Withbroe1988}
Withbroe, G.L., 1988, ``The temperature structure, mass, and energy flow in the
  corona and inner solar wind'', {\it Astrophys. J.\/}, {\bf 325}, 442--467.
  \newline ADS: \url{http://adsabs.harvard.edu/abs/1988ApJ...325..442W}

\bibitem[Withbroe {\it et~al.\/}(1982)]{WithbroeEtal1982}
Withbroe, G.L., Kohl, J.L., Weiser, H., Munro, R.H., 1982, ``Probing the solar
  wind acceleration region using spectroscopic techniques'', {\it Space Sci.
  Rev.\/}, {\bf 33}, 17--52. \newline ADS:
  \url{http://adsabs.harvard.edu/abs/1982SSRv...33...17W}

\bibitem[Woo(2005)]{Woo2005}
Woo, R., 2005, ``Relating White-Light Coronal Images to Magnetic Fields and
  Plasma Flow'', {\it Solar Phys.\/}, {\bf 231}, 71--85. \newline ADS:
  \url{http://adsabs.harvard.edu/abs/2005SoPh..231...71W}

\bibitem[Woo(2006)]{Woo2006}
Woo, R., 2006, ``Ultra-fine-Scale Filamentary Structures in the Outer Corona
  and the Solar Magnetic Field'', {\it Astrophys. J. Lett.\/}, {\bf 639},
  L95--L98. \newline ADS:
  \url{http://adsabs.harvard.edu/abs/2006ApJ...639L..95W}

\bibitem[Woo and Druckm\"{u}llerov\'{a}(2008)]{WooDruckmullerova2008}
Woo, R., Druckm\"{u}llerov\'{a}, H., 2008, ``Solar Eclipse Images and the Solar
  Wind'', {\it Astrophys. J. Lett.\/}, {\bf 678}, L149--L152. \newline ADS:
  \url{http://adsabs.harvard.edu/abs/2008ApJ...678L.149W}

\bibitem[Woo {\it et~al.\/}(1999)]{WooEtal1999}
Woo, R., Habbal, S.R., Howard, R.A., Korendyke, C.M., 1999, ``Extension of the
  Polar Coronal Hole Boundary into Interplanetary Space'', {\it Astrophys.
  J.\/}, {\bf 513}, 961--968. \newline ADS:
  \url{http://adsabs.harvard.edu/abs/1999ApJ...513..961W}

\bibitem[Woo {\it et~al.\/}(2004)]{WooEtal2004}
Woo, R., Habbal, S.R., Feldman, U., 2004, ``Role of Closed Magnetic Fields in
  Solar Wind Flow'', {\it Astrophys. J.\/}, {\bf 612}, 1171--1174. \newline
  ADS: \url{http://adsabs.harvard.edu/abs/2004ApJ...612.1171W}

\bibitem[Wood(2004)]{Wood2004}
Wood, B.E., 2004, ``Astrospheres and Solar-like Stellar Winds'', {\it Living
  Rev. Solar Phys.\/}, {\bf 1}, lrsp-2004-2. URL (cited on 1 July 2009):
  \newline\url{http://www.livingreviews.org/lrsp-2004-2}

\bibitem[Wood {\it et~al.\/}(1999)]{WoodEtal1999}
Wood, B.E., Karovska, M., Cook, J.W., Howard, R.A., Brueckner, G.E., 1999,
  ``Kinematic Measurements of Polar Jets Observed by the Large-Angle
  Spectrometric Coronagraph'', {\it Astrophys. J.\/}, {\bf 523}, 444--449.
  \newline ADS: \url{http://adsabs.harvard.edu/abs/1999ApJ...523..444W}

\bibitem[Wu and Yoon(2007)]{WuYoon2007}
Wu, C.S., Yoon, P.H., 2007, ``Proton heating via nonresonant scattering off
  intrinsic Alfv\'{e}nic turbulence'', {\it Phys. Rev. Lett.\/}, {\bf 99},
  075001. \newline ADS: \url{http://adsabs.harvard.edu/abs/2007PhRvL..99g5001W}

\bibitem[Wu and Yang(2006)]{WuYang2006}
Wu, D.J., Yang, L., 2006, ``Anisotropic and mass-dependent energization of
  heavy ions by kinetic Alfv\'{e}n waves'', {\it Astron. Astrophys.\/}, {\bf
  452}, L7--L10. \newline ADS:
  \url{http://adsabs.harvard.edu/abs/2006A&A...452L...7W}

\bibitem[Wu and Yang(2007)]{WuYang2007}
Wu, D.J., Yang, L., 2007, ``Nonlinear interaction of minor heavy ions with
  kinetic Alfv\'{e}n waves and their anisotropic energization in coronal
  holes'', {\it Astrophys. J.\/}, {\bf 659}, 1693--1701. \newline ADS:
  \url{http://adsabs.harvard.edu/abs/2007ApJ...659.1693W}

\bibitem[Yamauchi {\it et~al.\/}(2002)]{YamauchiEtal2002}
Yamauchi, Y., Suess, S.T., Sakurai, T., 2002, ``Relation between Pressure
  Balance Structures and polar plumes from Ulysses high latitude
  observations'', {\it Geophys. Res. Lett.\/}, {\bf 29}, 1383. \newline ADS:
  \url{http://adsabs.harvard.edu/abs/2002GeoRL..29j..21Y}

\bibitem[Yang {\it et~al.\/}(2009)]{YangEtal2009}
Yang, L.-H., Jiang, Y.-C., Ren, D.-B., 2009, ``Formation of Transient Coronal
  Holes during Eruption of a Quiescent Filament and its Overlying Sigmoid'',
  {\it Chinese J. Astron. Astrophys.\/}, accepted

\bibitem[Yoon and Fang(2008)]{YoonFang2008}
Yoon, P.H., Fang, T.-M., 2008, ``Parallel cascade of Alfv\'{e}n waves'', {\it
  Plasma Phys. Cont. Fusion\/}, {\bf 50}, 085007. \newline ADS:
  \url{http://adsabs.harvard.edu/abs/2008PPCF...50h5007Y}

\bibitem[Zangrilli {\it et~al.\/}(1999)]{ZangrilliEtal1999}
Zangrilli, L., Nicolosi, P., Poletto, G., Noci, G., Romoli, M., Kohl, J.L.,
  1999, ``Latitudinal properties of the Lyman alpha and O VI profiles in the
  extended solar corona'', {\it Astron. Astrophys.\/}, {\bf 342}, 592--600.
  \newline ADS: \url{http://adsabs.harvard.edu/abs/1999A&A...342..592Z}

\bibitem[Zhang {\it et~al.\/}(2003)]{ZhangEtal2003}
Zhang, J., Woch, J., Solanki, S.K., von Steiger, R., Forsyth, R., 2003,
  ``Interplanetary and solar surface properties of coronal holes observed
  during solar maximum'', {\it J. Geophys. Res.\/}, {\bf 108}, 1144. \newline
  ADS: \url{http://adsabs.harvard.edu/abs/2003JGRA..108.1144Z}

\bibitem[Zhang {\it et~al.\/}(2007)]{ZhangEtal2007}
Zhang, J., Richardson, I.G., Webb, D.F., Gopalswamy, N., Huttunen, E., Kasper,
  J.C., Nitta, N.V., Poomvises, W., Thompson, B.J., Wu, C.-C., Yashiro, S.,
  Zhukov, A.N., 2007, ``Solar and interplanetary sources of major geomagnetic
  storms (Dst $\leq$ --100 nT) during 1996--2005'', {\it J. Geophys. Res.\/},
  {\bf 112}, A10\,102. \newline ADS:
  \url{http://adsabs.harvard.edu/abs/2007JGRA..11210102Z}

\bibitem[Zhang(2003)]{Zhang2003}
Zhang, T.X., 2003, ``Preferential heating of particles by H-cyclotron waves
  generated by a global magnetohydrodynamic mode in solar coronal holes'', {\it
  Astrophys. J. Lett.\/}, {\bf 597}, L69--L72. \newline ADS:
  \url{http://adsabs.harvard.edu/abs/2003ApJ...597L..69Z}

\bibitem[Zhou and Matthaeus(1990)]{ZhouMatthaeus1990}
Zhou, Y., Matthaeus, W.H., 1990, ``Transport and turbulence modeling of solar
  wind fluctuations'', {\it J. Geophys. Res.\/}, {\bf 95}, 10\,291--10\,311.
  \newline ADS: \url{http://adsabs.harvard.edu/abs/1990JGR....9510291Z}

\bibitem[Zirin(1975)]{Zirin1975}
Zirin, H., 1975, ``The helium chromosphere, coronal holes, and stellar
  X-rays'', {\it Astrophys. J. Lett.\/}, {\bf 199}, L63--L66. \newline ADS:
  \url{http://adsabs.harvard.edu/abs/1975ApJ...199L..63Z}

\bibitem[Zirker(1977)]{Zirker1977}
Zirker, J.B. (Ed.), 1977, {\it Coronal Holes and High-speed Wind Streams\/},
  Colorado Assoc.\ Univ.\ Press, Boulder

\bibitem[Zurbuchen {\it et~al.\/}(2002)]{ZurbuchenEtal2002}
Zurbuchen, T.H., Fisk, L.A., Gloeckler, G., von Steiger, R., 2002, ``The solar
  wind composition throughout the solar cycle: A continuum of dynamic states'',
  {\it Geophys. Res. Lett.\/}, {\bf 29}, 1352. \newline ADS:
  \url{http://adsabs.harvard.edu/abs/2002GeoRL..29i..66Z}

\end{thebibliography}

\end{document}